\pdfoutput=1 
\documentclass[useAMS,usenatbib]{mn2e}
\usepackage{graphicx}
\usepackage{grffile}
\usepackage{amssymb}
\usepackage{amsmath}
\usepackage{natbib}
\usepackage[maxfloats=120]{morefloats}
\usepackage{ifpdf}
\title[Extended H$_2$ emission from MYSOs.]{A Survey of Extended H$_2$ Emission from Massive YSOs.}
\author[F. Navarete et al.]{F. Navarete$^{1}$\thanks{E-mail: navarete@usp.br (FN); damineli@iag.usp.br (AD)}, A. Damineli$^{1}$\footnotemark[1], C. L. Barbosa$^{2}$ and R. D. Blum$^{3}$\\
$^{1}$IAG-USP, Rua do Mat\~ao, 1226, 05508-900, S\~ao Paulo, SP Brazil\\
$^{2}$MCTI/Laborat\'orio Nacional de Astrof\'isica, Rua Estados Unidos 154, CEP 37504-364, Itajub\'a, MG, Brazil\\
$^{3}$NOAO, 950 N Cherry Ave., Tuczon, AZ 85719 USA}
\begin{document}

\date{Accepted 2015 April 22. Received 2015 April 22; in original form 2014 October 28}

\pagerange{\pageref{firstpage}--\pageref{lastpage}} \pubyear{2015}

\maketitle

\label{firstpage}

\begin{abstract}
    We present the results from a survey, designed to investigate the accretion process of massive young stellar objects (MYSOs) through near infrared narrow band imaging using the H$_2$ $\nu$=1-0 S(1) transition filter. 
    A sample of 353 Massive Young Stellar Object (MYSO) candidates was selected from the Red MSX Source survey using photometric criteria at longer wavelengths (infrared and submillimeter) and chosen with positions throughout the Galactic Plane.
    Our survey was carried out at the SOAR Telescope in Chile and CFHT in Hawaii covering both hemispheres.
    The data reveal that extended H$_2$ emission is a good tracer of outflow activity, which is a signpost of accretion process on young massive stars.
    Almost half of the sample exhibit extended H$_2$ emission and 74 sources (21\%) have polar morphology, suggesting collimated outflows.
    The polar-like structures are more likely to appear on radio-quiet sources, indicating these structures occur during the pre-UCH\,{\sc{ii}} phase.
    We also found an important fraction of sources associated with fluorescent H$_2$ diffuse emission that could be due to a more evolved phase.
    The images also indicate only $\sim$23\% (80) of the sample is associated with extant (young) stellar clusters.
    These results support the scenario in which massive stars are formed by accretion disks, since the merging of low mass stars would not produce outflow structures.
\end{abstract}


\section{Introduction}

    The scenario of formation of low mass stars (M $<$ 8 M$_\odot$), through disk accretion controlled by magnetic field is already  well understood \citep{Shu87}.
    For high masses (M $>$ 8 M$_\odot$), however, there are theoretical problems for applying the same scenario.
    From the observational point of view, there are too few bona fide candidates to test the models.
    This is due to the fact that massive young stellar objects (MYSOs) are short lived and occur inside large molecular clouds, subject to huge reddening.
    The relevant phases of massive stars formation are not seen directly at wavelengths shorter than the mid-infrared (MIR) and also to wavelengths longer than radio-mm.
    Shocked molecular outflows are ``smoking guns'' of the accretion process and can be traced outside the central heavily obscured zone.
    The aim of this work is to use the H$_2$ $\nu$=1-0 S(1) emission to constrain the possible models.
    
    Even after decades of progress both theoretical and observational fronts, formation of massive stars remains an open question.
    The short timescale for MYSO formation indicates that high rates of mass accretion are required.
    Two scenarios currently dominate the discussions,
    $a)$ in the first, high mass stars are formed by accretion through a disk \citep{Jijina96,Krumholz05};
    $b)$ in the second, high mass stars are formed via coalescence of low mass stars \citep{Bonnell98,Bonnell01,Bally05}.
    In the last decade, a number of observations have been presented favoring the first scenario, among them \citet{Bik04} and \citet{Blum04}. 

    Low and intermediate mass stars are formed by the gravitational collapse of the parental giant molecular cloud (GMC), followed by the accretion process \citep{Palla96}.
    During the accretion phase, material is ejected as well via collimated bipolar jets.
    However, when a YSO reaches 8 M$_\odot$, the radiative flux becomes so intense (using $\phi = L / 4 \pi d^2$, the ratio between the radiative fluxes of an O5 and a B3 star -- masses of $\sim 40$ and $\sim 8$ M$_\odot$, respectively -- is $\approx 250$) that it may interrupt the accretion flow.
    A process that constrains the outcoming radiation field to narrower angles may leave some room for the accretion process to continues in some directions. This seems to be the case for the outflows driven by young stars from a very broad mass range, as previous reported by several authors \citep{Shepherd96a,Bachiller96,Bontemps96,Beuther02,Wu04}.
    Outflows associated with high-mass objects are expected to be more energetic than the outflows observed in lower mass YSOs \citep{Beuther05,Zhang05,Lopez09}, with velocities greater than $\sim 100$ km s$^{-1}$ \citep{Marti98}. Some authors have found evidences that outflows associated with massive stars are scaled up versions of their low mass counterparts \citep{Vaidya11,Codella13} while other works have reported that no well-collimated outflows have been found toward MYSOs \citep{Shepherd03, Sollins04}. 
    Massive YSO outflows mapped in high-velocity CO lines have collimation factors R = length/width $\sim$ 2.05$\pm$0.96 as compared to R $\sim$ 2.81$\pm$2.16 for low mass stars \citep{Wu04}, indicating a weak tendency that outflows associated with massive stars are less collimated than those from low mass stars as previously thought \citep{Richer00}.
    Besides the degree of collimation, these massive outflows would work removing mass from the plane of the accretion disk, lowering the density on the plane and, therefore, facilitating the accretion flow to reach the stellar core as shown in the recent 3-D simulations presented by \citet{Krumholz09}. Although these authors have not included the outflow activity on their simulations, they argue that the presence of outflows would decrease the star formation efficiency from 70\% (considering purely radiation effects) to 50\%.

    In the second scenario, massive stars are formed by collision and coalescence of low mass stars in dense clusters \citep{Bonnell98,Stahler00}.
    The low mass stars are formed after the gravitational collapse of the GMC and through an accretion disk, interact with each other, colliding to form stars with greater masses.
    In this case, the massive stars would be formed in the highest density parts of clusters, and the low mass stars would be the dominant fraction of the entire cluster mass \citep{Bonnell98,Bally02}.
    Moreover, no massive stars would be expected in small stellar clusters.

    While there is no observational evidence for stellar merges in clusters, a growing number of observations supports the accretion disk scenario by answering questions such as the duration of accretion period \citep{Zhang05,Beuther02}, the required accretion rates to overcome the radiation pressure \citep{Kim06} and the confirmation of molecular jets ejected from these massive objects \citep{Bally05}.
    The existence of outflows driven by stars from a large mass range is a potentially interesting aspect, since stellar mergers cannot produce stable outflows; therefore, the observation of such flows can be used to discern between the two scenarios of massive star formation.

    Bipolar molecular jets associated with MYSO candidates indicate that massive stars are formed by accretion. However, previous samples were contaminated by low mass objects, mostly due to inaccurate photometry.
    Recently, MYSO candidates were observed in the H$_2$ $\nu$=1-0 S(1) transition and collimated jets were identified, suggesting accreting discs around these objects \citep{Varricatt10}.
    Although this work presents observational evidence for the accretion scenario, few MYSOs were confirmed in that sample. 
    
    To identify bonafide MYSO candidates, accurate photometry and well developed selection criteria are necessary to avoid contamination by normal H\,{\sc{ii}} regions, proto-planetary nebulae (PPNe) and cool stars.
    The Red MSX Source (RMS) survey \citep{Lumsden02,Mottram11a} contains a large list of objects measured in 2MASS, MSX, Spitzer, and reprocessed IRAS/IGAL images for objects along the Galactic plane, including ultra-compact (UC) H\,{\sc{ii}} regions and a large sample of MYSOs.
    Additional observations were used to separate the MYSOs from UCH\,{\sc{ii}}s, such as water masers \citep{Urquhart09b,Urquhart11a} and centimetric emission \citep{Urquhart07a,Urquhart09a}.
    Recently, the bolometric fluxes of those sources were derived by \citet{Mottram11a} using IR and submillimetric data and adopting the kinematic distances derived by CO observations from \citet{Urquhart07b,Urquhart08} and a spectral energy distribution (SED) model.
    They found there are few MYSOs ($\sim 20$) with L $\geq$ 10$^5$ L$_\odot$, and many MYSOs ($\sim 150$) with 10$^4$ L$_\odot$ $\leq$ L $\leq$ 10$^5$ L$_\odot$.

    There are a few examples of well documented accretion disks around massive (10-30 M$_\odot$) forming stars \citep{Davies10,Murakawa13} using keplerian motion in the CO lines.
    Disks have been identified in a few cases \citep{Bik04,Blum04} through profile fitting of emission features seen in high spectral resolution data, however, the accretion process might have already ceased and the disks are now passively evolving.
    Although the scenario of an accretion disk may apply for all massive stars, the details are lacking.
    Instead of doing a detailed study of a small number of potential candidates that might harbor a disk, we are moving toward a large statistical study which will point to accretion signatures (or not) of a well selected sample of MYSOs, selected by the RMS survey.

    A smaller survey based on the present motivation was conducted by \citet{Varricatt10}. These authors found that $76\%$ of the targets display H$_2$ emission, which is indicative of shocks, and in $50\%$ of the sample the H$_2$ features have some symmetry, suggesting collimated outflows and so accretion disks.
    Such signatures are more frequent in pre-UCH\,{\sc{ii}} than in UCH\,{\sc{ii}} phase.
    The most luminous YSOs in that sample correspond to early B or late O star progenitors.
    The lack of more massive stars may be due to several causes such as: insufficient number of targets (50); underestimation of luminosities due to errors in the kinematic distances or in the method of magnitude extraction.
    Problems with the kinematic distances of star forming regions have been pointed out by authors who derive spectrophotometric and trigonometric distances \citep[][and references therein]{Hanson97,Navarete11,Moises11}.
    Most non--kinematic distances are half the magnitude of the kinematic ones. As discussed in \citet{Moises11}, the kinematic distances are larger than their non-kinematic counterparts mostly due to two reasons: a) breaking the ambiguity in the far/near distance is not always a reliable procedure for the kinematic distance method, b) in directions close to the Galactic Center, where radial velocities are small, local perturbations in the gas produces velocity components of the same magnitude as the Galactic rotation. 
    If this result is reliable, than the luminosity of those stars would be $\approx$ 4 times smaller than predicted by the original data.
    Unfortunately, MYSOs in pre-UCH\,{\sc{ii}} phases have not been reported for massive star clusters, where spectrophotometric distances are accurate.

    In this paper, we present the results of our survey, which is organized as follows. In section 2, the observations and data reduction are described.
    In section 3 we present overall results from the survey. In section 4 we discuss the results. Finally, our conclusions are summarized in section 5.
    The images of the sources presenting extended H$_2$ emission are shown in the Appendix.    For those sources that do not display H$_2$ emission, the images are found in the Vizier Catalogue.

\section{Observations and Data Reduction}
\label{observations}

    The sample of 353 MYSO candidates comprises 135 sources located in the Northern hemisphere ($\delta>0^\circ$) and 219 from the Southern hemisphere ($\delta<0^\circ$).
    Most of the Northern subsample was observed with the Canada-France-Hawaii Telescope (CFHT) and the Southern subsample was observed using the Southern Astrophysical Research Telescope (SOAR).
    Details of the observing setup used on each telescope are described below.

    \subsection{Northern Sample}
    \label{obs_north}
    The data were obtained with the CFHT, Mauna Kea, Hawaii, using the Wide-field InfraRed Camera (WIRCam) between 2010 August and 2012 August.
    The WIRCam detector is a mosaic of four 2048 $\times$ 2048 Rockwell Scientific Hawaii-2RG HgCdTe arrays, with a $45\arcsec$ gap between detectors.
    The scale is $0.3\arcsec$/pixel so that a single exposure covers a field of view (FOV) of 21 $\times$ 21 arcmin$^2$. 
    The median seeing was $0.7\pm0.2\arcsec$ and the airmass of each pointing was $1.3\pm0.2$. WIRCam is described by \citet{Puget04}.

    The sample was observed through the H$_2$ ($\lambda = 2.122\,\mu$m, $\Delta\lambda = 0.032\,\mu$m) and {\it K}-band continuum ($\lambda = 2.218\,\mu$m, $\Delta\lambda = 0.033\,\mu$m) narrow band filters.
    Each target was observed using a four point dither pattern, defining a $60\arcsec \times 60\arcsec$ square on the sky, with exposure time of $t_{exp}=150$ seconds per frame, totalling 600 seconds per filter.
    The dither pattern was employed to cover the 45$\arcsec$ gap between the detectors and also to facilitate the identification and the suppression of bad pixels and cosmic rays.
    Using the observational setup described above, we reached an average 3$\sigma$ sensitivity of 19.7$\pm$0.5\,mag and 19.1$\pm$0.2\,mag on point sources in the H$_2$ and continuum filters, respectively.
    The equivalent 3$\sigma$ sensitivity per pixel is $3.1\times10^{-18}$\, W\,m$^{-2}$\,arcsec$^2$ and $5.9\times10^{-18}$\,W\,m$^{-2}$\,arcsec$^2$ for the H$_2$ and continuum filters, respectively.

    \subsection{Southern Sample}
    \label{obs_south}
    The Southern sample have been observed with the Southern Astrophysical Research Telescope (SOAR), Cerro Pach\'on, Chile, using the Ohio State Imager/Spectrometer (OSIRIS) camera between 2011 March and 2011 June and the Spartan camera between 2011 August and 2012 December.
    The OSIRIS camera has a 1024 $\times$ 1024 Rockewell Scientific Hawaii HgCdTe detector with a pixel scale of 0.139$\arcsec$. A single exposure covers a FOV of approximately 1.5 $\times$ 1.5\,arcmin$^2$. OSIRIS is described by \citet{Depoy93}.
    The narrow-band H$_2$ ($\lambda_c = 2.116\,\mu$m, $\Delta\lambda = 0.035\,\mu$m) and continuum ($\lambda_c = 2.139\,\mu$m, $\Delta\lambda = 0.050\,\mu$m) filters were used for the OSIRIS observations.
    The data taken with OSIRIS were observed using a three point dither pattern, separated by 20 arcseconds in the E-W direction, with exposure times of 180 s per frame, totaling 540 s per filter.
    Using the observational setup for OSIRIS, we reached an average 3$\sigma$ sensitivity of 16.8$\pm$0.6\,mag and 16.5$\pm$0.3\,mag on point sources in the narrow H$_2$ and continuum filters, respectively.
    The equivalent 3$\sigma$ sensitivity per pixel is $6.8\times10^{-17}$\,W\,m$^{-2}$\,arcsec$^2$ and $8.6\times10^{-17}$\,W\,m$^{-2}$\,arcsec$^2$ for the H$_2$ and continuum filters, respectively.

    The Spartan Infrared Camera \citep{Loh12} has four Hawaii--II arrays deployed in a 2 $\times$ 2 mosaic. The pixel scale is 0.067$\arcsec$  such that a single exposure covers a FOV of approximately 5 $\times$ 5\,arcmin$^2$.
    The narrow-band H$_2$ ($\lambda_c = 2.117\,\mu$m, $\Delta\lambda = 0.031\,\mu$m) and continuum filters from the Spartan's filter set ($\lambda_c = 2.140\,\mu$m, $\Delta\lambda = 0.030\,\mu$m) were used.
    The targets observed with Spartan used a four point dither pattern, defining a $60\arcsec \times 60\arcsec$ square on the sky, with exposure time of 150 seconds per frame, totaling 600 seconds per filter.
    Using the observational setup described above, we reached an average 3$\sigma$ sensitivity of 16.7$\pm$0.5\,mag and 17.0$\pm$0.4\,mag on point sources in the narrow H$_2$ and continuum filters, respectively.
    The equivalent 3$\sigma$ sensitivity per pixel was $2.6\times10^{-17}$\,W\,m$^{-2}$\,arcsec$^2$ and $3.6\times10^{-17}$\,W\,m$^{-2}$\,arcsec$^2$ for the H$_2$ and continuum filters, respectively.
\\    
    The central coordinates of the first 20 objects in the complete sample are listed in Table \ref{table:observations} (the full version of this table containing 353 entries is available online at the CDS). 

    Each H$_2$ processed image (see below) was continuum-subtracted by an associated {\it K}-band continuum image scaled to the same effective bandwidth.
    The usage of narrow filters avoided contamination by nebular features such as Brackett-$\gamma$ that would be present in a broad-band filter and so might mask H$_2$ emission.

\begin{table*}
	\caption{Log of observations.}
	\begin{tabular}{clrrrrcc}
\hline\hline
ID	& MSX name	        &	RA (J2000)	    &	Dec (J2000)	    &	d$_{kin}$ (kpc)	&$\log(L/L_\odot)$&	Telescope	&	Semester	\\
\hline\hline
001	&	G016.7981+00.1264	&	18:20:55.3	&	-14:15:30.8	&	14.6	&	5.14	&	SOAR	&	2011A	\\
002	&	G017.0332+00.7476	&	18:19:07.3	&	-13:45:23.6	&	13.8	&	4.64	&	SOAR	&	2011A	\\
003	&	G017.6380+00.1566	&	18:22:26.4	&	-13:30:12.0	&	 2.3	&	4.77	&	SOAR	&	2011A	\\
004	&	G018.6608+00.0372	&	18:24:50.2	&	-12:39:22.4	&	10.8	&	4.29	&	CFHT	&	2012A	\\
005	&	G019.8817-00.5347	&	18:29:14.7	&	-11:50:23.6	&	 3.5	&	3.94	&	CFHT	&	2012A	\\
006	&	G019.9386-00.2079	&	18:28:09.9	&	-11:38:22.9	&	 4.5	&	3.35	&	CFHT	&	2012A	\\
007	&	G020.5143+00.4936	&	18:26:43.5	&	-10:48:18.5	&	13.6	&	4.10	&	CFHT	&	2012A	\\
008	&	G020.7491-00.0898	&	18:29:16.4	&	-10:52:01.2	&	11.7	&	4.95	&	SOAR	&	2011A	\\
009	&	G022.3554+00.0655	&	18:31:44.2	&	 -9:22:17.0	&	 5.2	&	4.23	&	CFHT	&	2012A	\\
010	&	G023.2628+00.0713	&	18:33:24.8	&	 -8:33:50.0	&	 4.9	&	3.88	&	CFHT	&	2012A	\\
011	&	G023.3891+00.1851	&	18:33:14.3	&	 -8:23:57.4	&	 4.8	&	4.67	&	SOAR	&	2011A	\\
012	&	G023.4394-00.2394	&	18:34:51.3	&	 -8:32:55.0	&	 6.0	&	4.36	&	CFHT	&	2012A	\\
013	&	G023.6566-00.1273	&	18:34:51.6	&	 -8:18:21.6	&	 5.0	&	4.17	&	CFHT	&	2012A	\\
014	&	G023.8176+00.3841	&	18:33:19.5	&	 -7:55:37.8	&	 4.8	&	3.62	&	CFHT	&	2012A	\\
015	&	G023.8983+00.0648	&	18:34:37.2	&	 -8:00:12.4	&	 3.0	&	3.86	&	CFHT	&	2012A	\\
016	&	G024.0946+00.4565	&	18:33:34.9	&	 -7:38:53.5	&	 5.6	&	3.50	&	CFHT	&	2012A	\\
017	&	G024.6343-00.3233	&	18:37:22.7	&	 -7:31:41.5	&	12.3	&	4.90	&	CFHT	&	2012A	\\
018	&	G024.7320+00.1530	&	18:35:50.9	&	 -7:13:27.2	&	 9.1	&	4.26	&	CFHT	&	2012A	\\
019	&	G025.4118+00.1052	&	18:37:17.0	&	 -6:38:25.4	&	 5.7	&	3.97	&	CFHT	&	2012A	\\
020	&	G025.6498+01.0491	&	18:34:20.9	&	 -5:59:42.5	&	12.3	&	5.50	&	SOAR	&	2011A	\\
\hline\hline
	\end{tabular}
	\label{table:observations}
	
{\bf Notes.} The first 20 rows of the table is presented. The full version containing 354 entries is available online at the CDS.
The columns are -- 1: source's ID; 2: MSX PSC name; 3: right ascension; 4: declination; 5: kinematic distance from \citet{Urquhart07b,Urquhart08}; 6: logarithm of the bolometric luminosity \citep[using the bolometric flux from ][ and the kinematic distance from column 5]{Mottram11a}; 7: telescope; and 8: semester.
\end{table*}

    \subsection{Data Reduction}
    \label{reduction}

    The images were processed using THELI, an instrument-independent pipeline for automated reduction of astronomical images \citep{Erben05,Schirmer13}. The processing is accomplished through a series of tasks, briefly described as follows. First, the images are dark-subtracted and flat-field corrected. Then a model of the sky background is subtracted from the images. To perform the sky subtraction, THELI masks the point-like sources, creates a first-step model for the sky contribution and subtracts it from the flat-corrected images. Then, the pipeline creates a second-step model to improve the sky subtraction.
    The astrometric solution for the observed field was determined by cross-matching the sources in each observed field with those from the 2MASS catalog \citep{Skrutskie06}. For a good astrometric solution, each chip requires a minimum of 80-100 matches. THELI uses SExtractor \citep{Bertin96} to extract the position of the point sources on the images and then runs Scamp \citep{Bertin06} to derive the astrometric solution and the distortion of the instrument's field-of-view (that is, the pixel scale variation along the FOV). The images were resampled using the astrometric solution determined by THELI and were median combined, creating final H$_2$ and Continuum images. The astrometric solution was fit as a second order polynomial for each instrument.
    The residuals to the astrometric fit were $\sim0.3$ pixel RMS for all the three instruments.
    
    Post-processing routines were written using IDL to create the images presented in this paper. The first one crops the final images to smaller ones centered on the source. Next, the H$_2$ images were continuum-subtracted in order to produce pure H$_2$ emission maps.
    
    Differences in the filter transmission curves and bandpasses and non-zero interstellar extinction mean a simple subtraction is expected to leave systematic residuals in the point sources and extended emission (where the continuum is strong, at least). In fact the Spartan filters have similar peak transmission and bandpass, so that no large affect is expected based on the filter profile. For WIRCam, the filters have similar peak transmission, but the continuum filter is wider (the ratio of bandpasses is 0.7).

    For the strategy employed here of using only one continuum filter to the redward side of the H$_2$ line is that the interstellar extinction will be slightly higher in the line filter than for the continuum. All other things being equal, and making no extinction correction, there should be a net reduction in the emission signal in a typical source due to extinction.
    Considering the central wavelength of each filter used in this work (described in Section \ref{observations}), and assuming the \citet{Stead09} reddening law ($\alpha = 2.14^{+0.05}_{-0.04}$), we have $A_{H_2}/A_{Cont} \approx 1.099\pm0.002$ and $1.0234^{+0.0006}_{-0.0004}$ for WIRCam and Spartan, respectively.With no bandpass correction to the WIRCam data, this effect would be enhanced. In practice, residuals were identified on several images.
    We found a scale factor ranging from 0.63 to 1.75 was needed so that the H$_2$ and continuum fluxes of the point sources would be approximately equal, minimizing residuals in the final H$_2$ maps. In the end, we simply minimized the residuals to get a best H$_2$ map for each source. Since our main goal is to use the morphological information from the maps, this is sufficient for our purposes. The typical scale factor divided into the continuum image to match the line image was 1.14 $\pm$ 0.10. This is roughly consistent with expectations from the considerations of bandpass and reddening.

\section{Results}
\label{results}

\begin{figure}
    \resizebox{0.95\linewidth}{!}{\includegraphics{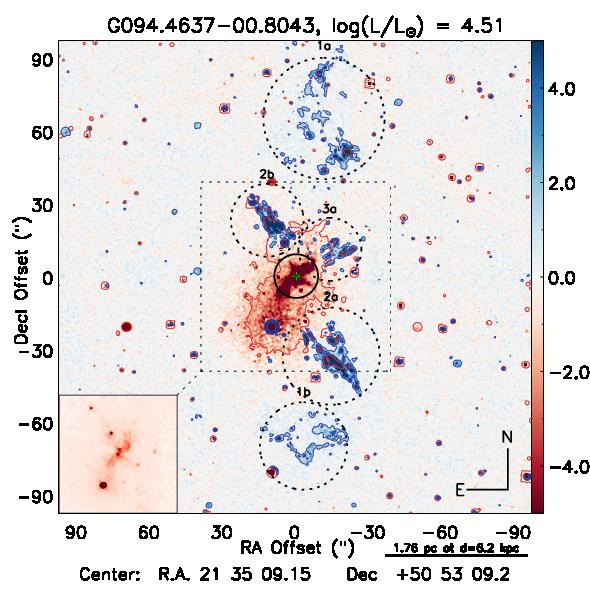}}
  \caption{Continuum-subtracted H$_2$ map of G094.4637-00.8043 scaled by the standard deviation of the continuum-subtracted background counts. Excess H$_2$ emission ($>$ 0) is shown in blue while excess in the continuum filter ($<$ 0) is shown in red. Radio quiet/loud RMS sources are represented by a green $\ast$ or a yellow $\diamond$, respectively (see text). The inset frame displays the details in the continuum filter of the region delimited by the dashed box in the main H$_2$ map. The inset frame displays the details of the region delimited by the dashed box in the continuum filter. For a complete description, see details in Section \ref{results}.}
  \label{figure:h2_map_example}
\end{figure}

The continuum-subtracted H$_2$ map of each source is presented in the Appendix. An illustrative example of one of these maps is shown in Figure \ref{figure:h2_map_example}.
The main frame corresponds to the continuum-subtracted H$_2$ map. The intensity of each pixel is shown as a multiple of the standard deviation of the continuum subtracted counts and a divergent color scale was used to highlight and increase the contrast of the extended emission: excess of H$_2$ emission ($> 0$) is shown in blue while excess of continuum ($< 0$) is shown in red. The colorbar at the right side of the main image indicates the scale adopted for the image, in units of $\sigma$ of the continuum subtracted counts. Each H$_2$ structure is labelled with a dashed circle or a line.
The position of the RMS source is indicated by the black circle and its diameter corresponds to the MSX resolution ($\sim 18\arcsec$). The coordinates of the MSX position are shown at the bottom of the main frame. We have adopted the green $\ast$ and yellow $\triangle$ symbols in order to distinguish between radio quiet or radio loud RMS sources. Radio loud sources are UCH\,{\sc{ii}} regions which harbor nascent O--type stars. MYSOs that are radio quiet must be extremely young if they are massive so that the UCH\,{\sc{ii}} region has not yet formed \citep[see][]{Churchwell02}. When available, the position of the 2MASS and IRAS counterparts of the RMS source are shown as black $\times$ and $\diamond$ symbols, respectively.
The secondary image located at one of the corners is shown in red and corresponds to the continuum image of the region delimited by the dashed box. It exhibits the point sources and continuum extended emission close to the MYSO at the center of the field.
Finally, we include a physical scale (in parsecs) based on the kinematic distance of the source (given in kpc). Both values are shown at the top of the horizontal ruler located at the bottom right side of the image. 

\subsection{Classification of Extended H$_2$ emission}
\label{class_h2}

\begin{figure}
    \resizebox{0.95\linewidth}{!}{\includegraphics{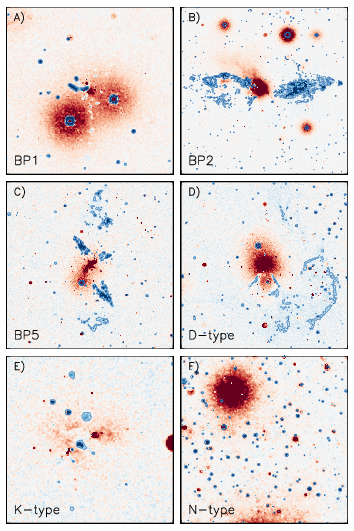}}
    \caption{Examples of the morphological classes based on the classification of the H$_2$ emission.
    The color scale of each image follows the same definition as Figure \ref{figure:h2_map_example}.
    The blue contours are placed at $\sigma=1.0$ and 3.0.
    Panels A: BP1-type; B: BP2-type; C: BP5-type; D: D-type; E: K-type; F: N-type.}
  \label{figure:h2_examples}
\end{figure}

The continuum-subtracted H$_2$ map of each source was analyzed and the emission morphology was classified according to the following classes (each type is illustrated in Figure \ref{figure:h2_examples}):
\begin{enumerate}
\item Polar emission (BPn): Extended H$_2$ emission likely to be polar and/or jet-like structures (the ``n'' corresponds to the number of polar structures associated: BP1 is monopolar emission, BP2 is bipolar emission and BP3 corresponds to a region with H$_2$ emission in three distinct areas/outflow lobes,  etc);
\item Knot emission (K): K-type morphology corresponds to those sources which display non-aligned knots of H$_2$ emission;
\item Diffuse emission (D): D-type sources are associated with diffuse or filamentary H$_2$ emission, possibly originating through fluorescent excitation of the H$_2$ molecules;
\item No emission (N): corresponds to non-detections.
\end{enumerate}

BP-type emission is expected to be more often associated with the central sources than K- or D-type ones. In addition, K-type emissions could
$i)$ correspond to the most brilliant region of a low surface brightness jet structure;
$ii)$ originate from the interaction of the radiation field with the interstellar medium surrounding the protostar or
$iii)$ be foreground structures on the line-of-sight of the MSX sources.

Each of the sources has received a primary classification regarding the four classes defined above, and in a few cases, a secondary classification was also provided.

\subsection{NIR Classification of the environment}
\label{class_nir}

The environment surrounding each RMS source was classified according to the following items:
\begin{enumerate}
\item presence of extended continuum emission; and 
\item association with a stellar cluster.
\end{enumerate}

Extended continuum emission could indicate free-free emission due to ionized gas or scattering by dust. The usage of a narrow band continuum filter could not resolve the origin of the positive detections, which requires additional observations in atomic transitions such as the Br$\gamma$ ($\lambda \approx 2.16\,\mu$m).
The association with stellar cluster candidates was based upon the identification of stellar overdensity around the position of each source. This was performed by analyzing the continuum image of each source ``by eye'' and distinguishing those which appears to be located within a group of stars (cases of sources \#099, \#104 and \#112, shown in Appendix A) and those which seems to be isolated sources (such like \#132, \#189 and \#287, also shown in Appendix A).

\subsection{Global Analysis of the Sample}
\label{main_results}

Each of the 353 sources were classified as described in the previous subsections and the results are summarized in Table \ref{table:structures}. The full version of this table containing 353 entries is available online at the CDS.

\begin{table*}
	\caption{General classification of the sources.}
    \begin{center}
	\begin{tabular}{cccccl}
\hline\hline
ID	&	Ext. Continuum	&	Stellar 	&	Primary H$_2$	&	Secondary H$_2$	&	Components	\\
	&	 Emission		&	Cluster	&	Classification	&	Classification		&				\\
\hline
001	&	Y	&	N	&	N	&	-	&	-	\\
002	&	N	&	N	&	BP2	&	D	&	BP2(1a,b), D(2a,b)	\\
003	&	Y	&	N	&	D	&	-	&	D(1)	\\
004	&	N	&	N	&	N	&	-	&	-	\\
005	&	N	&	N	&	N	&	-	&	-	\\
006	&	Y	&	N	&	BP2	&	K	&	BP2(1a-f),K(2,3,4)	\\
007	&	Y	&	N	&	K	&	-	&	K(1,2)	\\
008	&	N	&	N	&	N	&	-	&	-	\\
009	&	N	&	N	&	N	&	-	&	-	\\
010	&	N	&	N	&	N	&	-	&	-	\\
011	&	N	&	N	&	N	&	-	&	-	\\
012	&	N	&	N	&	N	&	-	&	-	\\
013	&	N	&	N	&	N	&	-	&	-	\\
014	&	N	&	N	&	N	&	-	&	-	\\
015	&	N	&	N	&	N	&	-	&	-	\\
016	&	N	&	N	&	N	&	-	&	-	\\
017	&	N	&	N	&	N	&	-	&	-	\\
018	&	Y	&	N	&	K	&	-	&	K(1,2)	\\
019	&	N	&	N	&	BP2	&	-	&	BP2(1a,b)	\\
020	&	N	&	N	&	N	&	-	&	-	\\
\hline\hline
	\end{tabular}
	\label{table:structures}
    \end{center}

	{\bf Notes.} First 20 rows of the table is presented. A full version containing 353 entries is available online at the CDS.
    The columns are -- 1: source's ID; 2: extended continuum emission (yes or no); 3: association with a stellar cluster (yes or no); 4: primary classification of the H$_2$ emission; 5: secondary (when given) classification of the H$_2$ emission; 6: structures classified in main components labeled by numbers (i.e., 1, 2, 3) and, when applicable, related components are given by numbers and letters (i.e., 1a, 1b). For example, a BP2(1a,b) indicates one bipolar structure with two components while a BP1(1a,b,c,d) means four distinct components within a single polar structure.
\end{table*}

\begin{figure}
    \resizebox{0.95\linewidth}{!}{\includegraphics{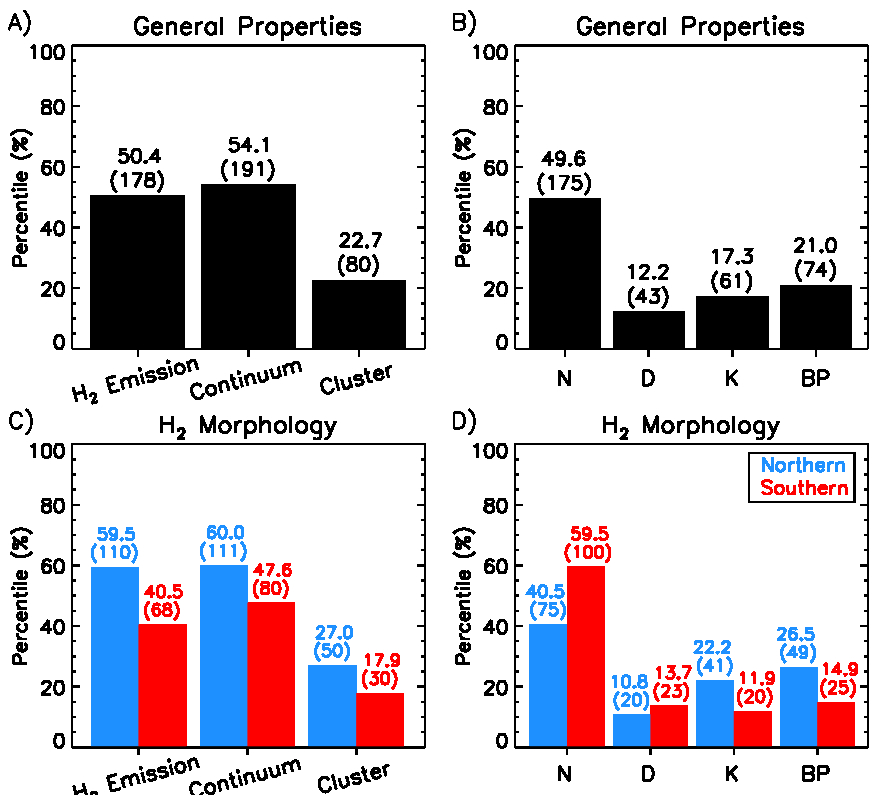}}
  \caption{Global properties of the sample. 
    Panel A: detection of extended H$_2$ and continuum emission and association with stellar clusters;
    Panel B: assigned morphology of the H$_2$ emission.
    Panels C and D: same as Panels A and B but shown for Northern (in blue) and Southern subsamples (in red) individually.
    The number of sources is shown in parenthesis at the top of each bar. Its corresponding percentile is shown right above it.}
  \label{figure:sample_statistics}
\end{figure}

The general statistics of the sample are shown in the Panel A from Figure \ref{figure:sample_statistics}.
We found extended H$_2$ emission towards $\approx50\%$ (178 sources) of the sample; $\sim54\%$ are associated with extended continuum emission (191); and only $\approx23\%$ of the sample (80) are apparently members of extant stellar clusters.
This result needs some further discussion.
Positive association occurs only for those stellar cluster candidates visible at near-infrared wavelengths. In some cases, the members of the cluster would not be visible or may not be formed yet. Therefore, is possible to say that, at least $\approx23\%$ of the sample have formed within high-density environments. 
The large number of cases of apparently isolated MYSOs ($\sim$ 77\%) suggests that massive stars require a formation mechanism that allows their formation even isolated from other stars. 

Panel B from Figure \ref{figure:sample_statistics} shows the morphological classification of the extended H$_2$ emission (defined in subsection \ref{class_h2}) as fractions of the total number of observed sources. The figure shows that $\approx49\%$ of the sample was classified as N-type objects (175 sources), about $\approx12\%$ of the sample displays diffuse H$_2$ emission (43), $\approx17\%$ of the sample was classified as K-type (61) and $\approx21\%$ of the sources (74) are associated with polar structures (monopolar, bipolar or multipolar) and, thus, classified as BP-type.
Sixty two (62) sources have received a secondary (or multiple secondary) H$_2$ classification (see Table \ref{table:structures}).

The formation of jets does not predict the existence of monopolar structures (BP1) driven by the young stars. BP1 is simply an observational classification.
In such cases, it is expected that a conjunction of factors (inclination angle, extinction, opacity of the interstellar medium [ISM] around the protostar) is obscuring the red counterpart of the observed structure.
On the other hand, the existence of multipolar structures could be due to the presence of multiple stars and/or a precessing central source with irregular or non-uniform accretion rates.
There are 10 BP1, 51 BP2 nine BP4, and one each of BP3, BP5, BP6 and BP8 sources among the total sample of BP-sources.

The magnitude limit of WIRCam observations is $\sim$ 19.5 mag (see section \ref{obs_north}) for both H$_2$ and continuum filters, and the magnitude limit for Spartan/OSIRIS is approximately equal to $\sim$ 16.7 mag (see subsection \ref{obs_south}). The sample can be divided into two subsamples: a northern subsample, containing 185 sources observed with WIRCam; and a Southern one, containing 169 targets observed with Spartan (125) and OSIRIS (43).
Panels C and D from Figure \ref{figure:sample_statistics} show the previous results but as a function of the number of targets from each subsample. Detection of H$_2$ emission corresponds to $\sim59.5\pm5.7\%$ of the Northern subsample and $\sim40.5\pm4.9\%$ of the Southern one (the uncertainties correspond to a random distribution error).
Comparing each H$_2$ emission type shown in Panel D, the Southern subsample has fewer K- and BP-type sources, but shows a similar fraction of D-type ones as compared to the Northern subsample.

These differences in subsamples do not appear significant, but several factors may explain them. The filter sets in WIRCam and Spartan/OSIRIS are fundamentally different. WIRCam uses wide, well separated filters, while the Spartan/OSIRIS filters are narrower but closer together (in fact they overlap to some extent).
For WIRCam, the filter transmission curve for the H$_2$ filter shows there might be some non-zero transmission in the red wing at the wavelength of Br$\gamma$. This suggests some strong Br$\gamma$ sources could produce false H$_2$ detections.
For Spartan/OSIRIS the concern is a strong Br$\gamma$ line in the wing of the continuum filter might cause a faint H$_2$ source to be undetected (due to over subtraction).
The other possibility is due to the flux limitations of the southern observations.
It is possible that we are unable to detect some faint H$_2$ emission in the south sample due to the flux limit of these images.
While the differences in the number and type of H$_2$ detections may be significant (2.8$\sigma$), we do not believe the difference in frequency between the two samples changes our main conclusions about the character of the star formation process.

\begin{figure}
    \resizebox{0.95\linewidth}{!}{\includegraphics{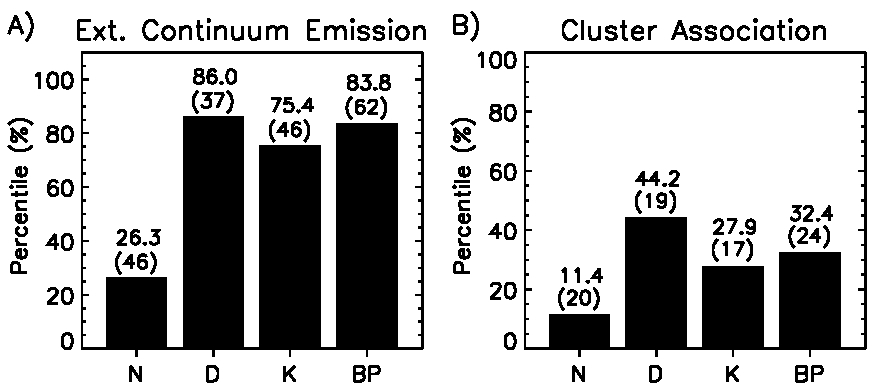}}
    \caption{Global properties of the sample as a function of the H$_2$ morphology.
    Each source was classified by the presence (or absence) of extended continuum emission (Panel A) and association with stellar clusters (Panel B).
    The bars correspond to the percentile of the sources classified as N-, D-, K- and BP-type, respectively (see text). 
    The number of sources is shown in parenthesis at the top of each bar. Its corresponding percentile is shown right above it.}
    \label{figure:sample_statistics_h2morph}
\end{figure}

Figure \ref{figure:sample_statistics_h2morph} summarizes the morphological classification defined in subsection \ref{class_nir} (N, D, K and BP).
Panel A displays a correlation between the sources associated with BP-, K- and D-type extended H$_2$ emission with those which display extended continuum emission (roughly $\sim80\%$ compared to $\sim26\%$ found for N-type sources).

This correlation with extended continuum emission could be due to two processes: $i)$ nebular emission or $ii)$ dust-scattered radiation.
The present observations do not allow us to resolve the nature of the continuum emission. Additional observations of atomic transitions (such as Brackett-$\gamma$ in $2.16\,\mu$m) could trace the gaseous emission associated with such regions.

Panel B from Figure \ref{figure:sample_statistics_h2morph} shows that most stellar clusters with MYSOs exhibit H$_2$ emission. 
Comparing stellar clusters and associated extended continuum emission, most of the sources associated with clusters also exhibit extended continuum emission (61 of 80). While clusters appear to be more likely to exhibit continuum emission, there is no preference for isolated MYSOs: 47\% of the sources that do not belong to stellar clusters exhibit extended continuum emission (139 of 274). 

\subsection{Radio quiet/loud phases and H$_2$ morphology}
The RMS sources were observed at 6\,cm by \citet{Urquhart07a,Urquhart09a} and were classified by those authors as radio quiet (YSO) or radio loud (H\,{\sc{ii}}) sources.  
Our sample consists of 282 radio quiet sources and 72 radio loud ones. 
The fraction of the radio loud sources is smaller than the radio quiet one, which may introduce a bias against UCH\,{\sc{ii}}s. Besides of that fact, the comparison between their NIR properties could provide useful information of the intrinsic properties of each phase.

\begin{figure}
    \resizebox{0.95\linewidth}{!}{\includegraphics{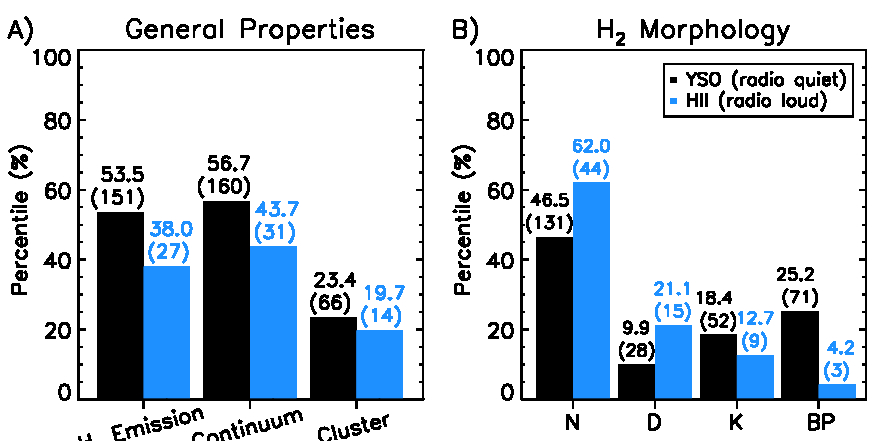}}
  \caption{Same data from Figure \ref{figure:sample_statistics} but separated into radio quiet (YSO, in black) and radio loud (H\,{\sc{ii}}, in blue) subsamples.
    The number of sources is shown in parenthesis at the top of each bar. Its corresponding percentile is shown right above it.}
  \label{figure:sample_statistics_radio}
\end{figure}

Figure \ref{figure:sample_statistics_radio} displays the global results of the survey as a function of the radio emission.
Panel A shows that the fraction of radio quiet sources (shown as black bars) associated with both extended H$_2$ and continuum emission is $\sim10\%$ larger than the values found for radio loud ones (shown in blue).
The fraction of radio quiet or radio loud sources associated with stellar clusters is almost the same ($\sim20\%$).
Panel B indicates that most of radio loud sources were classified as N-type sources (44 from 72). Comparing the other three classifications, the fraction of radio loud sources decreases from D-type ($\sim 21\%$) to K- ($\sim 13\%$) and BP-type sources ($\sim 4\%$). On the other hand, the fraction of radio quiet sources increases in the opposite direction: there is a minimum for D-type ($\sim 10\%$), and an increase for K- ($\sim18\%$) and BP-type ones ($\sim25\%$).
This result suggests that polar H$_2$ emission is found preferentially towards radio quiet sources (MYSOs) while diffuse H$_2$ emission is more often associated with radio loud ones, corresponding to more evolved phases (pre-UCH\,{\sc{ii}}).

\subsection{Luminosity vs. Distance diagrams}

\begin{figure}
    \resizebox{0.95\linewidth}{!}{\includegraphics{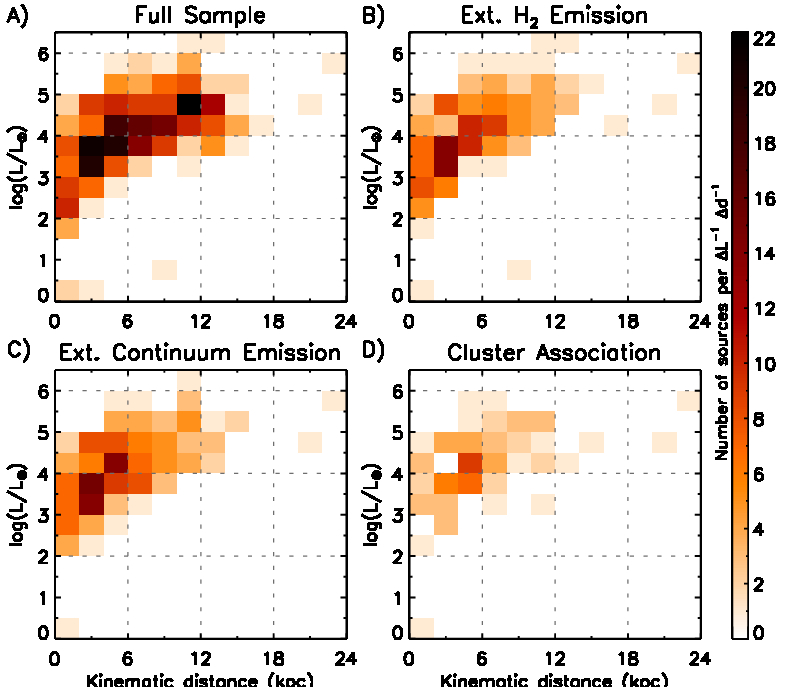}}
    \caption{Global properties and results of the survey as a function of luminosity vs. distance.
    Each `pixel' corresponds to a bin of $\Delta\log(L/L_\odot)=0.5$, $\Delta d = 2.0$\,kpc.
    The dashed lines are placed at $d=6$, 12 and 18\,kpc and $\log(L/L_\odot)=2.0$, 4.0 and 6.0.
    The colorbar at the right corresponds to the number of sources per each bin.
    Panel A: full sample;
    Panel B: subsample showing extended H$_2$ emission;
    Panel C: subsample showing extended continuum emission;
    Panel D: subsample associated with extant stellar clusters.}
    \label{figure:sample_analysis}
\end{figure}

Figure \ref{figure:sample_analysis} presents the distribution of MYSOs as a function of luminosity and distance.
The diagrams were constructed using a luminosity and kinematic distance binning of $\Delta\log(L/L_\odot)=0.5$ and $\Delta d = 2.0$\,kpc, respectively.
Since the RMS sources were selected by their infrared colors, fainter targets at larger distances may not be included in the final sample (our sample is flux limited). This is consistent with the lack of low luminosity sources beyond distances of $\sim$10 kpc.
The full sample distribution is shown in Panel A. Most of the sources are mid- to high-luminosity objects with $3.0 \lesssim \log(L/L_\odot) \lesssim 6.0$, located at $d \lesssim 14$\,kpc.
There are two overdensity regions at $i)$ $\log(L/L_\odot) \sim 3.5$ and $ d \sim 6.0$\,kpc; and $ii)$ $\log(L/L_\odot) \sim 5.0$ and $ d \sim 12.0$\,kpc.
This is consistent with the low star formation rate in the Solar neighborhood that comprises mainly low-mass stars \citep[the local IMF, $\xi(m)\propto m^\alpha$, is well constrained by assuming $\alpha \sim 1.3$ for m $<$ 0.5 M$_\odot$ and $\alpha \sim 2.2$ for m $>$ 0.5 M$_\odot$,][]{Kroupa07}.
Panel B displays the detections of extended H$_2$ emission. It peaks at mid- to high-luminosity sources ($3.0 \lesssim \log(L/L_\odot) \lesssim 4.5$) located at distances $ d \sim 3$\,kpc, consistent with the peak $i)$ from Panel A. The distribution extends up to the most luminous ($\log(L/L_\odot)\sim6.5$) and far sources ($d > 24$\,kpc) of the sample. The positive detections are mostly found toward sources with $d < 10$\,kpc (the second peak from Panel A was not identified in Panel B), indicating that the H$_2$ emission is detected for the most luminous sources of the sample ($\log(L/L_\odot) \sim 6.0$), although it is preferably found for less luminous ones located at closer distances.
The distribution of sources associated with extended continuum emission is shown in Panel C. It depicts a similar profile as that for the extended H$_2$ emission (Panel B).
Panel D presents the distribution of the sources associated with stellar clusters. The plot peaks at mid- to high-luminosity sources ($3.5 \lesssim \log(L/L_\odot) \lesssim 5.0$) located at $d \lesssim 6$\,kpc. This is not surprising, since most known clusters lie between the sun and Galactic center \citep{Moises11}.

\begin{figure}
    \centering
    \resizebox{0.95\linewidth}{!}{\includegraphics{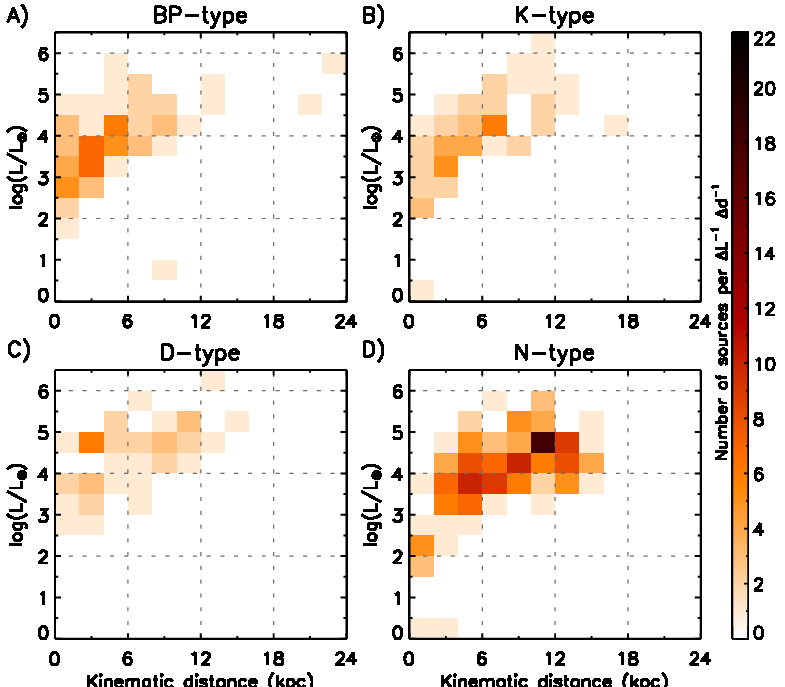}}
    \caption{Two-dimensional histograms characterizing the distribution for each H$_2$ morphological class.
    Each `pixel' corresponds to a bin of $\Delta\log(L/L_\odot)=0.5$, $\Delta d = 2.0$\,kpc.
    The colorbar at the right corresponds to the number of sources per each bin. 
    Panels A, B, C and D present the distribution for BP-, K-, D- and N-type sources, respectively.}
    \label{figure:sample_analysis_morphology}
\end{figure}

Figure \ref{figure:sample_analysis_morphology} displays the distribution for each H$_2$ morphological class defined in Section \ref{class_h2} in the luminosity vs. distance diagram.
Panel A shows that the most of the BP-type sources correspond to sources with $2.5 \lesssim \log(L/L_\odot) \lesssim 4.5$ within distances less than $d \sim 10$\,kpc. The maximum frequency of polar structures peaks at sources with $3.0 \lesssim \log(L/L_\odot) \lesssim 4.0$ located at $d \sim 4$\,kpc. This plot indicates that jet-like structures are associated with the low luminosity young stars, but not with the most luminous MYSOs. 
The K-type sources (Panel B) display a broader distribution towards high-luminosity sources and peak at high-luminosity sources ($\log(L/L_\odot) \sim 4.0$) located at distances closer than $d \lesssim 8$\,kpc. 
The distribution of D-type sources (Panel C) is more restricted in the luminosity and distance ranges as compared to the BP- and K-type sources. The diffuse H$_2$ emission is associated with relatively closer ( d $\lesssim$ 14\,kpc) mid- to high-luminosity sources [$\log(L/L_\odot) \gtrsim 2.5$]. That is, the radiative flux of the star must be sufficiently strong in order to generate diffuse emission. Diffuse emission has a lower surface brightness when compared to the knots or polar structures and it cannot be detected at larger distances.
The non-detections distribution is shown in Panel D. This is the complementary plot of the one shown in Panel B from Figure \ref{figure:sample_analysis}.
It indicates N-type sources peak towards high-luminosity sources at d $\sim$ 12 kpc.  Besides the peak toward larger distances, there are no significant differences between the distribution of detections and non-detections.

Finally, our data is not sufficient to prove the real nature of the increase of N-type sources towards larger distance when compared to other H$_2$ classes. It could be due to the fact that $i)$ N-type sources may correspond to a different kind of source or are related to a specific environment or stage of massive star formation, or even $ii)$ due to a systematic bias of the H$_2$ detection. It seems reasonable that N-type sources could include both radio quiet objects in the first stages of formation and even radio loud sources at the later stages of the pre-main sequence with no strong inflow/outflow activity.
The truly non-detections of H$_2$ extended emission may include farther sources associated with unresolved structures. Excluding those sources associated with wrong kinematic distance values and considering the largest pixel scale of our observations (WIRCam has a pixel size of 0.3 arcseconds, see Section \ref{observations}), it is possible to argue that we have clearly identified all the structures associated with the sources from our sample (located at distances up to $\sim$ 15 kpc) with projected lengths greater than $\sim$ 0.1 pc and brighter than the limiting sensitivity of our images (see details in Section \ref{observations}).

\subsection{Projected Length of the Structures}
The projected length ($\ell_{proj}$, in parsecs) of each structure was estimated as $\ell_{proj} \approx d \cdot \Delta\theta$, where $d$ is the kinematic distance and $\Delta\theta$ is the angular size of the structure, measured on the direction of the flow. 
Table \ref{table:struct_parameters} presents the properties of each H$_2$ component. It contains the projected length and the position angle (PA, measured in relation of the center of each field and rotation in degrees E of N) for every diffuse, knot or polar H$_2$ emission structure identified on the fields. The $\ell_{proj}$ values measured for H$_2$ knots correspond to the distance from the knot to the RMS source. The aspect ratio (R=length/width) of the polar features is presented in the last column of the table. In order to estimate the aspect ratio of each H$_2$ polar structure, their width were measured perpendicularly to the direction of the flow, corresponding to the maximum projected width of the structure.
The full version of this table containing 448 entries is available online at the CDS.

\begin{table}
	\caption{Properties of the H$_2$ structures identified on each field.}
	\begin{center}
	\begin{tabular}{ccccc}
\hline\hline
ID	&	Component	&	$\ell_{proj}$ (pc)	&	PA ($^\circ$)	&	R	\\
\hline
002	&	BP2(1a,b)	&	1.26, 0.91	&	50, 265	&	11.0, 8.0	\\
002	&	D(2a,b)	&	3.96	&	200-320	&	-	\\
003	&	D(1)	&	0.51	&	0-360	&	-	\\
006	&	BP2(1a-f)	&	0.68, 0.74	&	70, 285	&	5.5, 5.9	\\
006	&	K(2)	&	0.84	&	50	&	-	\\
006	&	K(3)	&	0.68	&	290	&	-	\\
007	&	K(1)	&	0.21	&	140	&	-	\\
007	&	K(2)	&	0.50	&	105	&	-	\\
006	&	K(4)	&	0.96	&	330	&	-	\\
018	&	K(1)	&	0.40	&	305	&	-	\\
018	&	K(2)	&	0.31	&	120	&	-	\\
019	&	BP2(1a,b)	&	0.28, 0.34	&	315, 135	&	1.6, 3.1	\\
022	&	D(1)	&	1.96	&	345-55	&	-	\\
026	&	BP2(1a-h)	&	2.27, 2.27	&	125, 295	&	3.8, 11.0	\\
034	&	K(1)	&	0.35	&	100	&	-	\\
034	&	K(2)	&	0.31	&	0	&	-	\\
034	&	K(3)	&	0.72	&	270	&	-	\\
034	&	K(4)	&	0.65	&	180	&	-	\\
034	&	K(5)	&	1.16	&	345	&	-	\\
045	&	BP2(1a,1b)	&	0.76; 0.71	&	125, 300	&	2.1, 6.0	\\
\hline\hline
	\end{tabular}
	\label{table:struct_parameters}
	\end{center}
	
	{\bf Notes.} First 20 rows of the table. A full version containing 448 entries is available online at the CDS. The columns are -- 1: source's ID; 2: type of the structure (H$_2$  emission or cluster); 3: projected length of the structure; 4: the position angle (in degrees, measured from N to E); 5: the aspect ratio of polar structures. In the case of K-type components, $\ell_{proj}$ corresponds to the distance from the H$_2$ emission to the RMS source. For structures with two or more components, the measurements follow the same order of each label. 
\end{table}

\subsection{Sources exhibiting H$_2$ polar emission}
\label{sec.polar_emission}

The identification of polar structures driven by MYSOs is an observational signature of the accretion by circumstellar disks proposed by \citep{Jijina96,Krumholz05}. Both the projected length ($\ell_{proj}$) and aspect ratio (R = length/width) were determined for every polar emission structure associated with each BP-type source. These values are shown in Table \ref{table:struct_parameters}.

The quantity $\ell_{proj}$ is important in order to $i)$ check any inconsistency on the determination of the kinematic distance associated with the source and $ii)$ to identify relationships between the morphology of the extended emission and intrinsic characteristics of the driving source. Since it is a projected measurement, $\ell_{proj}$ corresponds to a lower limit of the real dimension of the structure ($\ell_{proj} = \ell_{real} \sin i$, where $i$ is the inclination angle between the outflow axis and line-of-sight).
The first statement needs further discussion.
It is not expected to find $\ell_{proj}$ values greater than a few parsecs towards larger distances and, in these cases, there may be an inconsistency with the distance value. This seems to be the case of sources \#164 and \#165, located at the Galactic anti-centre ($\ell \sim 173^\circ$) and associated with a kinematic distance of 34.0 kpc. Table \ref{table:struct_parameters} displays a BP2-type emission located in the \#164 region has $\ell_{proj}$ = 25 pc and a D-type emission with $\ell_{proj}$ $\sim$ 35 pc in the same region. These values are much larger than the average projected length range of the structures found on this work ($\sim$ few parsecs).

The 74 BP-type sources are shown in Appendix \ref{appendix:bp_sources}. The H$_2$ maps for the other sources are available online at the CDS.

\subsection{Sources associated with stellar clusters}

Figure \ref{figure:sample_statistics} indicates that 80 MYSO candidates ($\sim 23\%$ of the sample) are associated with potential stellar clusters.
The presence of MYSOs in these clusters suggests they may be very young ($\sim10^5-10^6$\,yr for an O9 star) due to the short timescale related to the massive star formation.
The identification of clusters visible at near infrared wavelengths and associated with very young high mass objects is rare and the confirmation of such structures require follow up observations (e.g. deep J-, H- and K-band photometry) using high-angular resolution instruments in order to resolve the individual members of clusters located at distances of $\sim$ 10 kpc.
The list of cluster candidates is shown in Table \ref{table:clusters}. The table presents the central coordinates of the clusters, their angular size in arcseconds ($\Delta\theta$) and their projected diameter in parsecs (based on the kinematic distance of the RMS source in the same field). The full version of this table containing 80 entries is available at the CDS.

\begin{table}
	\caption{Properties of the potential stellar clusters identified on each field.}
	\begin{center}
	\begin{tabular}{cccccc}
\hline\hline
ID  & RA            &	Dec     & d     & $\ell_{proj}$ & $\Delta\theta$ \\
    & (J2000)       &	(J2000)	& (kpc) & (pc)          & ('') \\
\hline
22	&	18 36 47	&  -07 35 39	&	9.6	&	0.85	&	20	\\
34	&	18 43 46	&  -03 35 29	&	8.7	&	1.35	&	45	\\
39	&	18 47 36	&  -02 01 50	&	9.1	&	0.95	&	25	\\
66	&	19 07 01	&	08 18 44	&  11.2	&	0.47	&	20	\\
81	&	19 44 24	&	25 48 43	&  15.9	&	0.96	&	15	\\
85	&	20 21 55	&	39 59 45	&	4.5	&	0.70	&	35	\\
87	&	20 29 37	&	39 01 15	&	4.5	&	0.95	&	40	\\
88	&	20 19 40	&	40 56 33	&	3.8	&	1.08	&	60	\\
91	&	20 29 25	&	40 11 19	&	4.7	&	0.60	&	30	\\
99	&	20 38 56	&	42 22 41	&	3.7	&	0.62	&	40	\\
104	&	20 38 37	&	42 37 20	&	1.2	&	0.58	&  100	\\
107	&	20 42 34	&	42 56 51	&	1.0	&	0.26	&	55	\\
112	&	21 31 45	&	51 15 35	&	5.6	&	1.25	&	45	\\
115	&	21 15 55	&	54 43 30	&  10.7	&	0.73	&	20	\\
118	&	21 43 02	&	54 56 20	&	7.9	&	1.25	&	30	\\
120	&	21 52 57	&	56 39 54	&	7.4	&	0.52	&	15	\\
121	&	22 16 10	&	52 21 34	&	4.8	&	0.58	&	25	\\
124	&	22 55 29	&	57 09 24	&	6.1	&	0.80	&	30	\\
126	&	22 58 45	&	58 47 00	&	5.5	&	2.40	&	60	\\
127	&	22 59 03	&	59 28 25	&	5.3	&	0.88	&	55	\\

\hline\hline
	\end{tabular}
	\label{table:clusters}
	\end{center}
	
	{\bf Notes.} First 20 rows of the table. A full version containing 80 entries is available online at the CDS.
	The columns are as follows -- 1: source's ID; 2: right ascension (J2000); 3: declination (J2000); 4: kinematic distance (based on the RMS source); 5: projected length of the cluster based in the kinematic distance (column 4) in parsecs; 6: angular diameter of the cluster in arcseconds. 
\end{table}

\section{Discussion}
\label{discussion}

    \subsection{Near infrared colors}
The association of each MSX source and a 2MASS counterpart was performed by \citet{Lumsden02} in order to select the RMS sample. Thus, every positive association with a 2MASS counterpart should satisfy the criteria used by those authors, that is J - K$_{s}$ $>$ 2.0\,mag and F$_{K_s}$/F$_{8}$ $>$\,5, where F$_{8}$ corresponds to the flux measured at 8.3\,$\mu$m in the MSX images. Two other criteria used for the mid-infrared fluxes are: F$_{21}$/F$_{8}$ $>$ 2 and F$_{8}$ $<$ F$_{14}$ $<$ F$_{21}$.

\begin{figure*}
    \centering
    \resizebox{0.95\linewidth}{!}{\includegraphics{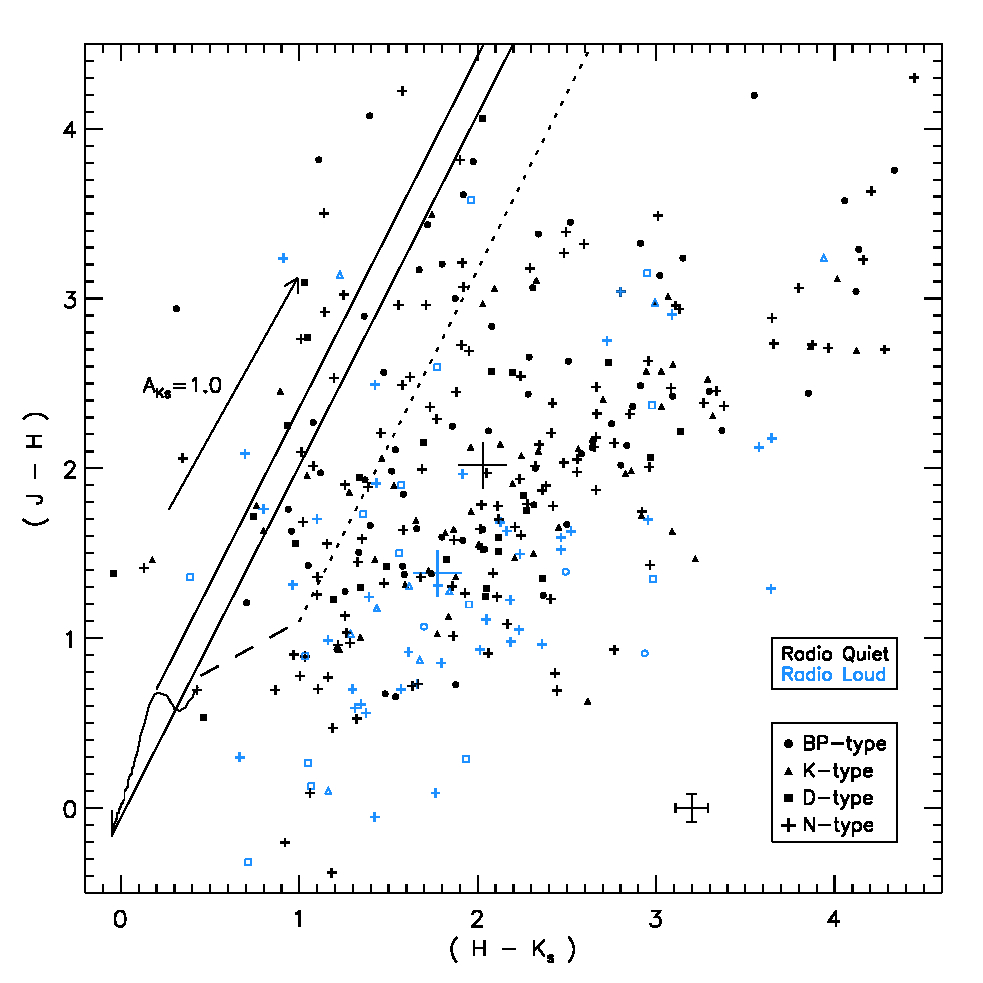}}
    \caption{Near infrared color-color (C-C) diagram of the MYSO candidates with 2MASS counterparts.
    The black curve denotes the locii of the main sequence stars with $A_{V} = 0$ mag. 
    The reddening vectors (shown as black lines) are from \citet{Stead09} reddening law. The upper line corresponds to M-type stars, the middle line to O-type and the dashed lines indicate the CTTS zone. 
    The arrow shown to the left of the M-type reddening line corresponds to A$_{K_s}=1.0$ mag.
    Objects to the right of the reddening vectors have excess emission in the K$_s-$band and are likely embedded YSOs.
    Data are shown as a function of H$_2$ emission morphology: BP-type ($\bullet$), K-type ($\blacktriangle$), D-type ($\blacksquare$) and N-type ($+$). Radio quiet and radio loud sources are shown as filled black and opened blue symbols.
    The average photometric error of each color is shown at the bottom right corner of the plot.
    The larger black and blue crosses indicate the median values of radio quiet and radio loud subsamples, respectively.}
    \label{figure:cc_nir_diagram}
\end{figure*}

Each positive association of a MSX source and a 2MASS counterpart was included in the near infrared color-color (C-C) diagram shown in Figure \ref{figure:cc_nir_diagram}. The data points are distinguished by each H$_2$ morphological type, defined in Section \ref{class_h2}. 
The locus of intrinsic colors for main sequence stars from \citet{Koornneef83} is shown as the continuous line.
The dashed black line shows the location of Classical T Tauri Stars (CTTS) from \citet{Meyer97}.
The black arrow corresponds to the reddening vector based on the \citet{Stead09} reddening law. The black lines indicate the reddening vectors for M-type stars, O-type and CTTS.
The region below the lower reddening line for the O-type stars is occupied by pre-main sequence (PMS) objects and the location of the most massive and young sources is located at higher H$-$K$_s$ values, to the right of the CTTS reddening line.
The photometric errors given by the 2MASS catalog were estimated using the photometric error of each considered filter and the average error was estimated in 0.081 and 0.091 for the J$-$H and H$-$K$_s$ colors, respectively.

Most of the objects in the NIR C-C diagram are located in the region occupied by PMS and MYSOs sources, that is, to the right of the O-type reddening line. This is a direct consequence of the MSX NIR color selection criteria from \citet[][J$-$K$_s$ $>$ 2.0 mag]{Lumsden02}.
Approximately $20$ sources are located to the left of the M-star reddening vector.
The existence of sources in this region of the C-C diagram may be due to the incomplete NIR color criteria used to select the RMS sources: it is not based on individual J$-$H and H$-$K$_s$ colors, but only on J$-$K$_s$.
Another $\sim 50$ sources lie between the M-star reddening line and the red limit of the CTTS sequence.
The last $\sim 280$ objects are located to the right of the CTTS reddening vector, displaying NIR colors expected for MYSO and low mass YSO candidates.

Figure \ref{figure:cc_nir_diagram} also displays the distribution of radio quiet/loud sources.
Most of the radio loud sources display J$-$H $\lesssim 2.0$ mag while radio quiet sources are preferably associated with J$-$H values greater than 1.0 mag. Only 15 sources ($\sim 20\%$ of the radio loud subsample) have greater H-band excess. 
The median NIR colors for the radio quiet and loud subsamples are shown as larger black and blue crosses in Figure \ref{figure:cc_nir_diagram}, respectively. The radio quiet subsample has J$-$H$_{med}=$2.02\,mag while the radio loud displays a smaller median excess, J$-$H$_{med}=$1.38\,mag. The median H$-$K$_s$ colors are less discrepant (2.03 and 1.77\,mag for the radio quiet and radio loud subsamples, respectively).
These results are consistent with the fact that radio loud sources are bluer NIR objects when compared to the bulk of radio quiet ones due to the dissipation of the circumstellar medium, reducing the reddening and color excess at near-infrared wavelengths.

Comparing the distribution of each H$_2$ class on the NIR C-C diagram, we found that both K- (shown as triangles) and BP-type (shown as circles) sources have NIR colors compatible with low to highly reddened sources (1.0 $\lesssim$ A$_K$ $\lesssim$ 3.5).
There are only three K-type sources to the left side of the M-star reddening vector and three BP-type ones located in the same region. Although these sources are not compatible with reddened MS stars, some of the BP-type sources display higher H-band excess (J$-$H $\gtrsim$ 3.0) when compared to K-type ones (J$-$H $\lesssim$ 3.0).
The presence of so many very reddened sources (H$-$K$_s \sim 4.0$) is consistent with the fact that most of the accretion occurs while the MYSO is still deeply embedded.
D-type sources (shown as squares) occupy a region compatible with less reddened objects (H$-$K$_s$ $\lesssim$ 3.0) when compared to the BP- and K-type ones.
There are only five D-type sources to the left of the M-star reddening vector.
Finally, N-type sources (shown as plus symbols) do not appear to fall into any preferred region on the C-C diagram.
The overall conclusion from Figure \ref{figure:cc_nir_diagram} is the absence of any well-defined region occupied by one of the classifications adopted in this work (either radio quiet/loud or H$_2$ classes).

\subsection{MSX and IRAS colors}

\begin{figure*}
    \centering
    \resizebox{0.95\linewidth}{!}{\includegraphics{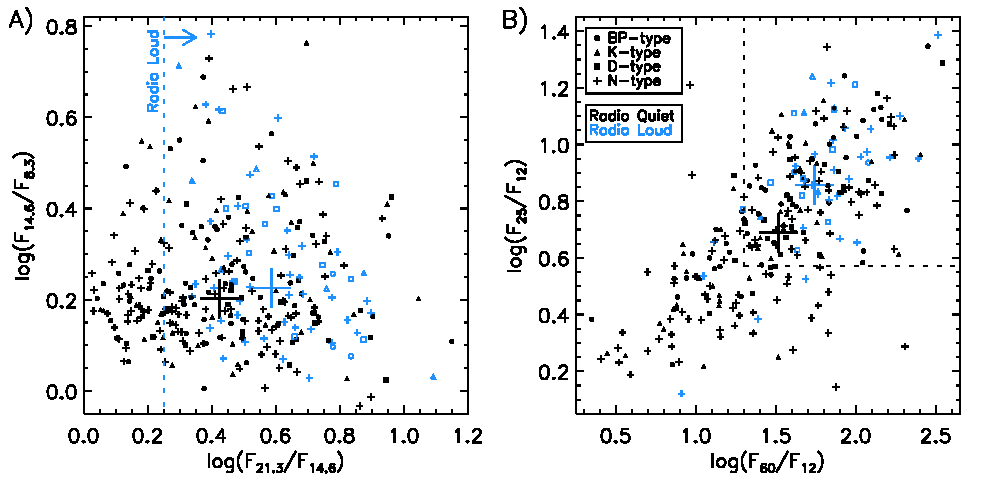}}
    \caption{Panel A: Mid-infrared C-C diagram of the MYSO candidates with MSX fluxes.
    Panel B: Far-infrared C-C diagram of the MYSO candidates with IRAS fluxes.
    The C-C diagrams are based on the logarithm of the flux ratio at different wavelengths.
    On both plots, data are shown as a function of H$_2$ emission morphology: BP-type ($\bullet$), K-type ($\blacktriangle$), D-type ($\blacksquare$) and N-type ($+$). Filled black symbols correspond to radio quiet sources while opened blue symbols are radio loud ones.
    The dashed red line shown in Panel A corresponds to the $\log(F_{21.3}/F_{14.6})$ value limit for the radio loud sources.
    The dashed lines shown in Panel B mark the IRAS color criteria for UCH\,{\sc{ii}} regions, defined by \citet{Wood89}.}
    \label{figure:cc_mir_fir_diagram}
\end{figure*}

The MSX and IRAS color C-C diagrams are based on the logarithm of the flux ratios at different wavelengths [i.e. $\log(F_{14.6}/F_{8.3})$ and $\log(F_{60}/F_{12})$]. The mid-infrared (MIR) and far-infrared (FIR) C-C diagrams are shown in Figure \ref{figure:cc_mir_fir_diagram}. Data are shown using the same representation from Figure \ref{figure:cc_nir_diagram}.

The MIR C-C diagram (Panel A) indicates most of the sample display $\log(F_{21.3}/F_{14.6})$ values between 0.0 and 1.2 and $\log(F_{14.6}/F_{8.3})$ between 0.0 and 0.8, being consistent with the MIR color criteria adopted by \citet{Lumsden02} for the RMS sample selection (F$_{8.3}$ $<$ F$_{14.6}$ $<$ F$_{21.3}$).
\citet{Varricatt10} claimed that young sources in later phases (UCH\,{\sc{ii}}) should occupy the upper region of the $\log(F_{14.6}/F_{8.3})$ vs. $\log(F_{21.3}/F_{14.6})$ diagram and have steeper SEDs than the ones with active accretion (MYSOs). However, the median values of MIR color indexes for the radio quiet/loud subsamples indicate that radio loud sources are preferably found towards larger $\log(F_{21.3}/F_{14.6})$ values, but display similar $\log(F_{14.6}/F_{8.3})$ values than those found for radio quiet ones. In addition, we found that every radio loud source has $\log(F_{21.3}/F_{14.6}) > 0.25$.

Regarding the H$_2$ classes, Panel A from Figure \ref{figure:cc_mir_fir_diagram} shows the ratio between BP- and K-type sources increases from larger $\log(F_{14.6}/F_{8.3})$ and $\log(F_{21.3}/F_{14.6})$ values towards smaller ones.
Sources associated with diffuse H$_2$ emission (shown as square symbols) have a similar distribution as compared to radio loud sources.
N-type sources do not display any preferred region in the Mid-IR C-C diagram.

The Far-IR C-C diagram (Panel B, Figure \ref{figure:cc_mir_fir_diagram}) indicates that all sources have positive FIR colors, indicating crescent SEDs at IRAS wavelengths (that is, the flux increases toward larger wavelengths).
Only 5 radio loud sources are located outside the region delimited by the IRAS color criteria for UCH\,{\sc{ii}} regions, defined by \citet{Wood89} while radio quiet sources do not show any preferred region on the plot.
As a function of the H$_2$ morphology, the Far-IR C-C diagram displays the fraction of sources outside the UCH\,{\sc{ii}} region \citep{Wood89} increases from the D-type (only 3) to BP- ($\sim19\%$), K- ($\sim26\%$) and N-type ones ($\gtrsim30\%$).
Most of the D-type sources are located inside the area of UCH\,{\sc{ii}} regions. This is consistent with the fact that the diffuse emission may be originating from more evolved sources in a phase after the major accretion period. The Panel B from Figure \ref{figure:cc_mir_fir_diagram} also indicates a correlation between D-type and radio loud sources.
Combining this information with the fact that the D-type sources display a larger fraction of associations with stellar clusters ($43.2\%$, Panel B from Figure \ref{figure:sample_statistics_h2morph}), this suggests diffuse H$_2$ emission is a result of the global environment of the cluster and not a result of a single star.

The photometric analysis of the sample leads to the conclusion that
$i)$ most of the sources are located to the right of the CTTS reddening line in the NIR C-C diagram, having colors compatible with young embedded objects. Radio quiet sources are preferably redder and dominate in the top part of the NIR C-C diagram (that is, they have greater J$-$H colors when compared to radio loud sources). No clearly NIR color criteria was found for any H$_2$ morphological class.
$ii)$ the MIR C-C diagram indicates the radio loud subsample is located toward $\log(F_{21.3}/F_{14.6})$ values greater than 0.25.
and $iii)$ most of the radio loud sources and those sources associated with diffuse H$_2$ emission are located inside the UCH\,{\sc{ii}} region in the Far-IR C-C diagram.

\subsection{Polar structures and comparison with other samples}

In order to check if we are dealing with some particular case of objects, we compared our results with the work from \citet{Varricatt10}, which present a smaller sample (50) of MYSO candidates. They found extended H$_2$ emission towards 38 of them (76\%) and 25 sources were associated with polar H$_2$ structures, suggesting collimated jets. The fraction of the sources associated with extended H$_2$ emission and those which present polar structures are 1.5 and 2.4 times the values found in the present work (50.4 and 21\%, respectively).
This could be due to a selection effect of their sample: Their selection is based on sources that display high-velocity CO emission, which is largely used to detect emission from molecular outflows associated with young stars. Thus, there is a high probability that the sources associated with CO outflows would also exhibit a H$_2$ counterpart. Our survey was unbiased with respect to a prior knowledge of morphology type.

\begin{figure}
    \centering
    \resizebox{0.95\linewidth}{!}{\includegraphics{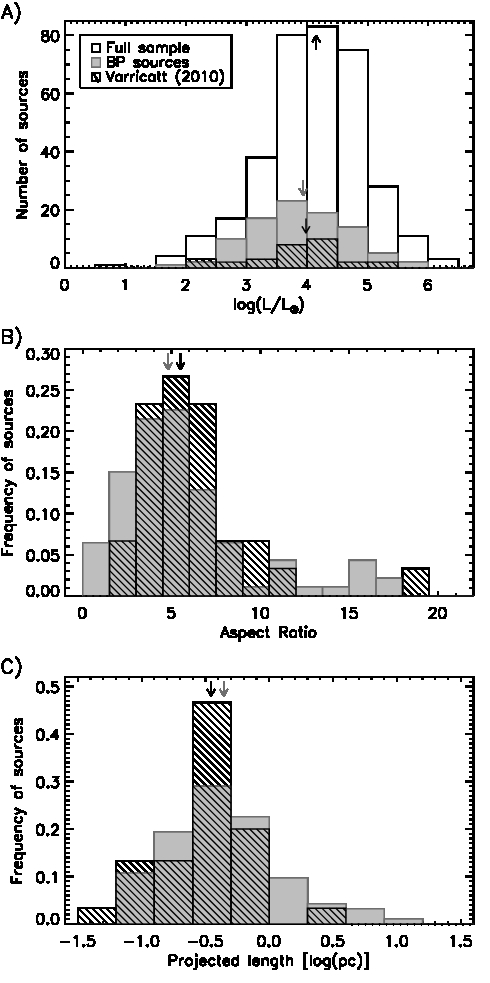}}
    \caption{Distribution of the full sample (black bars), sources classified as BP-type (grey bars) and sources associated with polar structures identified by \citet[][shown as black striped bars]{Varricatt10}.
    Panel A: histogram of the bolometric luminosity using a binning of 0.5\,dex;
    Panel B: distribution of the aspect ratio of each polar component assuming a binning of $\Delta R = 1.5$;
    Panel C: distribution of the projected length (in logarithmic units) of each structure using a binning of 0.3\,dex.
    The arrows shown in the plots indicate the median values of each sample.}
    \label{figure:histogram_samples}
\end{figure}

Figure \ref{figure:histogram_samples} compares the distribution of the bolometric luminosity of the sources associated with polar structures (Panel A), the aspect ratio (R, Panel B) and the projected length (Panel C) of the polar structures driven by the BP-type sources and those found by \citet{Varricatt10}. 
Panel A shows the distribution of the number of sources per 0.5 dex bin of $\log(L/L_\odot)$ for the complete sample (353 objects, shown as black bars), the BP-type objects (74, grey bars) and those 25 identified by \citet[][shown as black stripes]{Varricatt10}.
The present distribution and that of \citet{Varricatt10} appear similar in the range of luminosity for BP-types, confirming that such polar structures are driven by both high and low luminosity sources.
Panel B displays the aspect ratio of each polar component identified in both works. The data are shown as the fraction of the total number of structures (93 for the grey bars and 30 for the black striped bar, respectively). The histogram indicates the distribution of the R values for both samples is very similar.
The median of each distribution is indicated by arrows and corresponds to $R = 4.8$ and 5.5 for the BP-type sources and \citet{Varricatt10} sample, respectively.
Panel C shows the distribution of the projected length of each polar structure. 
The plot suggests those structures identified by \citet{Varricatt10} are relatively smaller than the ones found in the present work, although the shape of the distribution are very similar.
These plots reveal that although the samples have different sizes, their global properties are very alike.

\begin{figure}
    \resizebox{0.95\linewidth}{!}{\includegraphics{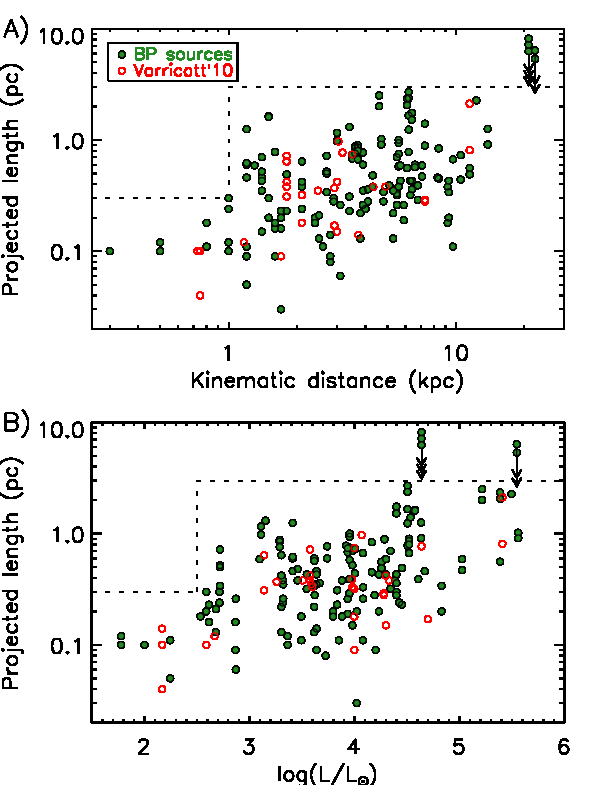}}
    \caption{Distribution of the polar structures associated with BP-type objects as a function of the kinematic distance (Panel A) and the bolometric luminosity of the associated RMS source (Panel B).
    Data from this work is shown as filled green circles while those from \citet{Varricatt10} is indicated by opened red circles.
    Filled black and red lines correspond to tentative fittings of our and Varricatt data, respectively.
    Dashed lines are placed at d$=1.0$\,kpc and $\log(L/L_\odot)=2.5$, indicating a changing in the upper limit of $\ell_{proj}$.
    Data associated with black arrows indicates the upper limit of the values due to problems on the kinematic distance determination (see text for details).}
    \label{figure:struct_bp}
\end{figure}

Figure \ref{figure:struct_bp} displays the distribution of the projected length of the polar structures as a function of the kinematic distance (Panel A) and the bolometric luminosity of their driving sources (Panel B). Data from this work are shown as filled green circles while those from \citet{Varricatt10} are shown as red opened circles. 
Panel A displays that smaller structures ($\ell_{proj} \sim 0.1$ pc) are found toward the entire distance range within the sample. A lack of larger structures was found towards closer distances than 1 kpc and a relative increase of the lower limit value of $\ell_{proj}$ was found for d $>$ 5 kpc.
The increasing of the lower limit is explained due to the difficult on identifying smaller structures towards farther distances and does not provide us any insight on the physical processes that are driving the outflows.
The upper limit of the structures reveals two different regions on the distance domain: $i)$ for d $<$ 1 kpc, indicating the incompleteness of our survey towards closer distances and due the absence of massive star formation in the Solar neighbourhood; and $ii)$ for distances greater than 1 kpc, which indicates a relatively constant upper limit of $\ell_{proj} \sim$ few pc.
There are 5 points at d $\sim$ 20 kpc with $\ell_{proj} \sim$ 10 pc. These points correspond to structures associated with sources \#162 and \#163 (shown in Appendix A), located at $\ell \sim 173^\circ$.
As discussed in Section \ref{sec.polar_emission}, six of our targets (\#161, \#162, \#163, \#164, \#165 and \#166) are located at $\ell \sim 170^\circ$. They are associated to distances farther than 16\,kpc, which is larger than the expected Galactic distances at this longitudinal region. As reported by \citet{Urquhart09a}, kinematic distances of sources located closer to the Galactic anti-centre are very uncertain.

Panel B displays the same data but as a function of the bolometric luminosity of the driven source. Those points associated with uncertain Galactic distances are placed at $\log(L/L_\odot) \approx 4.6$ and 5.5 and are excluded from the overall analysis.
The analysis of the Panel B is alike of that from the previous plot.
Small projected lengths are found towards the entire bolometric luminosity range and the lower limit of $\ell_{proj}$ does not provide any important information.
On the other hand, the upper limit of $\ell_{proj}$ traces two different regions: the first one, due to the incompleteness of our sample towards low-luminosity sources ($\log(L/L_\odot) \lesssim 2.5$), are limited by $\ell{proj} \lesssim$ 0.3 pc; for higher luminosities, the upper limit is roughly 10 times larger ($\ell{proj} \lesssim$ 3.0 pc).
\citet{Liu10} derives the statistical average value of the inclination angle of a randomly distribution of outflows and found $<i>=0.637$\,rad, or $<i>=36.5^\circ$. That is, most of the $\ell_{proj}$ values corresponds to physical lengths 1.68 times larger as the same as their aspect ratio (that is, the outflows are likely to be larger and more collimated than their measured values).

Data from \citet{Varricatt10} are within the range of the luminosity, distance and projected length of the structures identified in the present work, indicating that both samples are consistent.
The overall conclusion from Figure \ref{figure:struct_bp} is that the largest outflows associated to high-luminosity sources are limited to a few parsec of extension, suggesting the size of the jets may not scale with intrinsic characteristics of their driven source.

\begin{figure}
    \centering
    \resizebox{0.95\linewidth}{!}{\includegraphics{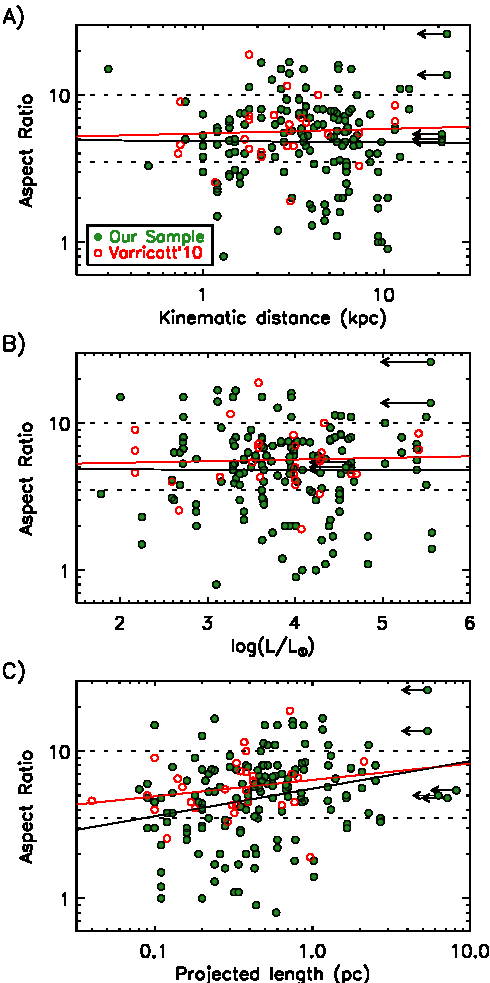}}
    \caption{Distribution of the aspect ratio of the polar emission associated with BP-type sources.
    The data are shown as a function of the logarithm of the kinematic distance (Panel A) and the luminosity of the associated RMS source (Panel B), and as a function of the projected length of each structure (Panel C).
    Data from this work is shown as filled green circles while those from \citet{Varricatt10} are shown as opened red circles.
    The horizontal dashed lines corresponds to R=3.5 and R=10.0.
    Filled black and red lines correspond to tentative fittings of our and Varricatt data, respectively.
    Data associated with black arrows indicates the upper limit of the values due to problems on the kinematic distance determination (see text for details).}
    \label{figure:struct_aspect}
\end{figure}

The distribution of the aspect ratio (length/width) of each polar structure is shown in Figure \ref{figure:struct_aspect}.
Panel A indicates that highly collimated ($R \gtrsim 10$) structures driven by BP-type sources are found over the entire range of luminosities while the \citet{Varricatt10} sample only displays three highly collimated structures driven by $3 \lesssim \log(L/L_\odot) \lesssim 4.5$ sources. Our sample presents several polar structures driven by high-luminosity ($\log(L/L_\odot) \gtrsim 4.0$) and less collimated ($R \lesssim 3.5$) sources than included in the sample of \citet{Varricatt10}. This may simply reflect the smaller sample size.
The Spearman correlation rank for data shown in Panel A (correlation factor of 0.0147) indicates a weak correlation between R values as a function of the luminosity of the driving source.
It is likely that R is a lower limit of the real aspect ratio (R $\lesssim$ R$_{real}$) since it is based on a projected value of the true length of the structures (that is, $\ell_{proj} \lesssim \ell_{real}$).

Panel B displays the distribution of R as a function of the kinematic distance of its driving source. In this plot, we found highly collimated outflows for distances greater than 2\,kpc while less collimated ones are found between 1 and 10\,kpc. The plot indicates a poor correlation between the aspect ratio and distance values (Spearman rank correlation factor of -0.02).
Panel C indicates that the projected length of these less collimated structures is more variable (larger width of the distribution) on scales from 0.1 to $\sim 1.0$\,pc, while the highly collimated outflows are associated with both smaller ($\sim0.1$\,pc) and larger structures ($\sim 10$\,pc).
There are no data between with $\ell_{proj} \gtrsim 1.0$\,pc and $R \lesssim 3$.
The absence of structures with these properties indicates that the mechanism that produces wider outflows is not able to sustain its structure towards larger length scales.

The overall analysis of Figure \ref{figure:struct_aspect} indicates that both samples contains analogous structures, although several of those identified in this work are less collimated than those identified by \citet{Varricatt10}.
On the other hand, there is no correlation between the aspect ratio of the polar emission and the bolometric luminosity of their driving sources.

\subsection{Masses of the outflows}

An outflow with length equal to its projected length ($\ell_{proj}$) and its width given by $\Delta w \approx \ell_{proj} / R$) could be approximated by a conical structure with volume $V = \pi \cdot  \Delta w^2 \cdot \ell_{proj}/3 = \pi \ell_{proj}^3 / 3 R^2 $. Assuming a typical density of an outflow associated with a massive young star \citep[$n = 10^3 - 10^4 cm^{-3}$][]{Codella13}, one can estimate the outflow mass, $M_{out} = V \cdot n \cdot m_{H_2}$, where $m_{H_2}$ corresponds to the H$_2$ molecular mass. In terms of the observable quantities, we have $M_{out} = 4.3 \left(\frac{\ell_{proj}}{1.0\,pc}\right)^3 \left(\frac{R}{3.0}\right)^{-2} \left(\frac{n}{10^3 cm^{-3}}\right)\,M_\odot$.

\begin{figure}
    \centering
    \resizebox{0.95\linewidth}{!}{\includegraphics{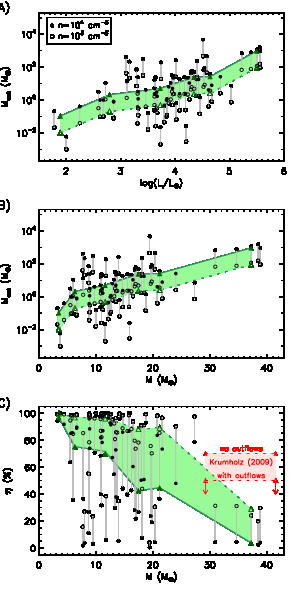}}
    \caption{Distribution of the outflow mass associated with each  as a function of the bolometric luminosity (Panel A) and the central mass (Panel B) for the BP-type sources.
    Panel C displays the star formation efficiency as a function of the central mass.
    Filled and opened circles correspond to the maximum and minimum particle density of the outflows, respectively.
    Median values of the mass of the out flow for each mass range (0-4, 4-8, 8-15, 15-20, 20-30 and 30-40\,M$_\odot$) are shown as green triangles and are connected by green lines (filled/solid for the higher density).
    Data from \citet{Krumholz09} are shown as red triangles and connected by dashed red lines. Filled symbols indicate the results without considering outflow effects while the opened symbols indicate the efficiency if outflow activity was included.}
    \label{figure:struct_mass}
\end{figure}

Figure \ref{figure:struct_mass} displays the outflow mass (M$_{out}$) in solar mass units as a function of the bolometric luminosity (Panel A) and the mass of its driving source (assuming $L \propto M^{3.5}$, Panel B).
M$_{out}$ was estimated assuming the minimum (shown as opened circles) and maximum density (shown as filled circles) from \citet[][$n=10^3-10^4 cm^{-3}$]{Codella13}. The maximum and minimum values determined for a single point are connected by a gray line. In case of multiple components, the mass of each component was summed and, thus, M$_{out}$ corresponds to the total outflow mass ejected by the source.
The plots also display the median mass out flow values for the following mass ranges (and corresponding luminosity ranges): 0-4, 4-8, 8-15, 15-20, 20-30 and 30-40\,M$_\odot$ (and for both values of density). The green shadowed region is delimited by the maximum  (shown as filled triangles and connected by the filled green line) and minimum masses (shown as open triangles and connected by the dashed green line).
Panel A shows M$_{out}$ ranges from 10$^{-2}$ to 10$^5$\,M$_\odot$. The data have a larger dispersion between $\log(L/L_\odot) \sim 3$ and $\sim 5$, where most of the BP-sources are located, but the median values clearly indicate a tendency for M$_{out}$ to increase with the bolometric luminosity of the driving source (M$_{out} \propto L^\alpha$, with $\alpha \approx 1.15$).
Panel B displays the outflow mass as a function of the mass of its driving source. This plot indicates a slight increase of the M$_{out}$ value as a function of the central mass star.
It is expected that most of the mass not transferred to the central star is ejected in an outflow. In another words, one can estimate the efficiency of the star formation assuming that $\eta = M_{star} / ( M_{star} + M_{out} )$.
Panel C displays the star formation efficiency as a function of the central mass. The median values indicate that $\eta$ decreases toward high mass stars. Most of the stars with $M \lesssim 15\,_\odot$ display $\eta$ values higher than 70\%, even considering the highest density for the outflows. The efficiency starts to decrease to 50\% for M$\sim15-20$\,M$_\odot$, and displays the lowest values for M $\sim30-40$\,M$_\odot$. That is, giant molecular clouds with masses $\sim 1000 M_\odot$ are necessary to form very massive stars. 
\citet{Krumholz09} have simulated the formation of a binary system with 41.5 and 29.2\,M$_\odot$ from an initial cloud with M=100\,M$_\odot$. The process ended with $\eta = 70\%$ without considering any mass loss by outflow effects. The authors suggests that if one considers outflow activity, this value should drop to $\eta \sim 50\%$. These points are shown in Panel C as red triangles. Comparing with our observations, the efficiency determined by \citet{Krumholz09} is reasonable for the $\sim30$\,M$_\odot$ star but is higher than observed for the $\sim40$\,M$_\odot$ one.

\subsection{Collimation effects at different luminosity ranges}
\begin{figure}
    \centering
    \resizebox{0.95\linewidth}{!}{\includegraphics{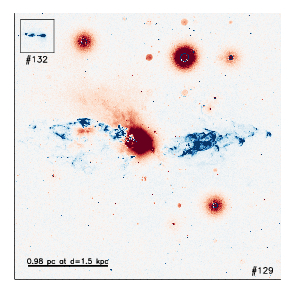}}
    \caption{Comparison between bipolar outflows driven by a MYSO (Cepheus A, source \#129, $\log(L/L_\odot) = 4.58$) and a YSO (Cepheus E, source \#132, $\log(L/L_\odot) = 2.59$), both located in the Cepheus region at the same kinematic distance (d=1.5\,kpc). Both H$_2$-maps are within the same scaling, indicated by the horizontal ruler in the bottom left side of the image. Cep E inset has been rotated through 90 deg, so N is to the left.}
    \label{figure:polar_comparison}
\end{figure}

The last sections have shown that young stellar objects in different luminosity ranges drive outflows with indistinguishable aspect ratios, although the largest outflows are found for the most luminous sources. In order to study the intrinsic properties of the outflows, we must compare a high-luminosity MYSO with a low-luminosity one, located nearby each other and relatively close to the sun.
G109.8715+02.1156 (\#129, Fig. \ref{figure:h2_map_A129}), also known as Cepheus A \citep{Rodriguez94,Torrelles96}, is a high-luminosity source ($\log(L/L_\odot)=4.58$) located at d=1.5\,kpc. G110.4771+01.4803 (\#132, Fig. \ref{figure:h2_map_A132}), known as Cepheus E \citep{Lefloch96} is an intermediate mass star ($\log(L/L_\odot)=2.59$) located in the same region of the sky and with the same kinematic distance.
Both sources shares the same physical scale shown in Figure \ref{figure:polar_comparison}.
The high-luminosity source Cepheus A displays a bipolar outflow oriented in the E-W direction. The east lobe has a projected length of $\ell_{proj} = 1.64$\,pc and its aspect ratio was estimated as R=4.8.
The lower-luminosity source Cepheus E exhibits a bipolar outflow oriented in the N-S direction. The southern lobe has a projected length of $\ell_{proj} = 0.20$\,pc and its aspect ratio is $R=3.1$.

Figure \ref{figure:polar_comparison} displays both sources aligned in the same direction and sharing the same physical scale.
The H$_2$ map of Cepheus A exhibits several inner components within the main structure of the bipolar outflow, suggesting either the driving source has a considerable precessing motion or even there are multiple sources producing outflows in the same region. This effect was not observed for Cepheus E, which exhibits a well defined single outflow structure.
Previously, we showed that larger structures are driven by high luminosity sources and that there are no structures larger than $\sim 1.0$\,pc with aspect ratio values smaller than R$\sim 3.0$.
Thus, both precession and the presence of multiple and active sources driving outflows would produce an enlargement of the global aspect ratio of the observed structure, such as the case for Cepheus A.
Recently, \citet{Peters14} have reproduced the global properties of the Cepheus A outflow on a three-dimensional simulation considering individual jets from multiple embedded stellar sources.
This suggests some of the BP features may be classified as having low collimation when in fact the individual flows are highly collimated but enlarged by the effect of multiple sources and so appearing to have a broader opening morphology.

\section{Summary and Conclusions}

We have presented observations of 353 MYSO candidates observed through H$_2$ $\nu$=1-0 S(1) and continuum narrow-band filters. We use the angularly resolved images to characterize the nature of massive young stellar objects (MYSOs) with an eye to understanding the formation mechanism of massive stars. This work confirms and extends the results of \citet{Varricatt10}. 

The analysis of our observations lead to the following conclusions: 

\begin{enumerate}
  \item{178 of the sources ($\approx 50\%$) display extended H$_2$ emission: $21\%$ exhibit polar structures, $17\%$ are associated with knots or amorphous H$_2$ emission and $\approx 12\%$ corresponds to diffuse H$_2$ emission;}
  \item{74 of MYSOs with H$_2$ emission ($\approx 21\%$) exhibit collimated structures and bipolar outflows and most of these are so called radio quiet sources (i.e. not compact or extended H{\,\sc{ii}} regions). The association with collimated outflows is taken as observational evidence for the accretion scenario and suggests that MYSOs are formed through disc accretion (as for low mass YSOs) and most of the accretion occurs on timescales less than $10^5$ years;}
  \item{the data suggest that larger outflow structures are found toward high-luminosity sources. The low-luminosity sources cannot drive very large outflows or even be associated with diffuse emission, which requires higher radiative fluxes;}
  \item{the data do not display a clearly correlation between the size of the outflow structures and the properties of their driven source. The upper limit of the projected lengths of the outflows is roughly constant (a few parsecs) for a large range of bolometric luminosity;}
  \item{the low-luminosity sources exhibit smaller outflows ($\sim 0.1$ pc) and are not associated with diffuse H$_2$ emission, which requires higher radiative fluxes;}
  \item{there is no correlation between the aspect ratio (length/width of the flow) of the polar structures and the bolometric luminosity of their driving sources;}
  \item{ our data show a lack of poorly collimated large outflows, suggesting that the structure of the outflows cannot be maintained/supported by an ejection flux driven at very broad opening angles;}
  \item{Considering the mass enclosed by the outflows, we found that the star formation efficiency decreases towards high-mass stars, indicating they expel up to $\sim10$ times the amount of mass accreted into the central source;}
  \item{Only $\approx 25\%$ of the sample (89) is associated with infrared-visible stellar clusters. Two of these sources are associated with polar H$_2$ emission.}
\end{enumerate}

\section*{Acknowledgments}

FN is grateful to CNPq and FAPESP support. 
AD and CLB acknowledge FAPESP for continuous financial support.
We thank Sergio Scarano Jr., Cristina Armond and Tiago Ribeiro, who dedicated much effort to obtaining high-quality imaging for SOAR programs and Eder Martioli, Karun Thanjavur for the same effort on CFHT programs.
We acknowledge all the discussion and suggestions given by the TERAPIX team which were very helpful.

We also thank the comments and suggestions made by the anonymous referee, which improved the quality of the revised version of this article.

Based on observations obtained with WIRCam, a joint project of CFHT, Taiwan, Korea, Canada, France, and the Canada-France-Hawaii Telescope (CFHT) which is operated by the National Research Council (NRC) of Canada, the Institute National des Sciences de l'Univers of the Centre National de la Recherche Scientifique of France, and the University of Hawaii.

Based on observations obtained at the Southern Astrophysical Research (SOAR) telescope, which is a joint project of the Minist\'erio da Ci\^encia, Tecnologia, e Inova\c{c}\~ao (MCTI) da Rep\'ublica Federativa do Brasil, the U.S. National Optical Astronomy Observatory (NOAO), the University of North Carolina at Chapel Hill (UNC), and Michigan State University (MSU).

This research has made use of the VizieR catalogue access tool, CDS, Strasbourg, France.

\bibliographystyle{mn2e}

\clearpage

\appendix

\section{Individual comments on the sources with extended H$_2$ emission}
\label{appendix:bp_sources}

The mass-luminosity relation $L \propto M^{3.5}$ was adopted for an estimation of the central mass. The spectral type was inferred using the relations given by \citet{Martins05} and \citet{Panagia73} for the O-type stars on the MS and the B-type stars on the Zero Age Main Sequence (ZAMS). The orientations are given by North (N), East (E), South-West (SW) and so on. The caption of each image displays the following information: MSX name of the source, $\log(L/L_\odot)$, distance, Spectral type classification and structures present on the field shown as ``Morphological classification'' (label): $\ell_{proj}$; PA; R (if available)' (i.e. BP1(1): 0.91\,pc; 8.5; 60$^\circ$). A full description of the images is presented in Figure \ref{figure:h2_map_example}.

\setcounter{figure}{001}
\begin{figure}
  \resizebox{0.95\linewidth}{!}{\includegraphics{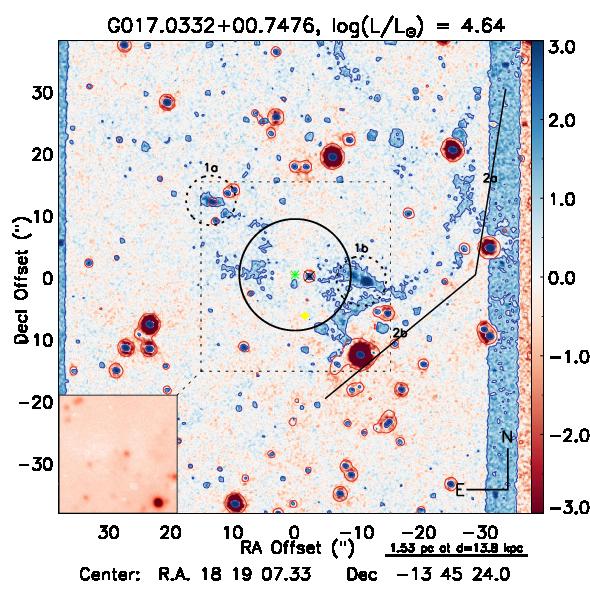}}
  \caption{G017.0332+00.7476, $\log(L/L_\odot) = 4.64$, d=13.8\,kpc, Sp. Type: O9.5\,V, H-K$_s$ = 1.26\,mag.
           BP2(1a,1b): 1.26, 0.91\,pc; 50$^\circ$, 265$^\circ$; R=11.0, 8.0; 
           D(2a,2b): 3.96\,pc; 200-320$^\circ$.          }
  \label{figure:h2_map_A002}
\end{figure}

\setcounter{figure}{005}
\begin{figure}
  \center
  \resizebox{0.95\linewidth}{!}{\includegraphics{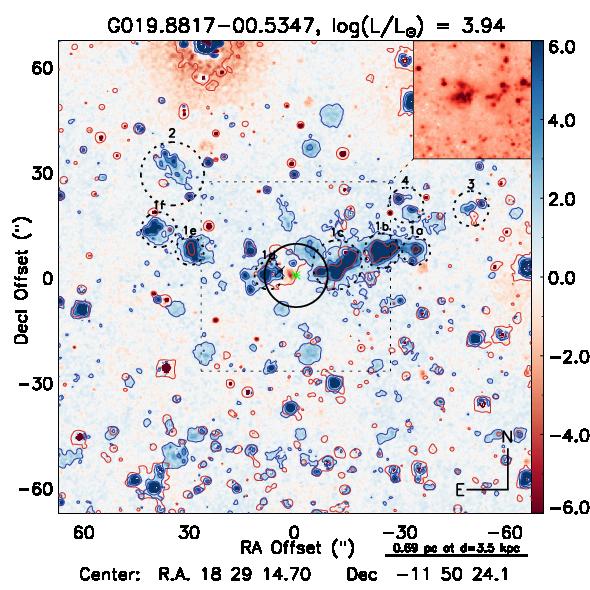}}
  \caption{G019.8817-00.5347, $\log(L/L_\odot)=3.94$, d=3.5\,kpc, Sp. Type: B1\,V$_0$, H-K$_s$ = 1.63\,mag.
           BP2(1a,1f): 0.68, 0.74\,pc; 70$^\circ$, 285$^\circ$; R=5.5, 5.9;
           K(2): 0.84\,pc; 50$^\circ$;
           K(3): 0.68\,pc; 290$^\circ$;
           K(4): 0.96\,pc; 330$^\circ$.          }
  \label{figure:h2_map_A006}
\end{figure}

\setcounter{figure}{018}
\begin{figure}
  \center
  \resizebox{0.95\linewidth}{!}{\includegraphics{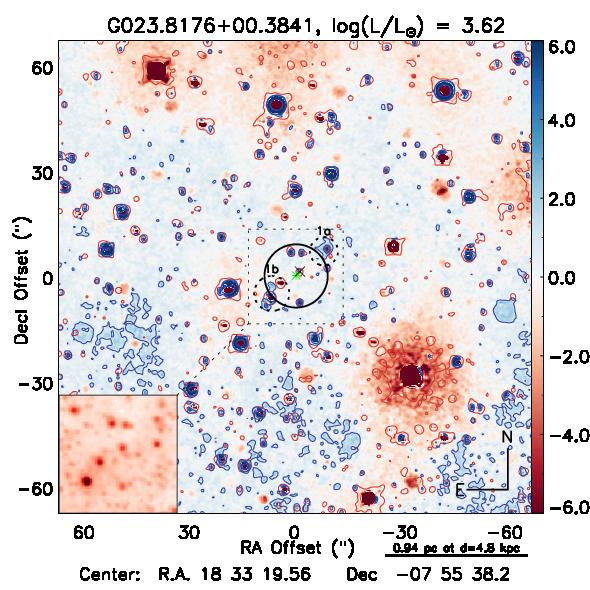}}
  \caption{G023.8176+00.3841, $\log(L/L_\odot)=3.62$, d=4.8\,kpc, Sp. Type: B2\,V$_0$, H-K$_s$ = 0.67\,mag.
           BP2(1a,1b): 0.28, 0.34\,pc; 315$^\circ$, 135$^\circ$; 1.6, 3.1.          }
  \label{figure:h2_map_A019}
\end{figure}

\setcounter{figure}{025}
\begin{figure}
  \resizebox{0.95\linewidth}{!}{\includegraphics{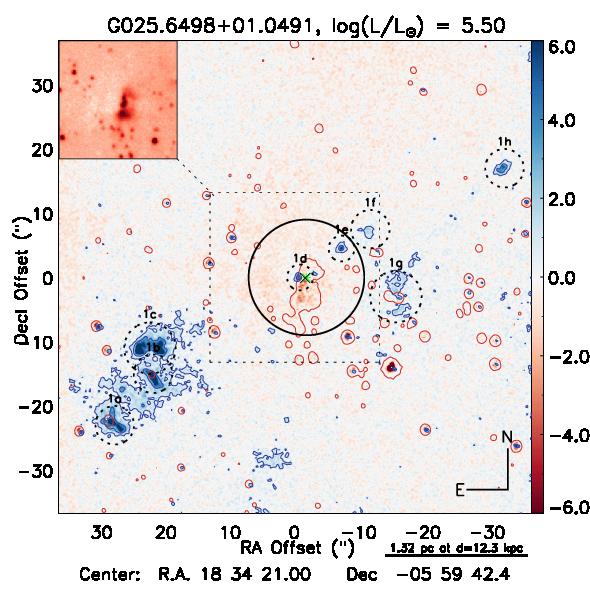}}
  \caption{G025.6498+01.0491, $\log(L/L_\odot)=5.50$, d=12.3\,kpc, Sp. Type: O5.5\,V, H-K$_s$ = 3.61\,mag.
           BP2(1a-1h): 2.27, 2.27\,pc; 125$^\circ$, 295$^\circ$; 3.8, 11.0.          }
  \label{figure:h2_map_A026}
\end{figure}

\setcounter{figure}{044}
\begin{figure}
  \resizebox{0.95\linewidth}{!}{\includegraphics{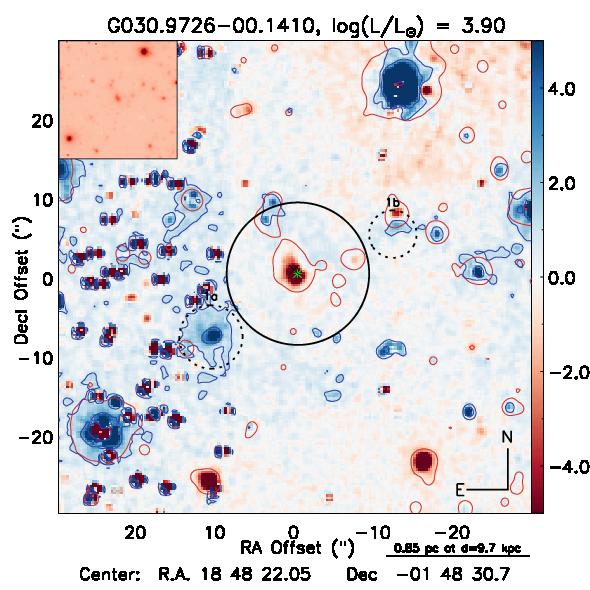}}
  \caption{G030.9726-00.1410, $\log(L/L_\odot)=3.90$, d=9.7\,kpc, Sp. Type: B1\,V$_0$, H-K$_s$ = 3.03\,mag.
           BP2(1a,1b): 0.76, 0.71\,pc; 125$^\circ$, 300$^\circ$; 2.1, 6.0.          }
  \label{figure:h2_map_A045}
\end{figure}

\setcounter{figure}{054}
\begin{figure}
  \resizebox{0.95\linewidth}{!}{\includegraphics{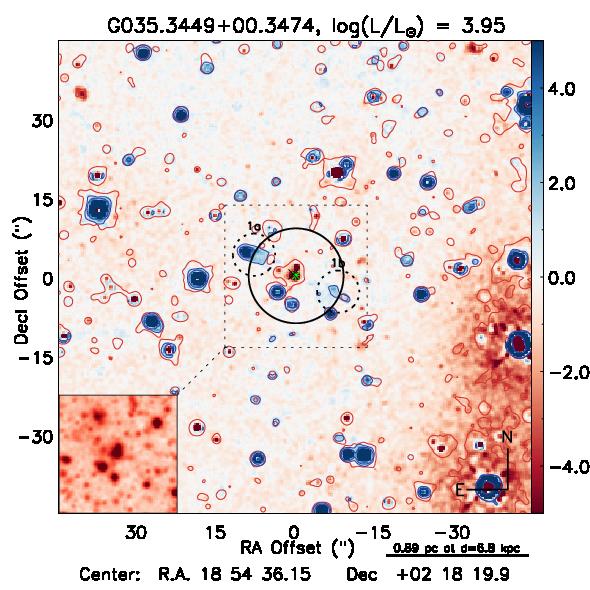}}
  \caption{G035.3449+00.3474, $\log(L/L_\odot)=3.95$, d=6.8\,kpc, Sp. Type: B1\,V$_0$, H-K$_s$ = 1.52\,mag.
           BP2(1a,1b): 0.39, 0.29\,pc; 60$^\circ$, 250$^\circ$; 4.5, 5.0.          }
  \label{figure:h2_map_A055}
\end{figure}

\setcounter{figure}{089}
\begin{figure}
  \resizebox{0.95\linewidth}{!}{\includegraphics{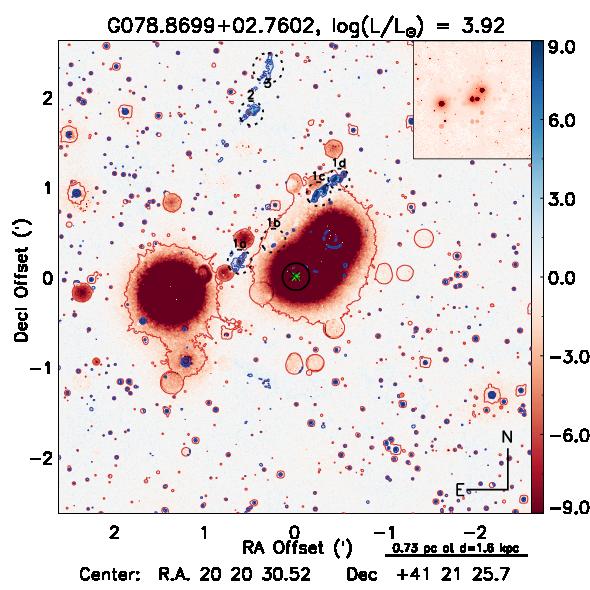}}
  \caption{G078.8699+02.7602, $\log(L/L_\odot)=3.92$, d=1.6\,kpc, Sp. Type: B1\,V$_0$, H-K$_s$ = 2.20\,mag.
           BP2(1a,1b,1c,1d): 0.78\,pc; 80$^\circ$, 35$^\circ$, 345$^\circ$, 335$^\circ$; 7.5;
           K(2): 0.96\,pc; 5$^\circ$;
           K(3): 1.08\,pc; 15$^\circ$.           }
  \label{figure:h2_map_A090}
\end{figure}

\setcounter{figure}{090}
\begin{figure}
  \resizebox{0.95\linewidth}{!}{\includegraphics{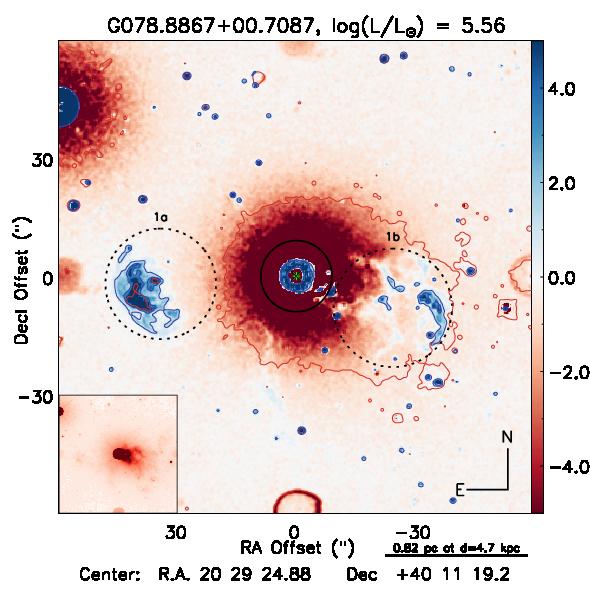}}
  \caption{G078.8867+00.7087, $\log(L/L_\odot)=5.56$, d=4.7\,kpc, Sp. Type: O5\,V, H-K$_s$ = 4.20\,mag.
           BP2(1a,1b): 1.02, 0.91\,pc; 100$^\circ$, 250$^\circ$; 1.8, 1.4.          }
  \label{figure:h2_map_A091}
\end{figure}

\setcounter{figure}{091}
\begin{figure}
  \resizebox{0.95\linewidth}{!}{\includegraphics{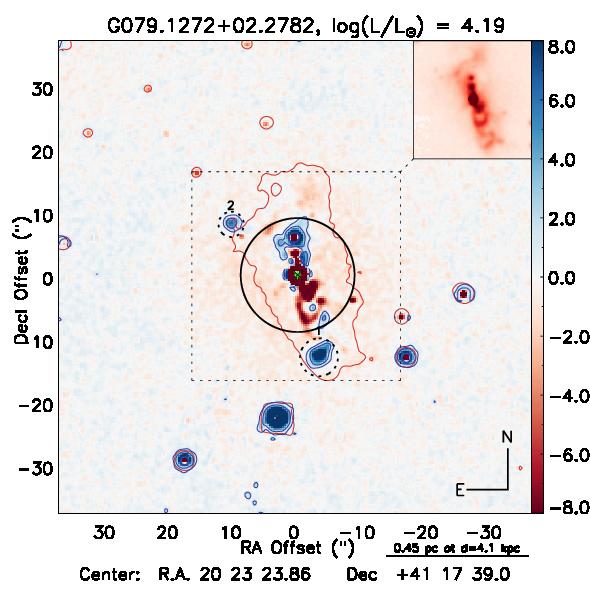}}
  \caption{G079.1272+02.2782, $\log(L/L_\odot)=4.19$, d=4.1\,kpc, Sp. Type: B0.5\,V$_0$, H-K$_s$ = 2.60\,mag.
           K(1): 0.28\,pc; 60; 
           BP2(2): 0.28\,pc; 0$^\circ$, 195$^\circ$; 3.4.          }
  \label{figure:h2_map_A092}
\end{figure}

\setcounter{figure}{096}
\begin{figure}
  \resizebox{0.95\linewidth}{!}{\includegraphics{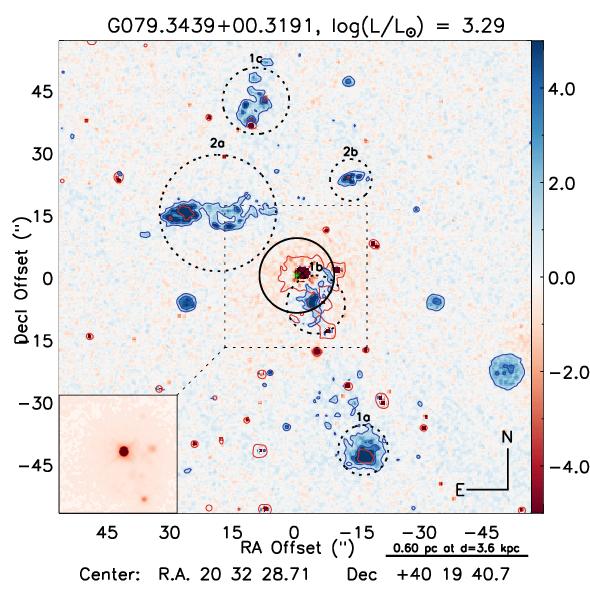}}
  \caption{G079.3439+00.3191, $\log(L/L_\odot) = 3.29$, d=0.60\,kpc, Sp. Type: B3\,V$_0$, H-K$_s$ = 2.45\,mag.
           BP2(1a,1b): 0.85, 0.76\,pc; 200$^\circ$, 15$^\circ$; 4.8, 5.2.          }
  \label{figure:h2_map_A097}
\end{figure}

\setcounter{figure}{097}
\begin{figure}
  \resizebox{0.95\linewidth}{!}{\includegraphics{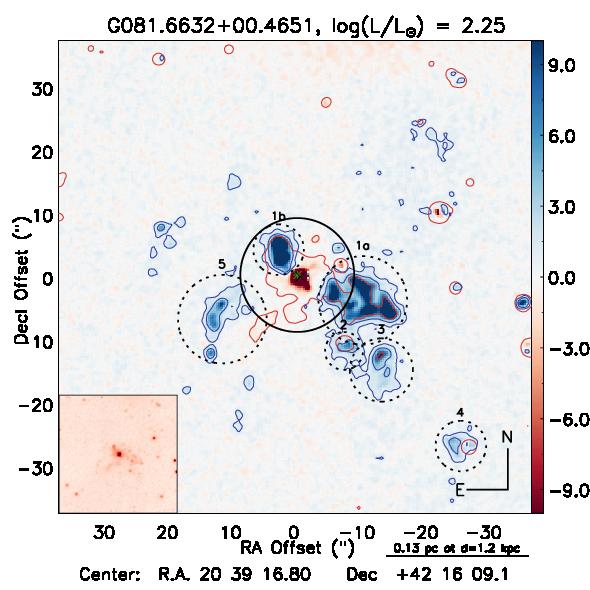}}
  \caption{G081.6632+00.4651, $\log(L/L_\odot) = 2.25$, d=1.2\,kpc, Sp. Type: $<$B3\,V$_0$, H-K$_s$ = 3.82\,mag.
           BP2(1a,1b): 0.11, 0.05\,pc; 250$^\circ$, 30$^\circ$; 2.3, 1.5;
           K(2): 0.09\,pc; 215$^\circ$;
           K(3): 0.14\,pc; 225$^\circ$;
           K(4): 0.23\,pc; 225$^\circ$;
           K(5): 0.10\,pc; 115$^\circ$.          }
  \label{figure:h2_map_A098}
\end{figure}

\setcounter{figure}{098}
\begin{figure}
  \resizebox{0.95\linewidth}{!}{\includegraphics{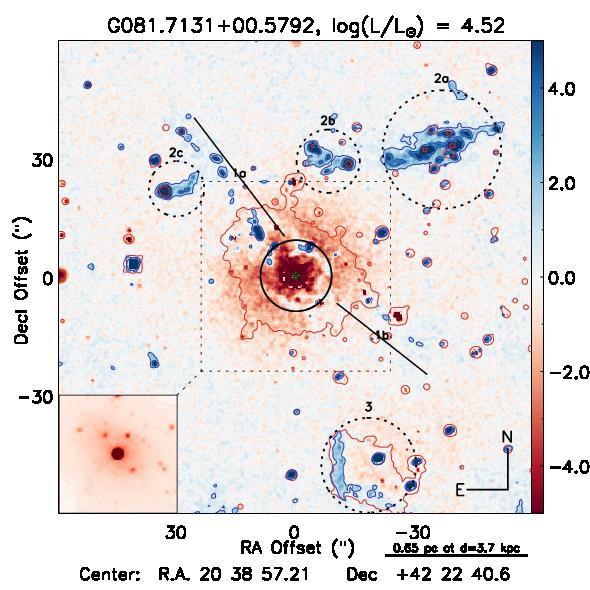}}
  \caption{G081.7131+00.5792, $\log(L/L_\odot) = 4.52$, d=3.7\,kpc, Sp. Type: B0\,V, H-K$_s$ = 2.13\,mag.
           BP2(1a,1b): 0.90, 0.67\,pc; 40$^\circ$, 235$^\circ$; 11.2, 8.4;
           BP2(2a,2b): 0.86, 0.80\,pc; 50$^\circ$, 310$^\circ$; 4.5, 8.3;
           D(3): 1.12\,pc; 195$^\circ$.          }
  \label{figure:h2_map_A099}
\end{figure}

\setcounter{figure}{101}
\begin{figure}
  \resizebox{0.95\linewidth}{!}{\includegraphics{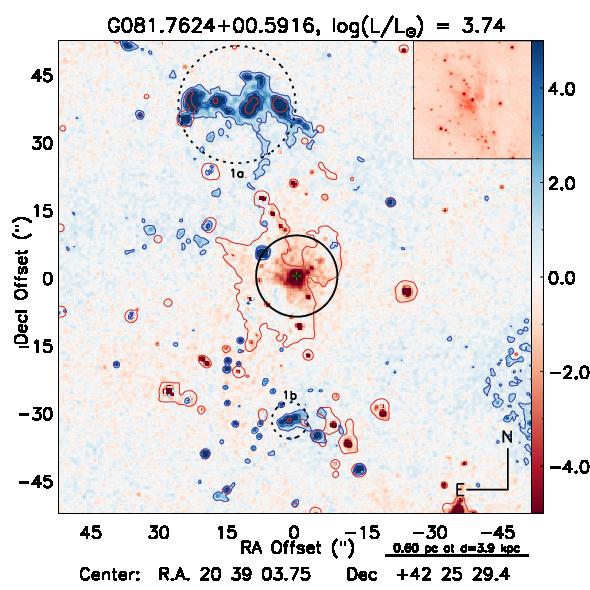}}
  \caption{G081.7624+00.5916, $\log(L/L_\odot)=3.74$, d=3.9\,kpc, Sp. Type: B1\,V$_0$, H-K$_s$ = 3.81\,mag.
           BP2(1a,1b): 0.77, 0.60\,pc; 20$^\circ$, 175$^\circ$; 1.8, 7.0.          }
  \label{figure:h2_map_A102}
\end{figure}

\setcounter{figure}{102}
\begin{figure}
  \resizebox{0.95\linewidth}{!}{\includegraphics{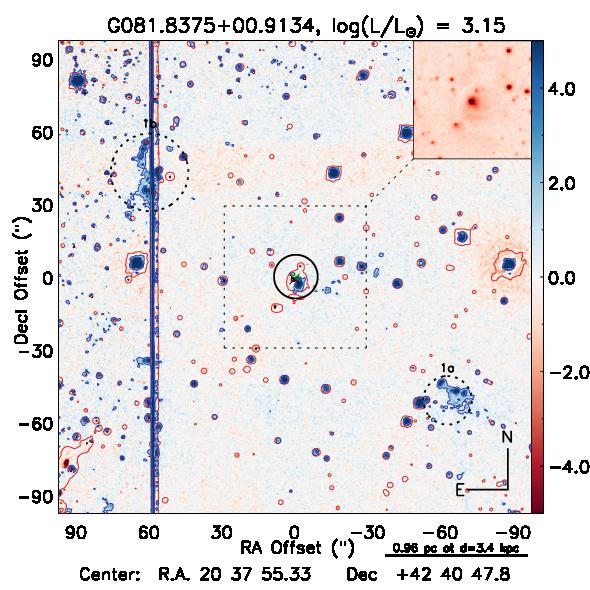}}
  \caption{G081.8375+00.9134, $\log(L/L_\odot) = 3.15$, d=3.4\,kpc, Sp. Type: B3\,V$_0$, H-K$_s$ = 3.20\,mag.
           BP2(1a,1b): 1.31, 1.31\,pc; 55$^\circ$, 235$^\circ$; 4.0, 9.3.          }
  \label{figure:h2_map_A103}
\end{figure}

\setcounter{figure}{103}
\begin{figure}
  \resizebox{0.95\linewidth}{!}{\includegraphics{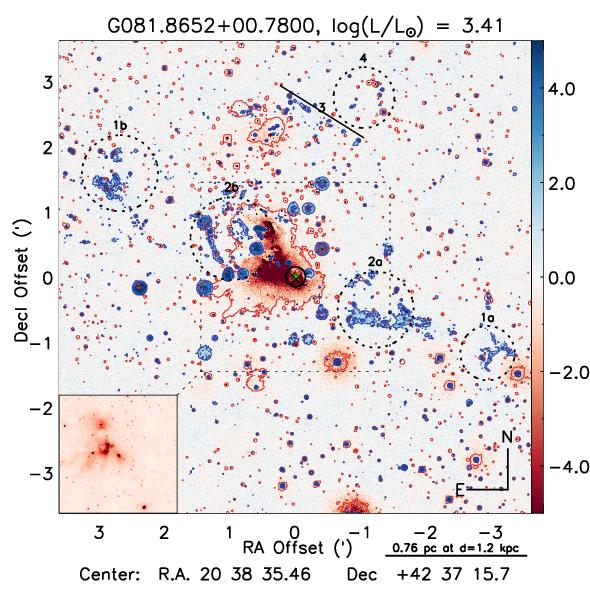}}
  \caption{G081.8652+00.7800, $\log(L/L_\odot)=3.41$, d=1.2\,kpc, Sp. Type: B3\,V$_0$, H-K$_s$ = 3.29\,mag.
           BP2(1a,1b): 1.25, 1.25\,pc; 60$^\circ$, 250$^\circ$; 4.6, 5.7;
           BP2(2a,2b): 0.60, 0.62\,pc; 60$^\circ$, 250$^\circ$; 2.2, 2.5;
           BP2(3): 0.46\,pc; 350$^\circ$; 5.6;
           K(4): 1.14\,pc; 340$^\circ$.          }
  \label{figure:h2_map_A104}
\end{figure}

\setcounter{figure}{104}
\begin{figure}
  \resizebox{0.95\linewidth}{!}{\includegraphics{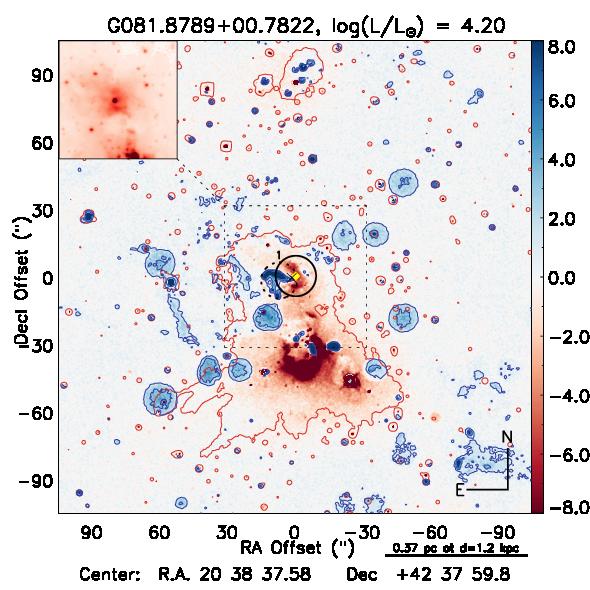}}
  \caption{G081.8789+00.7822, $\log(L/L_\odot) = 4.20$, d=1.2\,kpc, Sp. Type: B0.5\,V$_0$, H-K$_s$ = 1.39\,mag.
           BP1(1): 0.09\,pc; 105$^\circ$; 6.0.          }
  \label{figure:h2_map_A105}
\end{figure}

\setcounter{figure}{110}
\begin{figure}
  \resizebox{0.95\linewidth}{!}{\includegraphics{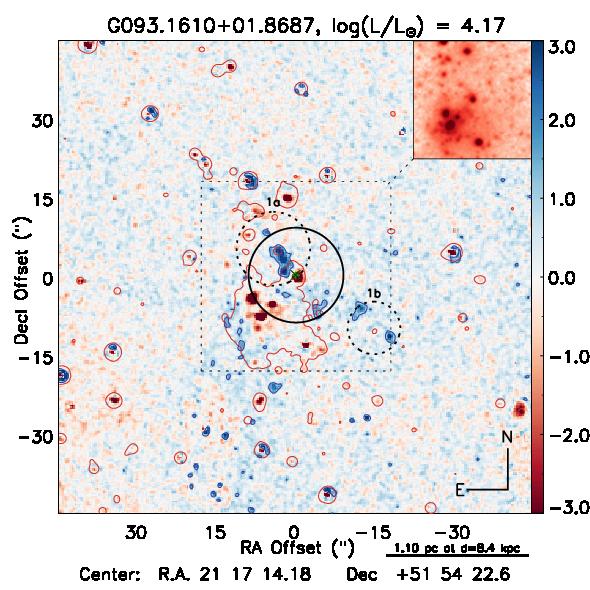}}
  \caption{G093.1610+01.8687, $\log(L/L_\odot)=4.17$, d=8.4\,kpc, Sp. Type: B0.5\,V$_0$, H-K$_s$ = 0.73\,mag.
           BP2(1a,1b): 0.45, 0.84\,pc; 45$^\circ$, 240$^\circ$; 3.5, 7.4.          }
  \label{figure:h2_map_A111}
\end{figure}

\setcounter{figure}{111}
\begin{figure}
  \resizebox{0.95\linewidth}{!}{\includegraphics{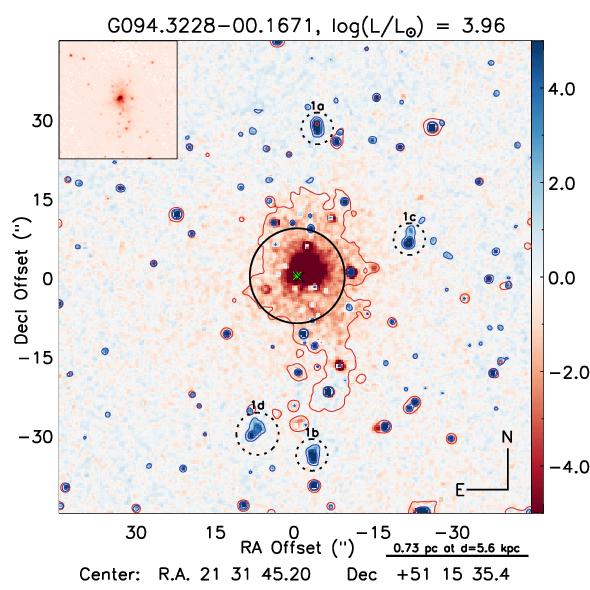}}
  \caption{G094.3228-00.1671, $\log(L/L_\odot)=3.96$, d=5.6\,kpc, Sp. Type: B1\,V$_0$, H-K$_s$ = 2.22\,mag.
           BP4(1a,1b,1c,1d): 0.75, 1.00, 0.58, 0.94\,pc; 355$^\circ$, 185$^\circ$, 290$^\circ$, 165$^\circ$; 16.0, 15.0, 6.8, 11.0. }
  \label{figure:h2_map_A112}
\end{figure}

\setcounter{figure}{112}
\begin{figure}
  \resizebox{0.95\linewidth}{!}{\includegraphics{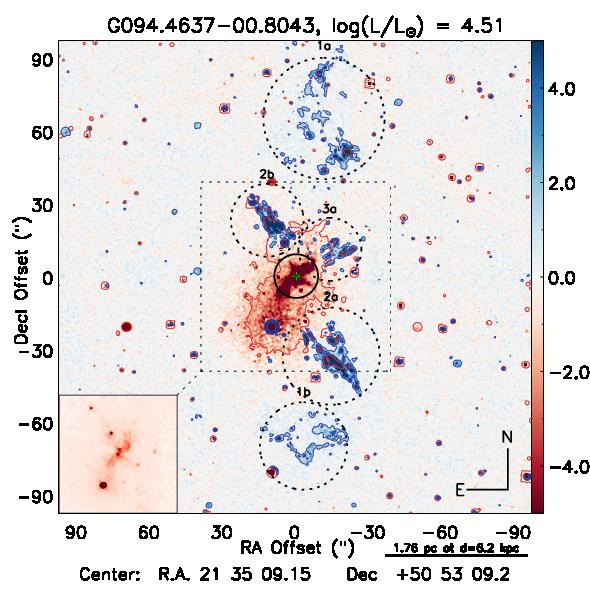}}
  \caption{G094.4637-00.8043, $\log(L/L_\odot)=4.51$, d=6.2\,kpc, Sp. Type: B0\,V$_0$, H-K$_s$ = 2.90\,mag.
           BP5(1a,1b): 2.70, 2.38\,pc; 355$^\circ$, 180$^\circ$; 3.5, 3.3;
           BP5(2a,2b): 1.66, 1.24\,pc; 210$^\circ$, 30$^\circ$; 3.2, 3.0;
           BP5(3a): 0.78\,pc; 300$^\circ$; 2.5.          }
  \label{figure:h2_map_A113}
\end{figure}

\setcounter{figure}{118}
\begin{figure}
  \resizebox{0.95\linewidth}{!}{\includegraphics{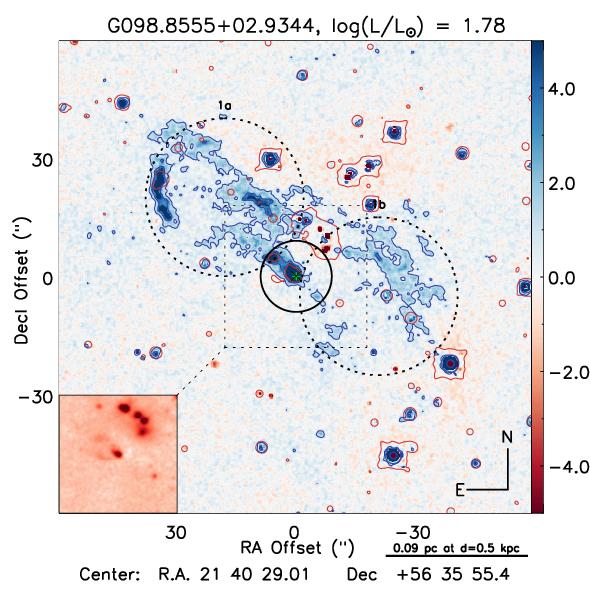}}
  \caption{G098.8555+02.9344, $\log(L/L_\odot)=1.78$, d=0.5\,kpc, Sp. Type: $<$B3\,V$_0$, H-K$_s$ = 1.51\,mag.
           BP2(1a,1b): 0.12, 0.10\,pc; 45$^\circ$, 260$^\circ$; 3.3, -.          }
  \label{figure:h2_map_A119}
\end{figure}

\setcounter{figure}{121}
\begin{figure}
  \resizebox{0.95\linewidth}{!}{\includegraphics{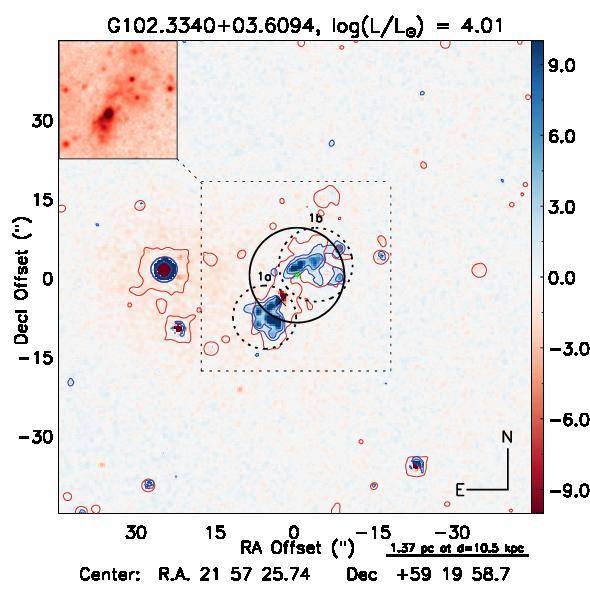}}
  \caption{G102.3340+03.6094, $\log(L/L_\odot)=4.01$, d=10.5\,kpc, Sp. Type: B1\,V$_0$, H-K$_s$ = 1.51\,mag.
           BP2(1a,1b): 0.44, 0.73\,pc; 135$^\circ$, 315$^\circ$; 0.9, 2.0.          }
  \label{figure:h2_map_A122}
\end{figure}

\setcounter{figure}{124}
\begin{figure}
  \resizebox{0.95\linewidth}{!}{\includegraphics{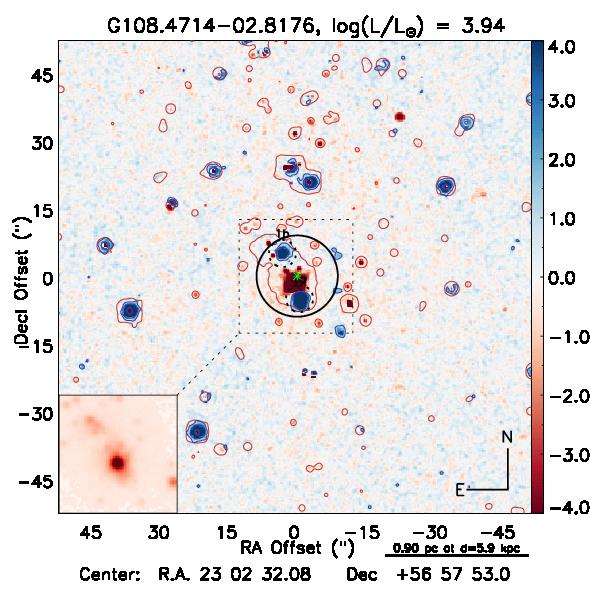}}
  \caption{G108.4714-02.8176, $\log(L/L_\odot)=3.94$, d=5.9\,kpc, Sp. Type: B1\,V$_0$, H-K$_s$ = 2.16\,mag.
           BP2(1a,1b): 0.19, 0.26\,pc; 20$^\circ$, 195$^\circ$; 2.0, 4.5.          }
  \label{figure:h2_map_A125}
\end{figure}

\setcounter{figure}{125}
\begin{figure}
  \resizebox{0.95\linewidth}{!}{\includegraphics{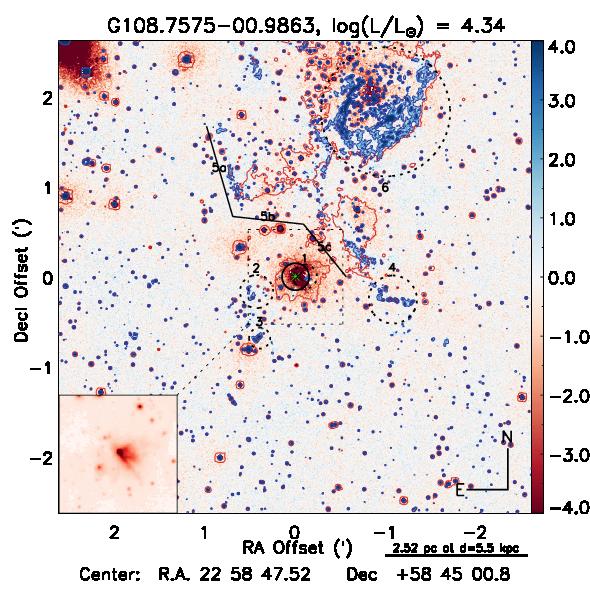}}
  \caption{G108.7575-00.9863, $\log(L/L_\odot)=4.35$, d=5.5\,kpc, Sp. Type: B0.5\,V$_0$, H-K$_s$ = 5.05\,mag.
           BP1(1): 0.32\,pc; 270$^\circ$; 1.1;
           K(2): 0.98\,pc; 110$^\circ$;
           K(3): 1.30\,pc; 150$^\circ$;
           BP1(4): 0.70\,pc; 255$^\circ$; 5.7;
           D(5a,5b,5c): 4.10\,pc; 275$^\circ < \theta <$ 30$^\circ$;
           D(6): 2.40\,pc; 50$^\circ$, 240$^\circ$.          }
  \label{figure:h2_map_A126}
\end{figure}

\setcounter{figure}{126}
\begin{figure}
  \resizebox{0.95\linewidth}{!}{\includegraphics{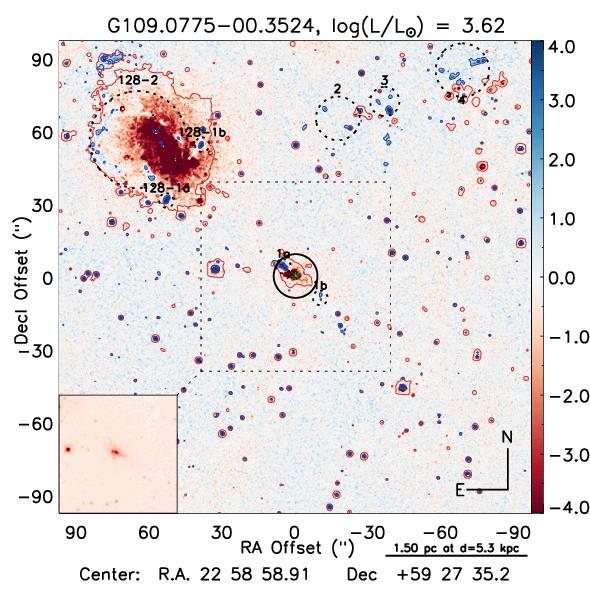}}
  \caption{G109.0775-00.3524, $\log(L/L_\odot)=3.62$, d=5.3\,kpc, Sp. Type: B2\,V$_0$, H-K$_s$ = 3.01\,mag.
           BP2(1a,1b): 0.22, 0.35\,pc; 60$^\circ$, 230$^\circ$; 3.5, 7.0;
           BP2(128-1a,128-1b): -; 40$^\circ$, 60$^\circ$; 5.2, 7.8;
           D(128-2): -; 50$^\circ$;
           K(2): 1.80\,pc; 345$^\circ$;
           K(3): 2.10\,pc; 335$^\circ$;
           K(4): 2.81\,pc; 320$^\circ$.           }
  \label{figure:h2_map_A127}
\end{figure}

\setcounter{figure}{127}
\begin{figure}
  \resizebox{0.95\linewidth}{!}{\includegraphics{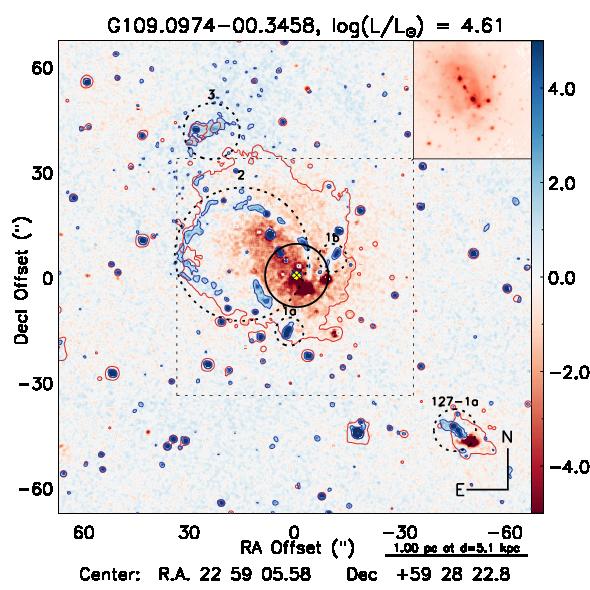}}
  \caption{G109.0974-00.3458, $\log(L/L_\odot) = 4.61$, d=5.1\,kpc, Sp. Type: B0\,V$_0$, H-K$_s$ = 1.07\,mag.
           BP2(1a,1b): 0.39, 0.39\,pc; 150$^\circ$, 320$^\circ$; 5.2, 7.8;
           D(2): 0.93\,pc; 55$^\circ$;
           K(3): 1.25\,pc; 35$^\circ$.          }
  \label{figure:h2_map_A128}
\end{figure}

\setcounter{figure}{128}
\begin{figure}
  \resizebox{0.95\linewidth}{!}{\includegraphics{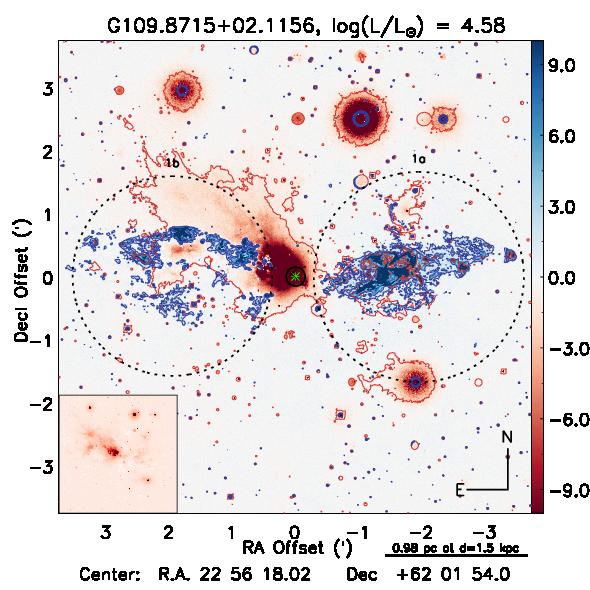}}
  \caption{G109.8715+02.1156, $\log(L/L_\odot)=4.58$, d=1.5\,kpc, Sp. Type: B0\,V$_0$, H-K$_s$ = 2.44\,mag.
           BP2(1a,1b): 1.64, 1.61\,pc; 75$^\circ$, 275$^\circ$; 4.8, 5.0.          }
  \label{figure:h2_map_A129}
\end{figure}

\setcounter{figure}{129}
\begin{figure}
  \resizebox{0.95\linewidth}{!}{\includegraphics{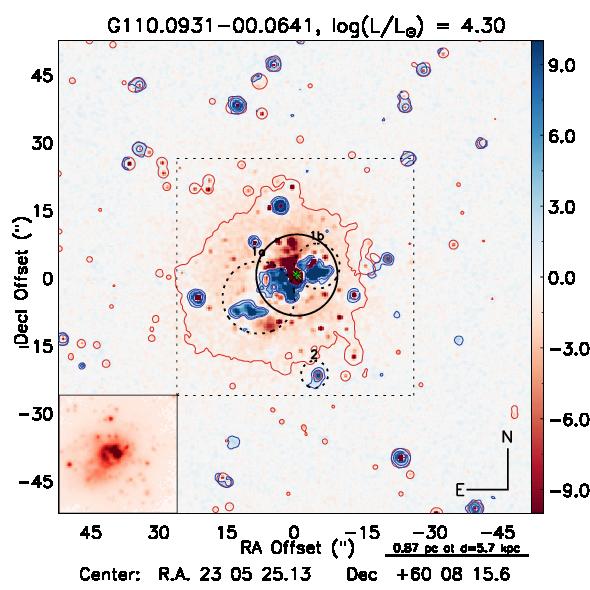}}
  \caption{G110.0931-00.0641, $\log(L/L_\odot)=4.30$, d=5.7\,kpc, Sp. Type: B0.5\,V$_0$, H-K$_s$ = 1.60\,mag.
           BP2(1a,1b): 0.43, 0.59\,pc; 125$^\circ$, 275$^\circ$; 1.6, 3.6;
           K(2): 0.64\,pc; 195$^\circ$.          }
  \label{figure:h2_map_A130}
\end{figure}

\setcounter{figure}{131}
\begin{figure}
  \resizebox{0.95\linewidth}{!}{\includegraphics{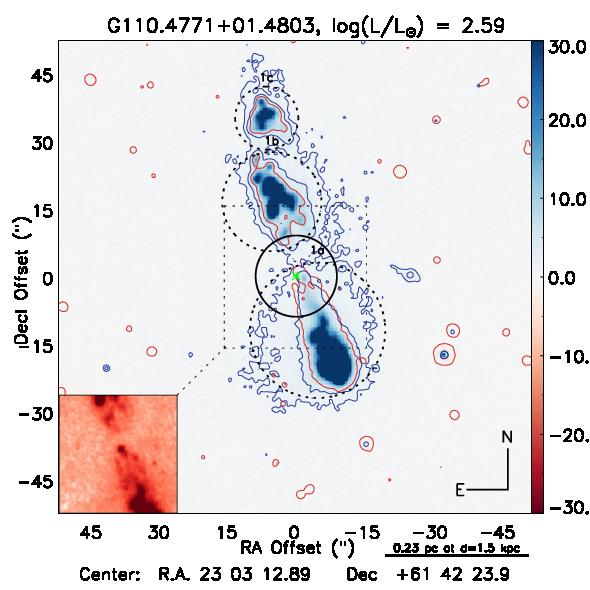}}
  \caption{G110.4771+01.4803, $\log(L/L_\odot)=2.59$, d=1.5\,kpc, Sp. Type: B3\,V$_0$.
           BP2(1a,1b,1c): 0.20, 0.30\,pc; 15$^\circ$, 205$^\circ$; 3.1, 4.1.          }
  \label{figure:h2_map_A132}
\end{figure}

\setcounter{figure}{135}
\begin{figure}
  \resizebox{0.95\linewidth}{!}{\includegraphics{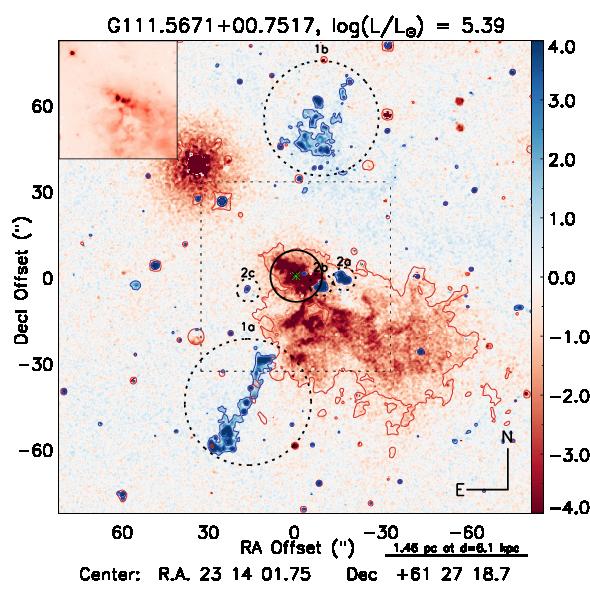}}
  \caption{G111.5671+00.7517, $\log(L/L_\odot)=5.39$, d=6.1\,kpc, Sp. Type: O6\,V, H-K$_s$ = 3.76\,mag.
           BP4(1a,1b): 2.06, 2.35\,pc; 155$^\circ$, 350$^\circ$; 7.2, 5.6;
           BP4(2a,2b,2c): 0.56, 0.56\,pc; 95$^\circ$, 270$^\circ$; 4.8, 6.7.          }
  \label{figure:h2_map_A136}
\end{figure}

\setcounter{figure}{138}
\begin{figure}
  \resizebox{0.95\linewidth}{!}{\includegraphics{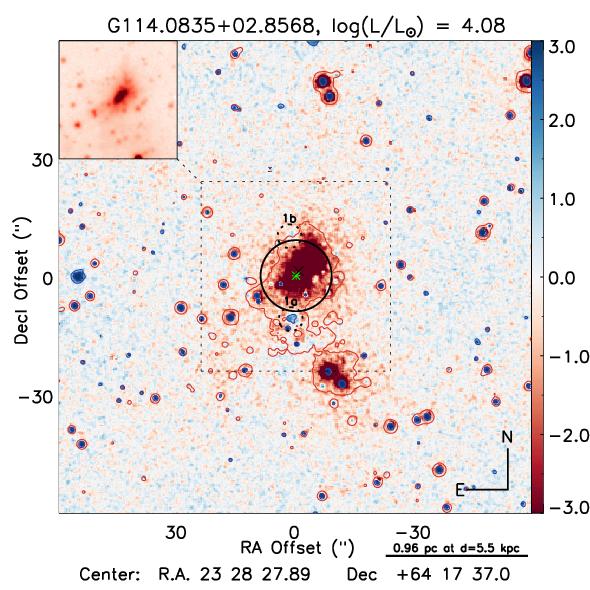}}
  \caption{G114.0835+02.8568, $\log(L/L_\odot)=4.08$, d=5.5\,kpc, Sp. Type: B0.5\,V$_0$, H-K$_s$ = 3.00\,mag.
           BP2(1a,1b): 0.35, 0.26\,pc; 160$^\circ$, 15$^\circ$; 4.8, 7.5.          }
  \label{figure:h2_map_A139}
\end{figure}

\setcounter{figure}{139}
\begin{figure}
  \resizebox{0.95\linewidth}{!}{\includegraphics{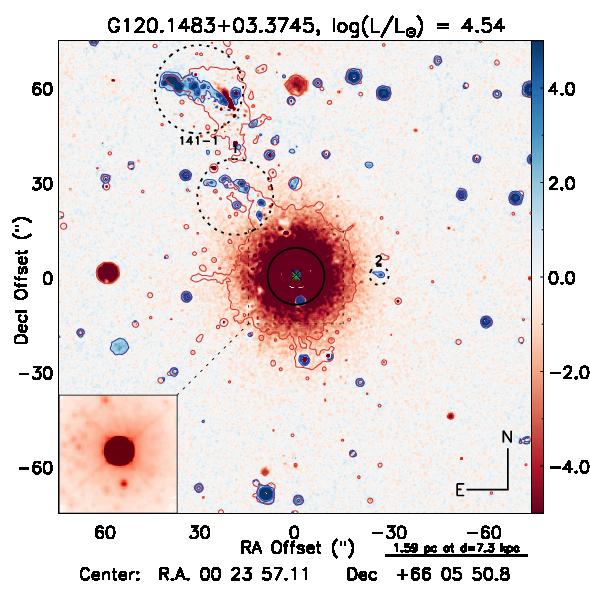}}
  \caption{G120.1483+03.3745, $\log(L/L_\odot)=4.54$, d=7.3\,kpc, Sp. Type: B0\,V$_0$, H-K$_s$ = 2.02\,mag.
           BP1(1): 1.40\,pc; 65$^\circ$; 6.5;
           K(2):   0.94\,pc; 270$^\circ$.
           BP1(141-1): -; 30$^\circ$; 4.6.          }
  \label{figure:h2_map_A140}
\end{figure}

\setcounter{figure}{140}
\begin{figure}
  \resizebox{0.95\linewidth}{!}{\includegraphics{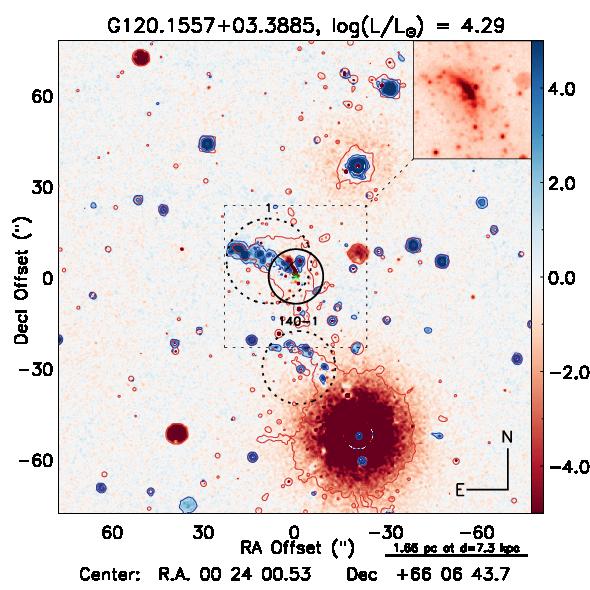}}
  \caption{G120.1557+03.3885, $\log(L/L_\odot) = 4.29$, d=7.3\,kpc, Sp. Type: B0.5\,V$_0$, H-K$_s$ = 2.27\,mag.
           BP1(1): 0.89\,pc; 65$^\circ$; 4.6.
           BP1(140-1): -; 185$^\circ$; 6.5.          }
  \label{figure:h2_map_A141}
\end{figure}

\setcounter{figure}{141}
\begin{figure}
  \resizebox{0.95\linewidth}{!}{\includegraphics{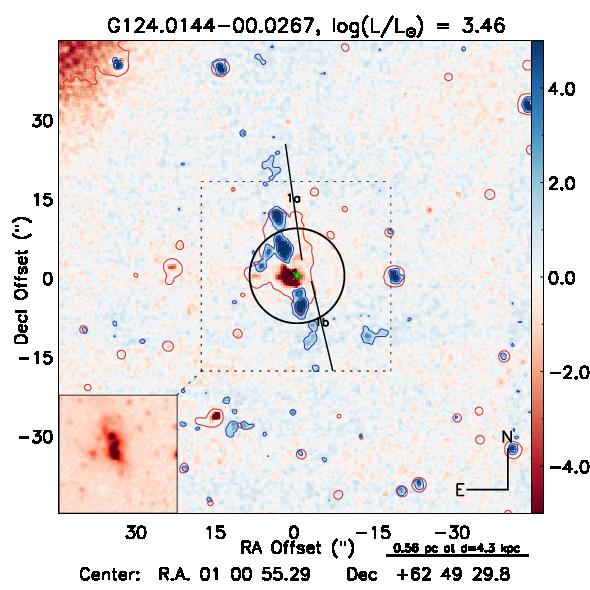}}
  \caption{G124.0144-00.0267, $\log(L/L_\odot)=3.46$, d=4.3\,kpc, Sp. Type: B2\,V$_0$, H-K$_s$ = 1.38\,mag.
           BP2(1a,1b): 0.48, 0.38\,pc; 10$^\circ$, 200$^\circ$; 8.0, 6.3.          }
  \label{figure:h2_map_A142}
\end{figure}

\setcounter{figure}{144}
\begin{figure}
  \resizebox{0.95\linewidth}{!}{\includegraphics{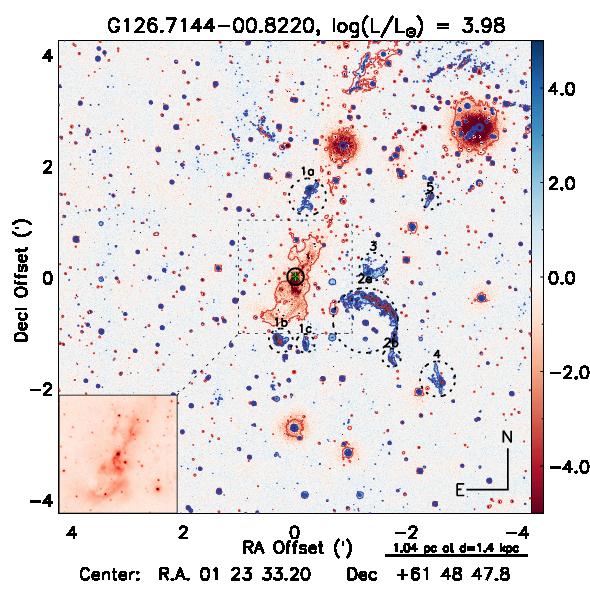}}
  \caption{G126.7144-00.8220, $\log(L/L_\odot)=3.98$, d=1.4\,kpc, Sp. Type: B1\,V$_0$, H-K$_s$ = 4.08\,mag.
           BP2(1a,1b,1c): 0.70, 0.52, 0.52\,pc; 350$^\circ$, 160$^\circ$, 185$^\circ$; 6.5, 4.9, 5.6;
           BP2(2a,2b): 0.52, 0.15\,pc; 245$^\circ$, 230$^\circ$;
           K(3): 0.57\,pc; 280$^\circ$; 
           K(4): 1.35\,pc; 235$^\circ$; 
           K(5): 1.13\,pc; 305$^\circ$.           }
  \label{figure:h2_map_A145}
\end{figure}

\setcounter{figure}{145}
\begin{figure}
  \resizebox{0.95\linewidth}{!}{\includegraphics{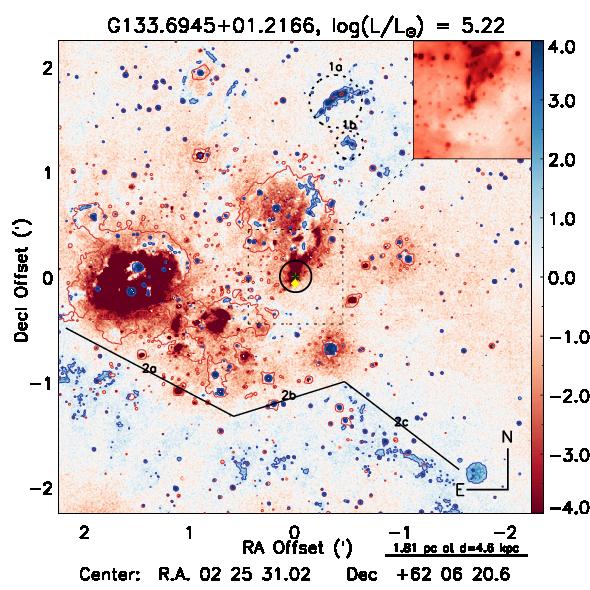}}
  \caption{G133.6945+01.2166, $\log(L/L_\odot)=5.22$, d=4.6\,kpc, Sp. Type: O6.5\,V, H-K$_s$ = 3.38\,mag.
           BP1(1a,1b): 2.51, 2.01\,pc; 350$^\circ$, 340$^\circ$; 6.3, 8.0;
           D(2a,2b,2c): 5.24\,pc; 100-225$^\circ$.          }
  \label{figure:h2_map_A146}
\end{figure}

\setcounter{figure}{148}
\begin{figure}
  \resizebox{0.95\linewidth}{!}{\includegraphics{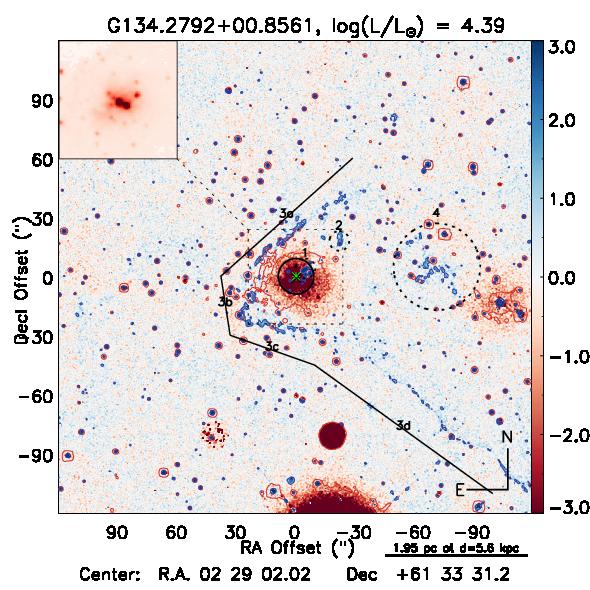}}
  \caption{G134.2792+00.8561, $\log(L/L_\odot)=4.39$, d=5.6\,kpc, Sp. Type: B0.5\,V$_0$, H-K$_s$ = 1.85\,mag.
           BP1(1): 0.38\,pc; 310$^\circ$; 2.0;
           K(2): 0.92\,pc; 320$^\circ$.
           D(3a,3b,3c,3d): 4.88\,pc; 320$^\circ$.
           D(2): 1.90\,pc; 275$^\circ$.          }
  \label{figure:h2_map_A149}
\end{figure}

\setcounter{figure}{149}
\begin{figure}
  \resizebox{0.95\linewidth}{!}{\includegraphics{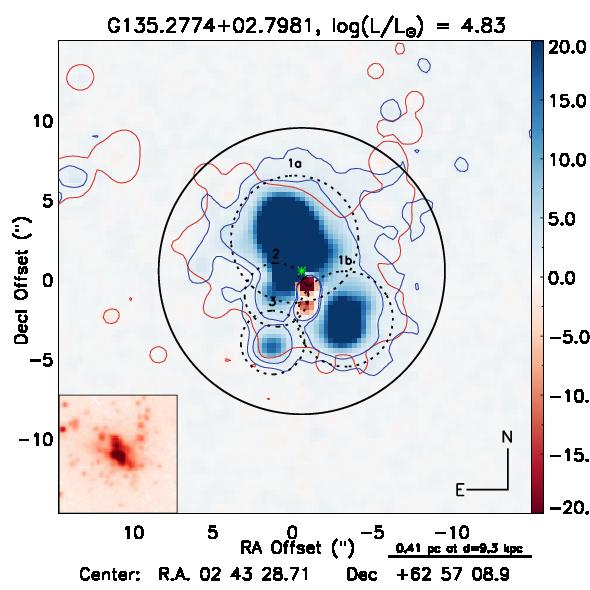}}
  \caption{G135.2774+02.7981, $\log(L/L_\odot)=4.83$, d=9.3\,kpc, Sp. Type: O8.5\,V, H-K$_s$ = 1.64\,mag.
           BP2(1a,1b): 0.34, 0.20\,pc; 15$^\circ$, 240$^\circ$; 1.7, 1.1;
           K(2): 0.17\,pc; 135$^\circ$.          }
  \label{figure:h2_map_A150}
\end{figure}

\setcounter{figure}{152}
\begin{figure}
  \resizebox{0.95\linewidth}{!}{\includegraphics{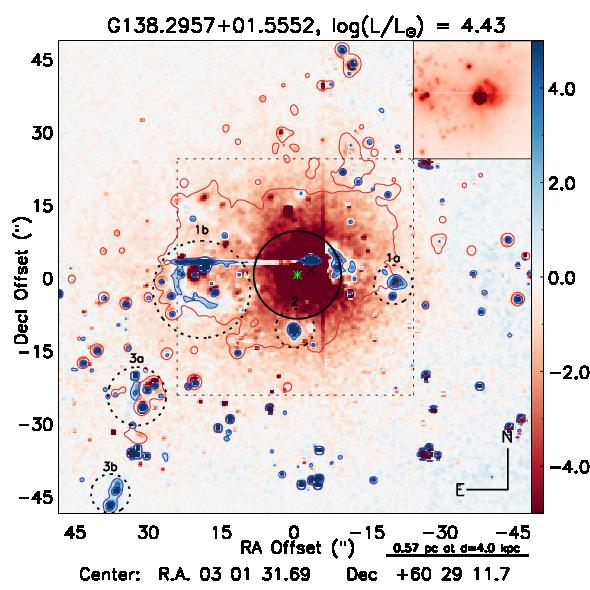}}
  \caption{G138.2957+01.5552, $\log(L/L_\odot)=4.43$, d=4.0\,kpc, Sp. Type: B0\,V, H-K$_s$ = 2.66\,mag.
           BP2(1a,1b): 0.24, 0.24\,pc; 270$^\circ$, 90$^\circ$; 1.7, 1.3; 
           K(2): 0.28\,pc; 180$^\circ$; 
           BP1(3a,3b): 0.28, 0.28\,pc; 125$^\circ$, 140$^\circ$; 7.0, 7.0.          }
  \label{figure:h2_map_A153}
\end{figure}

\setcounter{figure}{154}
\begin{figure}
  \resizebox{0.95\linewidth}{!}{\includegraphics{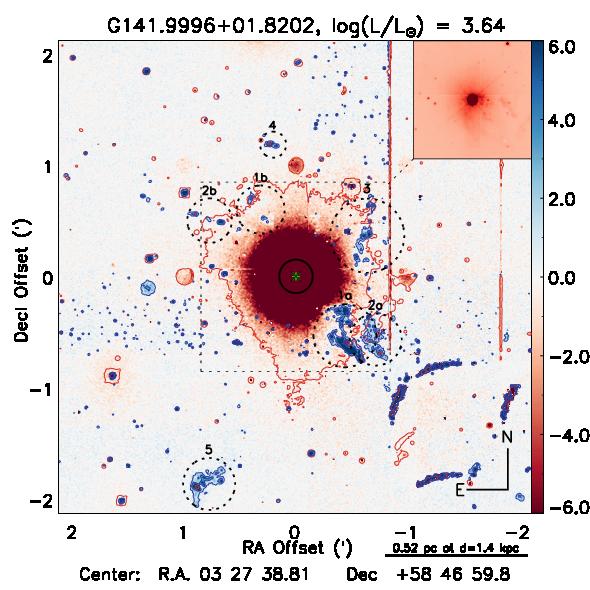}}
  \caption{G141.9996+01.8202, $\log(L/L_\odot)=3.64$, d=1.4\,kpc, Sp. Type: B2\,V$_0$, H-K$_s$ = 1.93\,mag.
           BP4(1a,1b): 0.35, 0.35\,pc; 230$^\circ$, 30$^\circ$; 2.9, 2.8;
           BP4(2a,2b): 0.46, 0.43\,pc; 235$^\circ$, 60$^\circ$; 4.0, 6.8;
           K(3): 0.38\,pc; 300$^\circ$; 
           K(4): 0.49\,pc; 15$^\circ$; 
           K(5): 0.31\,pc; 155$^\circ$.           }
  \label{figure:h2_map_A155}
\end{figure}

\setcounter{figure}{161}
\begin{figure}
  \resizebox{0.95\linewidth}{!}{\includegraphics{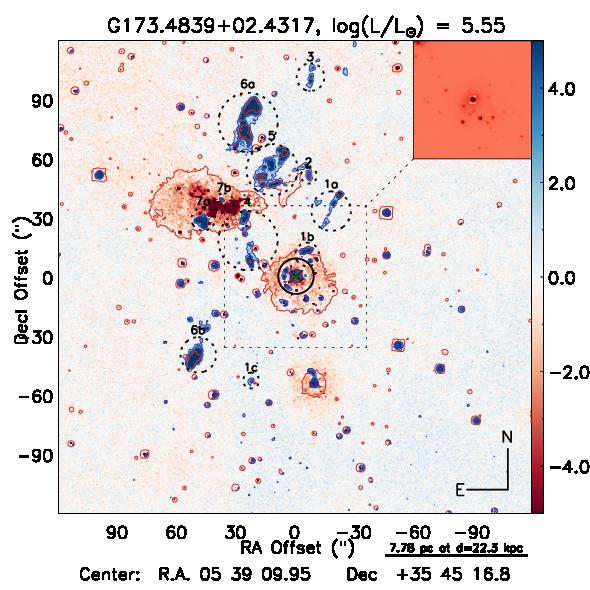}}
  \caption{G173.4839+02.4317, $\log(L/L_\odot)=5.55$, d=22.3\,kpc, Sp. Type: O5\,V, H-K$_s$ = 1.38\,mag.
           BP8(1a,1b,1c): 5.36, 6.38\,pc;  330$^\circ$, 150$^\circ$; 13.7, 26.0; 
           BP8(4): -; 50$^\circ$; 
           BP8(6a,6b): -; 20$^\circ$, 125$^\circ$;
           BP8(7a,7b): -; 60$^\circ$, 50$^\circ$;
           K(2): 5.99\,pc; 355$^\circ$;
           K(3): 11.60\,pc; 355$^\circ$; 
           K(5): -; 15$^\circ$.          }
  \label{figure:h2_map_A162}
\end{figure}

\setcounter{figure}{162}
\begin{figure}
  \resizebox{0.95\linewidth}{!}{\includegraphics{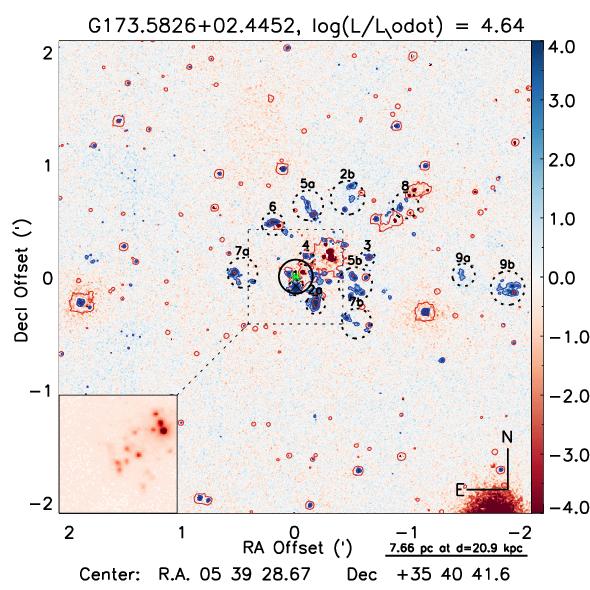}}
  \caption{G173.5826+02.4452, $\log(L/L_\odot)=4.64$, d=20.9\,kpc, Sp. Type: O9.5\,V, H-K$_s$ = 1.76\,mag.
           K(1): 0.72\,pc; 180$^\circ$;
           BP6(2a,2b): 7.16\,pc; 210$^\circ$, 330$^\circ$; 4.8;
           BP6(5a,5b): 6.28\,pc; 0$^\circ$, 265$^\circ$; 5.0;
           BP6(7a,7b): 8.16\,pc; 90$^\circ$, 135$^\circ$;  5.4;
           K(3): 3.89\,pc; 285$^\circ$;
           K(4): 0.88\,pc; 345$^\circ$;
           K(6): 3.39\,pc; 30$^\circ$;
           K(9a): 7.66\,pc; 270$^\circ$;
           K(9b): 10.17\,pc; 270$^\circ$.          }
  \label{figure:h2_map_A163}
\end{figure}

\setcounter{figure}{166}
\begin{figure}
  \resizebox{0.95\linewidth}{!}{\includegraphics{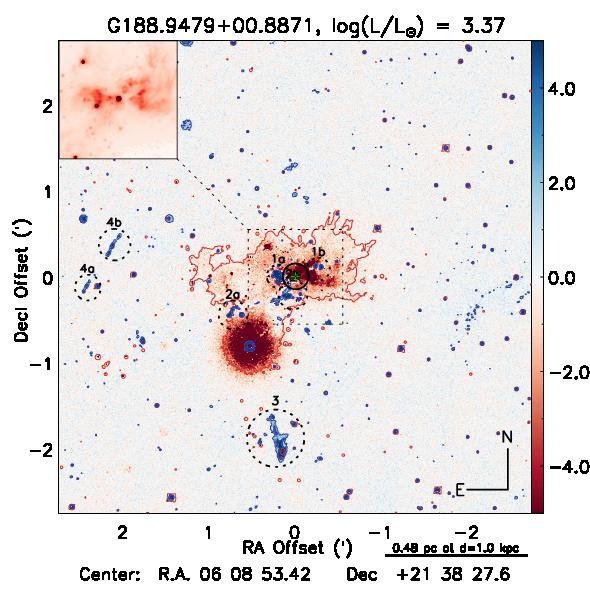}}
  \caption{G188.9479+00.8871, $\log(L/L_\odot)=3.37$, d=1.0\,kpc, Sp. Type: B3\,V$_0$, H-K$_s$ = 2.42\,mag.
           BP4(1a,1b): 0.10, 0.10\,pc; 105$^\circ$, 285$^\circ$; 3.0, 4.5;
           BP4(2a,2b): 0.12, 0.12\,pc; 120$^\circ$, 135$^\circ$; 3.8, 3.8;
           BP1(3): 0.48\,pc;   165$^\circ$; 8.2;
           BP2(4a,4b): 0.13, 0.13\,pc; 85$^\circ$; 12.5, 12.5.          }
  \label{figure:h2_map_A167}
\end{figure}

\setcounter{figure}{167}
\begin{figure}
  \resizebox{0.95\linewidth}{!}{\includegraphics{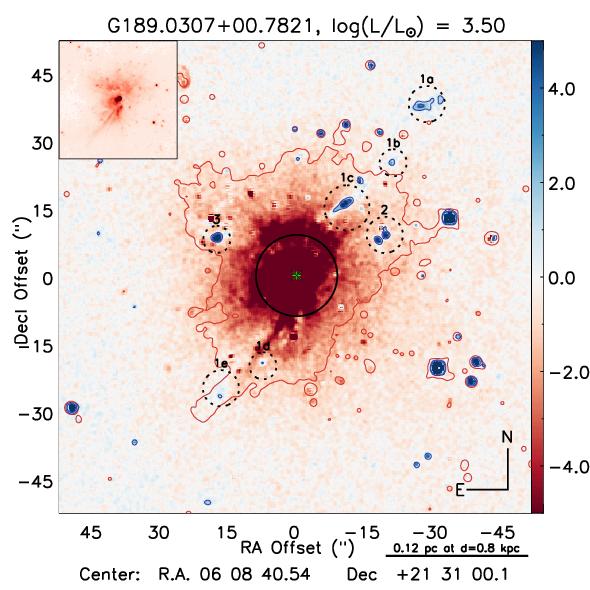}}
  \caption{G189.0307+00.7821, $\log(L/L_\odot)=3.50$, d=0.8\,kpc, Sp. Type: B2\,V$_0$, H-K$_s$ = 4.06\,mag.
           BP2(1a-1e): 0.18, 0.11\,pc; 320$^\circ$, 140$^\circ$; 9.0, 5.0.          }
  \label{figure:h2_map_A168}
\end{figure}

\setcounter{figure}{170}
\begin{figure}
  \resizebox{0.95\linewidth}{!}{\includegraphics{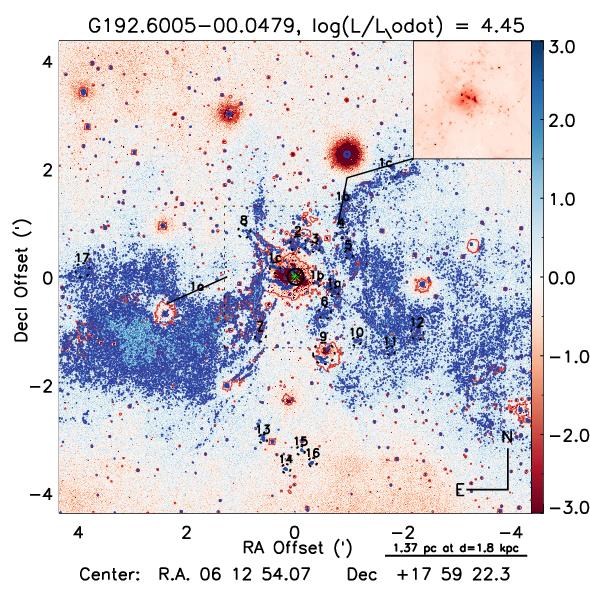}}
  \caption{G192.6005-00.0479, $\log(L/L_\odot)=4.45$, d=1.8\,kpc, Sp. Type: B0\,V, H-K$_s$ = 2.63\,mag.
           BP2(1a,1b,1c): 0.49, 0.23\,pc; 245$^\circ$, 70$^\circ$; 11.3, 8.0;
           K(2-12);
           D(13a,13b,13c): 3.15\,pc.           }
  \label{figure:h2_map_A171}
\end{figure}

\setcounter{figure}{175}
\begin{figure}
  \resizebox{0.95\linewidth}{!}{\includegraphics{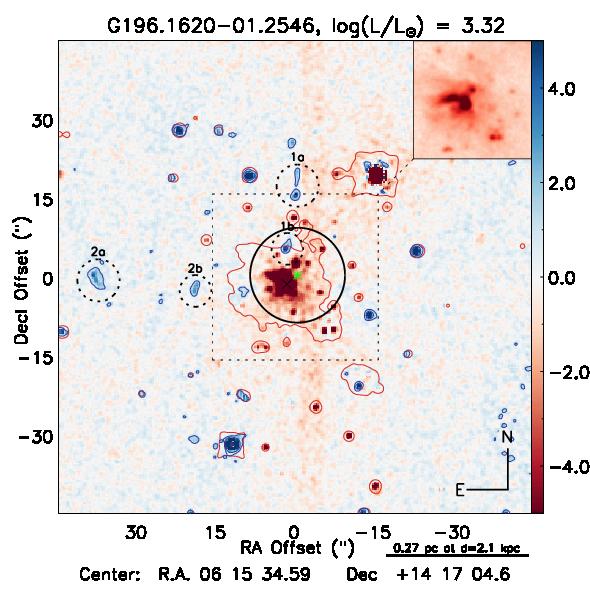}}
  \caption{G196.1620-01.2546, $\log(L/L_\odot)=3.32$, d=2.1\,kpc, Sp. Type: B0\,V$_0$, H-K$_s$ = 2.09\,mag.
           BP2(1a,1b): 0.24\,pc; 0$^\circ$; 16.6;
           BP2(2a,2b): 0.38\,pc; 90$^\circ$, 270$^\circ$; 8.0.          }
  \label{figure:h2_map_A176}
\end{figure}

\setcounter{figure}{178}
\begin{figure}
  \resizebox{0.95\linewidth}{!}{\includegraphics{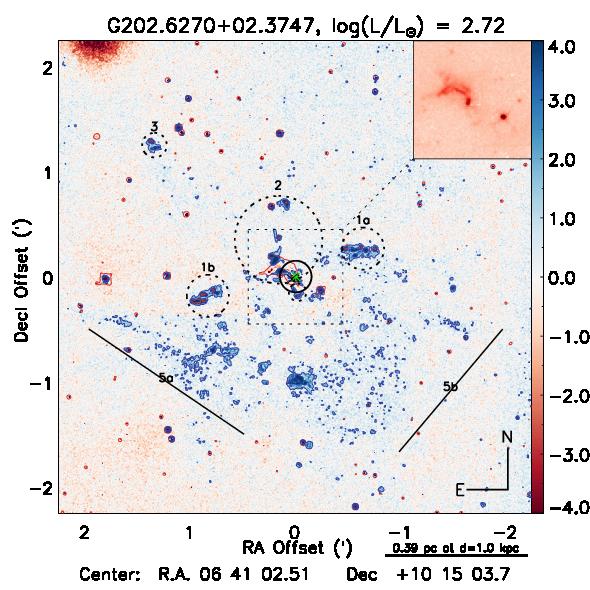}}
  \caption{G202.6270+02.3747, $\log(L/L_\odot)=2.72$, d=1.0\,kpc, Sp. Type: $<$B3\,V$_0$, H-K$_s$ = 2.94\,mag.
           BP3(1a,1b): 0.24, 0.30\,pc; 105$^\circ$, 285$^\circ$; 5.3, 7.4; 
           BP3(2): 0.22\,pc; 10$^\circ$; 6.3;
           K(3): 0.56\,pc; 50$^\circ$; 
           K(4): 0.04\,pc; 185$^\circ$; 
           D(5a,5b): 1.00\,pc; 100-260$^\circ$.          }
  \label{figure:h2_map_A179}
\end{figure}

\setcounter{figure}{180}
\begin{figure}
  \resizebox{0.95\linewidth}{!}{\includegraphics{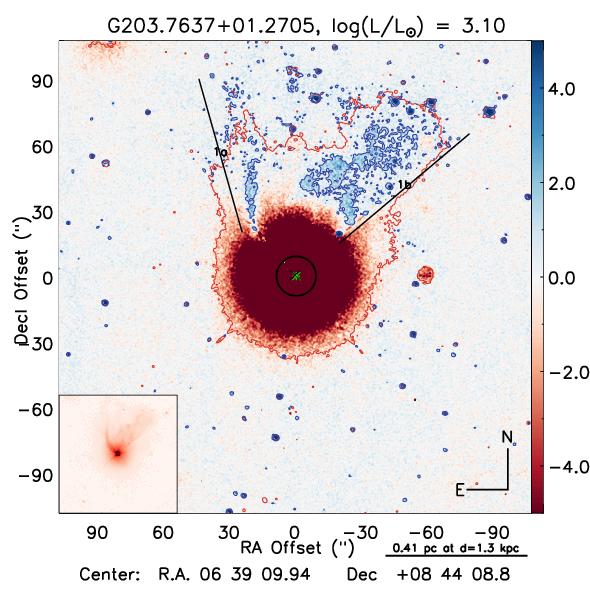}}
  \caption{G203.7637+01.2705, $\log(L/L_\odot)=3.10$, d=1.3\,kpc, Sp. Type: B3\,V$_0$, H-K$_s$ = 1.64\,mag.
           BP1(1a,1b): 0.59\,pc; $<30^\circ$, $>310^\circ$; 0.8.          }
  \label{figure:h2_map_A181}
\end{figure}

\setcounter{figure}{182}
\begin{figure}
  \resizebox{0.95\linewidth}{!}{\includegraphics{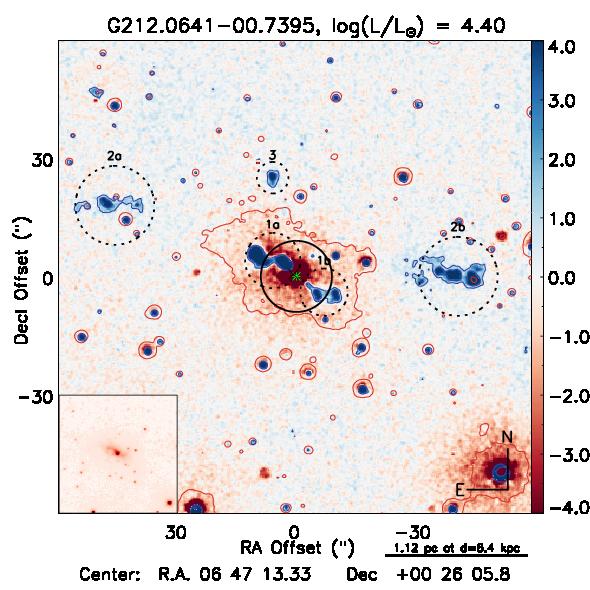}}
  \caption{G212.0641-00.7395, $\log(L/L_\odot)=4.40$, d=6.4\,kpc, Sp. Type: B0\,V, H-K$_s$ = 2.00\,mag.
           BP4(1a,1b): 0.43, 0.36\,pc; 60$^\circ$, 240$^\circ$; 2.2, 2.2; 
           BP4(2a,2b): 1.75, 1.52\,pc; 65$^\circ$, 270$^\circ$; 7.6, 7.7; 
           K(3): 0.89\,pc; 15$^\circ$.          }
  \label{figure:h2_map_A183}
\end{figure}

\setcounter{figure}{186}
\begin{figure}
  \resizebox{0.95\linewidth}{!}{\includegraphics{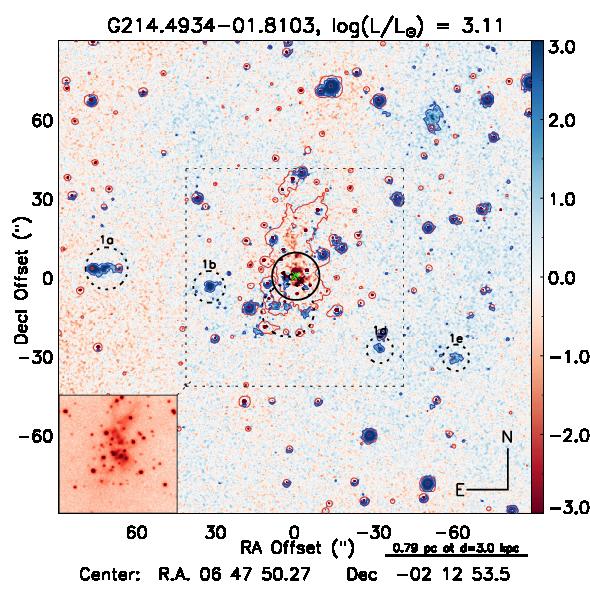}}
  \caption{G214.4934-01.8103, $\log(L/L_\odot)=3.11$, d=3.0\,kpc, Sp. Type: B3\,V$_0$, H-K$_s$ = 0.66\,mag.
           BP2(1a-1e): 1.16, 0.98\,pc; 90$^\circ$, 240$^\circ$; 16.7, 14.0.          }
  \label{figure:h2_map_A187}
\end{figure}

\setcounter{figure}{188}
\begin{figure}
  \resizebox{0.95\linewidth}{!}{\includegraphics{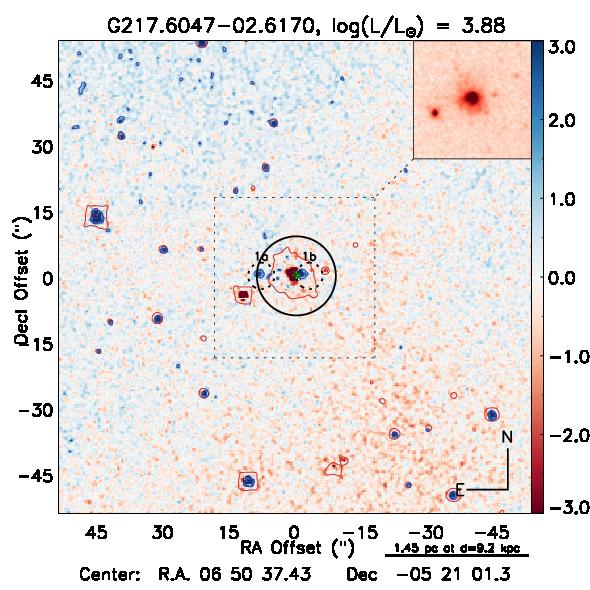}}
  \caption{G217.6047-02.6170, $\log(L/L_\odot)=3.88$, d=9.2\,kpc, Sp. Type: B1\,V$_0$, H-K$_s$ = 1.93\,mag.
           BP2(1a,1b): 0.38, 0.18\,pc; 90$^\circ$, 270$^\circ$; 4.0, 2.0.          }
  \label{figure:h2_map_A189}
\end{figure}

\setcounter{figure}{192}
\begin{figure}
  \resizebox{0.95\linewidth}{!}{\includegraphics{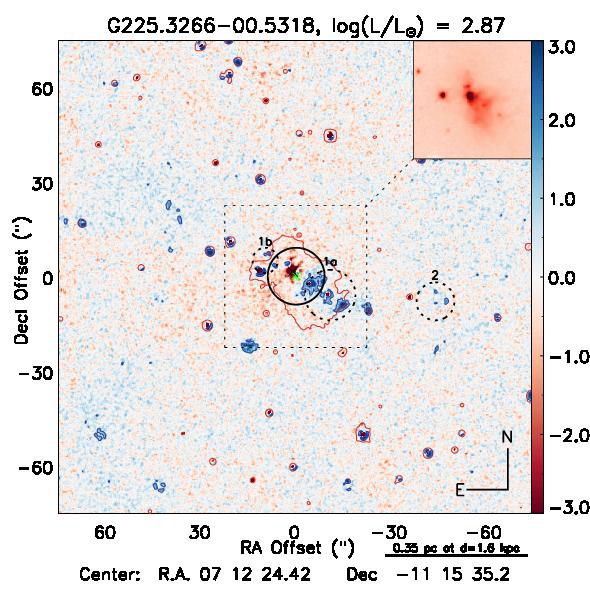}}
  \caption{G225.3266-00.5318, $\log(L/L_\odot)=2.87$, d=1.6\,kpc, Sp. Type: $<$B3\,V$_0$, H-K$_s$ = 2.44\,mag.
           BP2(1a,1b): 0.16, 0.09\,pc; 230$^\circ$, 50$^\circ$; 2.8, 2.5;
           K(2): 0.38\,pc; 260$^\circ$.          }
  \label{figure:h2_map_A193}
\end{figure}

\setcounter{figure}{195}
\begin{figure}
  \resizebox{0.95\linewidth}{!}{\includegraphics{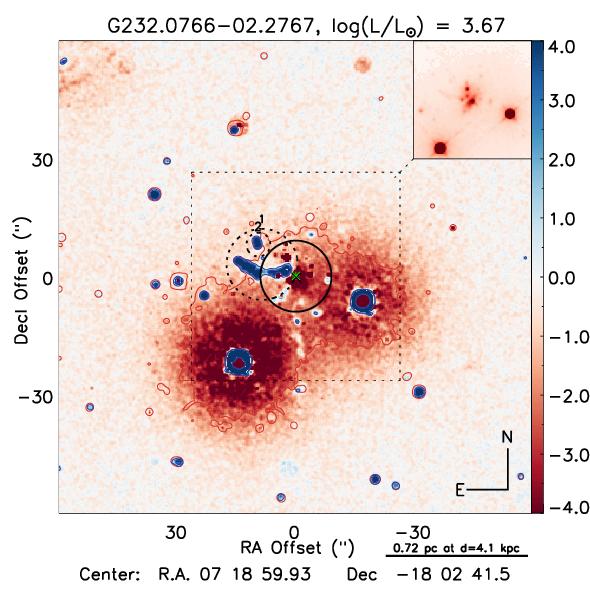}}
  \caption{G232.0766-02.2767, $\log(L/L_\odot)=3.67$, d=4.1\,kpc, Sp. Type: B2\,V$_0$, H-K$_s$ = 1.66\,mag.
           BP1(1): 0.36\,pc; 85$^\circ$; 6.0;
           K(2): 0.30\,pc; 50$^\circ$.          }
  \label{figure:h2_map_A196}
\end{figure}

\setcounter{figure}{203}
\begin{figure}
  \resizebox{0.95\linewidth}{!}{\includegraphics{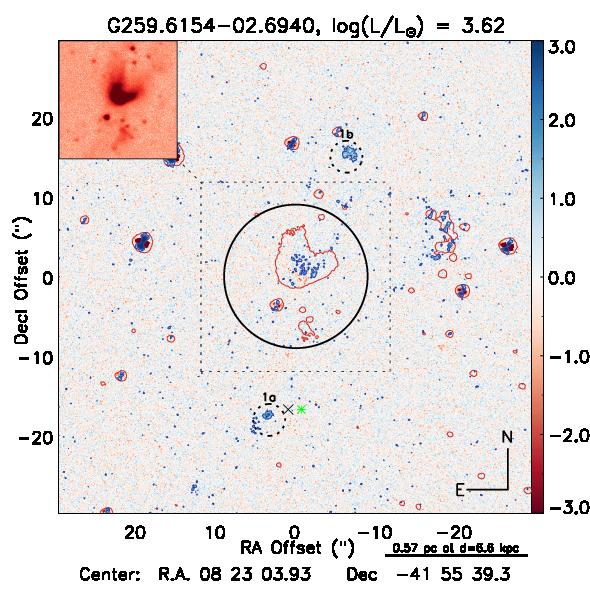}}
  \caption{G259.6154-02.6940, $\log(L/L_\odot)=3.62$, d=6.6\,kpc, Sp. Type: B2\,V$_0$, H-K$_s$ = 1.58\,mag.
           BP2(1a,1b): 0.6, 0.57\,pc; 170$^\circ$, 340$^\circ$; 15.0, 8.0.          }
  \label{figure:h2_map_A204}
\end{figure}

\setcounter{figure}{205}
\begin{figure}
  \resizebox{0.95\linewidth}{!}{\includegraphics{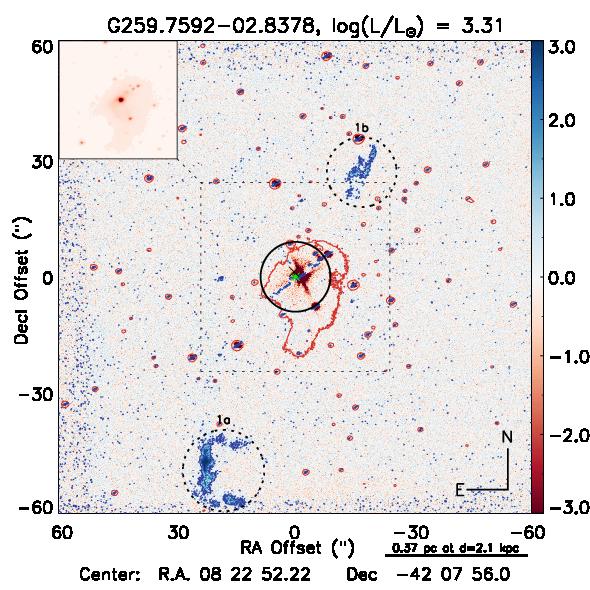}}
  \caption{G259.7592-02.8378, $\log(L/L_\odot)=3.31$, d=2.1\,kpc, Sp. Type: B3\,V$_0$, H-K$_s$ = 3.04\,mag.
           BP2(1a,1b): 0.64, 0.42\,pc; 160$^\circ$, 330$^\circ$; 3.9, 7.0.          }
  \label{figure:h2_map_A206}
\end{figure}

\setcounter{figure}{206}
\begin{figure}
  \resizebox{0.95\linewidth}{!}{\includegraphics{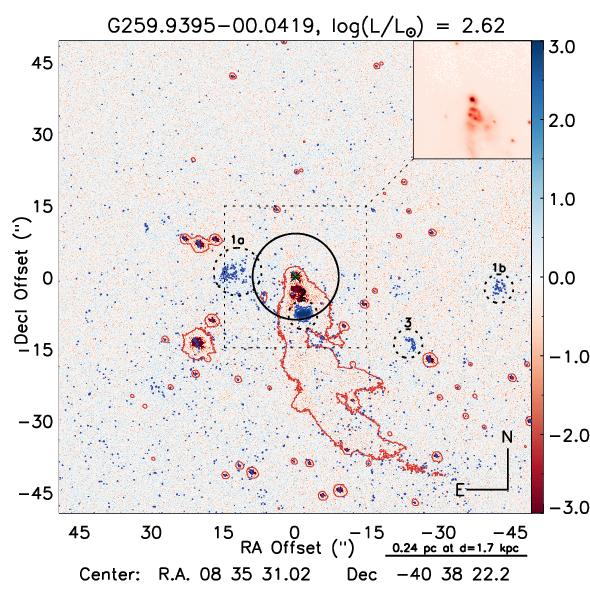}}
  \caption{G259.9395-00.0419, $\log(L/L_\odot)=2.62$, d=1.7\,kpc, Sp. Type: $<$B3\,V$_0$, H-K$_s$ = 1.79\,mag.
           BP2(1a,1b): 0.16, 0.23\,pc; 90$^\circ$, 240$^\circ$; 3.0, 6.3; 
           K(2): 0.07\,pc; 200$^\circ$;
           K(3): 0.23\,pc; 240$^\circ$.          }
  \label{figure:h2_map_A207}
\end{figure}

\setcounter{figure}{208}
\begin{figure}
  \resizebox{0.95\linewidth}{!}{\includegraphics{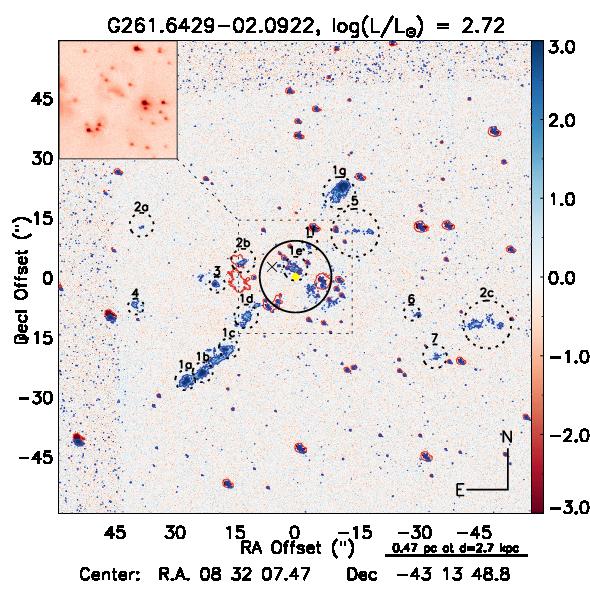}}
  \caption{G261.6429-02.0922, $\log(L/L_\odot)=2.72$, d=2.7\,kpc, Sp. Type: $<$B3\,V$_0$, H-K$_s$ = 0.91\,mag.
           BP4(1a-1g): 0.50, 0.34\,pc; 135$^\circ$, 330$^\circ$; 15.0, 6.7;
           BP4(2a-2c): 0.52, 0.72\,pc; 75$^\circ$, 255$^\circ$; 11.0, 8.0;
           K(3): 0.26\,pc; 90$^\circ$;
           K(4): 0.56\,pc; 100$^\circ$;
           K(5): 0.30\,pc; 320$^\circ$;
           K(6): 0.43\,pc; 255$^\circ$;
           K(7): 0.52\,pc; 240$^\circ$.          }
  \label{figure:h2_map_A209}
\end{figure}

\setcounter{figure}{211}
\begin{figure}
  \resizebox{0.95\linewidth}{!}{\includegraphics{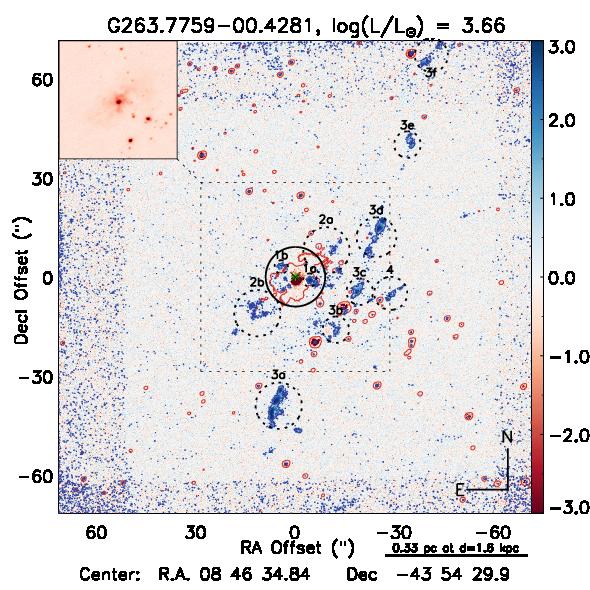}}
  \caption{G263.7759-00.4281, $\log(L/L_\odot)=3.66$, d=1.6\,kpc, Sp. Type: B2\,V$_0$, H-K$_s$ = 3.24\,mag.
           BP4(1a,b): 0.04, 0.05\,pc;  65, 245$^\circ$;
           BP4(2a,b): 0.30\,pc;	315, 130$^\circ$;
           BP2(3a-f): 0.38, 0.58\,pc; 170, 230, 260, 265, 320, 330$^\circ$;	7.5, 11.8.	}
  \label{figure:h2_map_A211}
\end{figure}

\setcounter{figure}{219}
\begin{figure}
  \resizebox{0.95\linewidth}{!}{\includegraphics{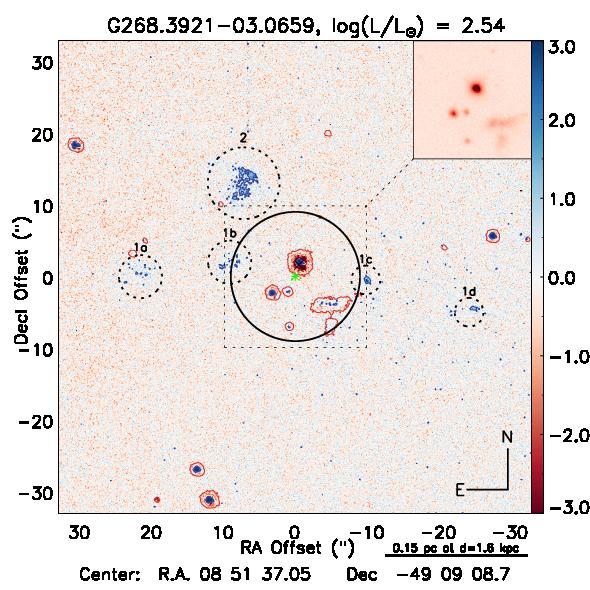}}
  \caption{G268.3921-03.0659, $\log(L/L_\odot)=2.54$, d=1.6\,kpc, Sp. Type: $<$B3\,V$_0$, H-K$_s$ = 0.71\,mag.
           BP2(1a,b,c,d): 0.18, 0.18\,pc;
           K(2): 0.12\,pc.          }
  \label{figure:h2_map_A220}
\end{figure}

\setcounter{figure}{227}
\begin{figure}
  \resizebox{0.95\linewidth}{!}{\includegraphics{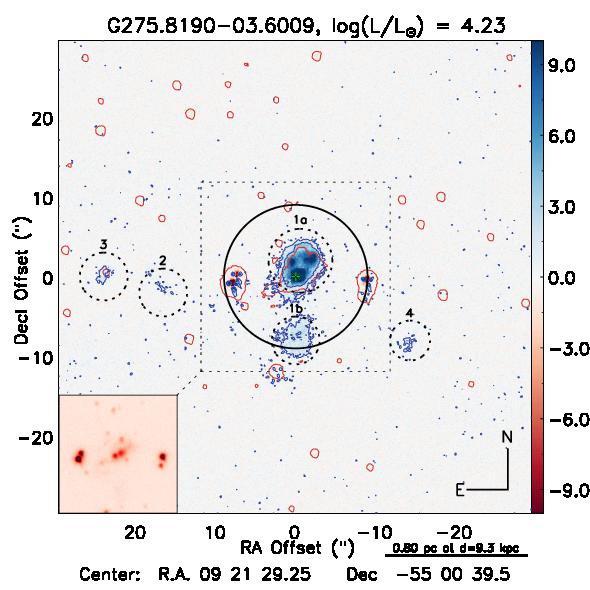}}
  \caption{G275.8190-03.6009, $\log(L/L_\odot)=4.23$, d=9.3\,kpc, Sp. Type: B0.5\,V$_0$, H-K$_s$ = 2.57\,mag.
           BP2(1a,1b): 0.2, 0.43\,pc;  0$^\circ$, 180$^\circ$; 1.0, 3.0;
           K(2): 0.74\,pc; 85$^\circ$; 
           K(3): 1.09\,pc; 85$^\circ$; 
           K(4): 0.71\,pc; 250$^\circ$.          }
  \label{figure:h2_map_A228}
\end{figure}

\setcounter{figure}{230}
\begin{figure}
  \resizebox{0.95\linewidth}{!}{\includegraphics{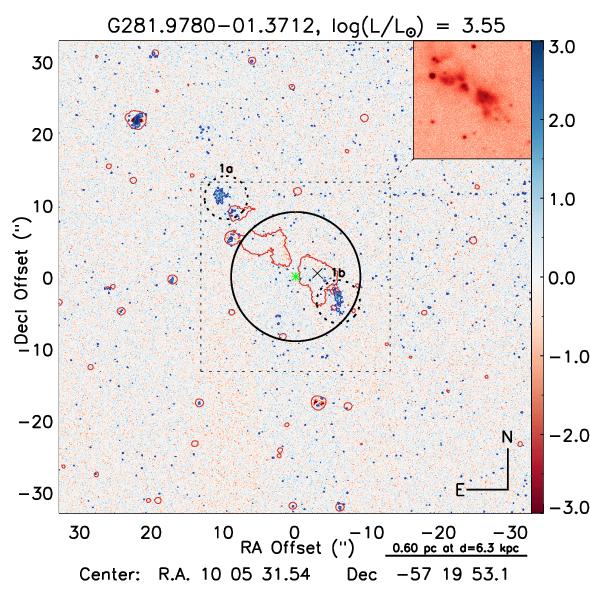}}
  \caption{G281.9780-01.3712, $\log(L/L_\odot)=3.55$, d=6.3\,kpc, Sp. Type: B2\,V$_0$, H-K$_s$ = 2.84\,mag.
           BP2(1a,1b): 0.43, 0.32\,pc; 50$^\circ$, 230$^\circ$; 5.7, 6.0.          }
  \label{figure:h2_map_A231}
\end{figure}

\setcounter{figure}{232}
\begin{figure}
  \resizebox{0.95\linewidth}{!}{\includegraphics{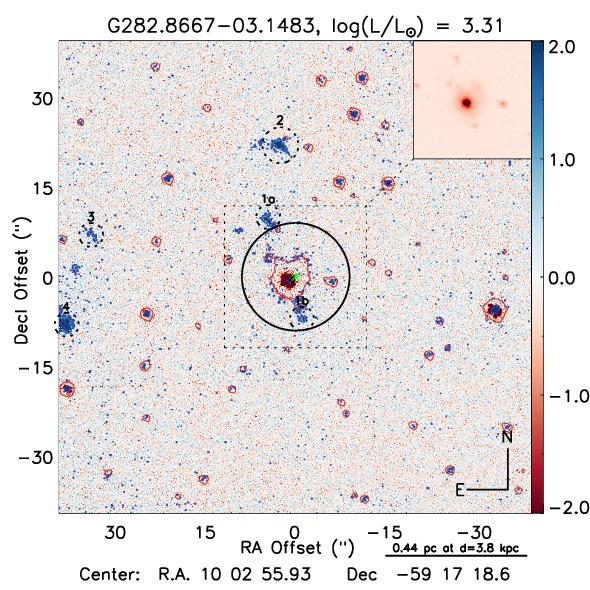}}
  \caption{G282.8667-03.1483, $\log(L/L_\odot)=3.85$, d=2.6\,kpc, Sp. Type: B3\,V$_0$ , H-K$_s$ = 3.45\,mag.
           BP2(1a,1b): 0.22, 0.13\,pc; 20$^\circ$, 200$^\circ$; 7.0, 5.3;
           K(2): 0.42\,pc; 5$^\circ$; 
           K(3): 0.63\,pc; 75$^\circ$;
           K(4): 0.69\,pc; 110$^\circ$.          }
  \label{figure:h2_map_A233}
\end{figure}

\setcounter{figure}{233}
\begin{figure}
  \resizebox{0.95\linewidth}{!}{\includegraphics{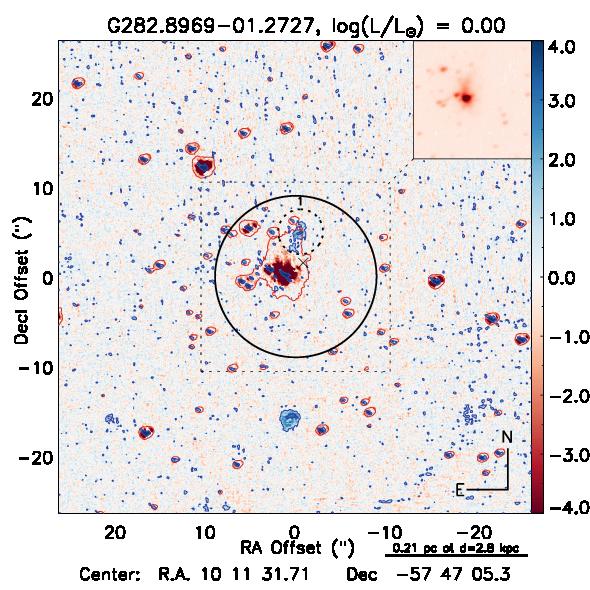}}
  \caption{G282.8969-01.2727, d=2.8\,kpc, H-K$_s$ = 2.12\,mag.
           BP1(1): 0.08\,pc; 350$^\circ$; 5.5.          }
  \label{figure:h2_map_A234}
\end{figure}

\setcounter{figure}{236}
\begin{figure}
  \resizebox{0.95\linewidth}{!}{\includegraphics{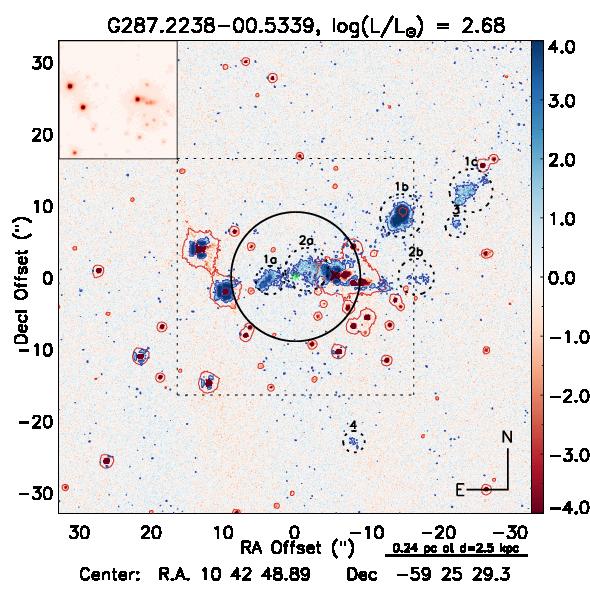}}
  \caption{G287.2238-00.5339, $\log(L/L_\odot)=2.68$, d=2.5\,kpc, Sp. Type: $<$B3\,V$_0$, H-K$_s$ = 1.43\,mag.
           BP4(1a,1b,1c): 0.21, 0.21\,pc;  115$^\circ$, 295$^\circ$; 6.3, 6.3;
           BP4(2a,2b): 0.13, 0.13\,pc; 85$^\circ$, 265$^\circ$; 3.8, 3.8;
           K(3): 0.23\,pc; 295$^\circ$;
           K(4): 0.29\,pc; 200$^\circ$.          }
  \label{figure:h2_map_A237}
\end{figure}

\setcounter{figure}{237}
\begin{figure}
  \resizebox{0.95\linewidth}{!}{\includegraphics{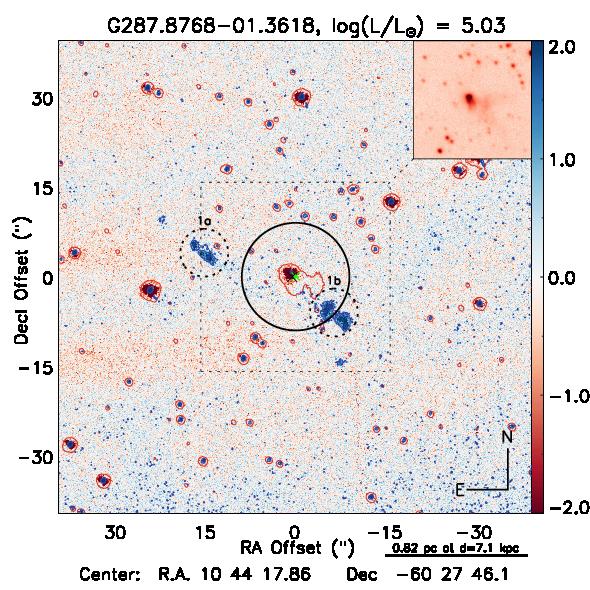}}
  \caption{G287.8768-01.3618, $\log(L/L_\odot)=3.85$, d=2.6\,kpc, Sp. Type: O7.5\,V, H-K$_s$ = 3.33\,mag.
           BP2(1a,1b): 0.47, 0.59\,pc; 65$^\circ$, 245$^\circ$; 5.3, 10.0.          }
  \label{figure:h2_map_A238}
\end{figure}

\setcounter{figure}{254}
\begin{figure}
  \resizebox{0.95\linewidth}{!}{\includegraphics{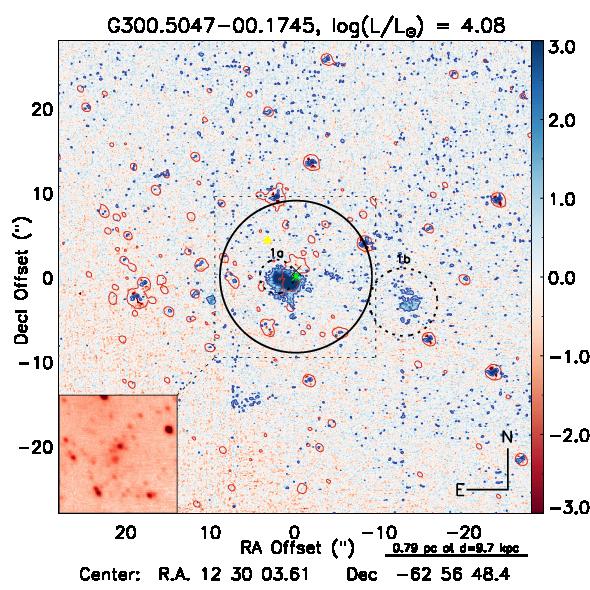}}
  \caption{G300.5047-00.1745, $\log(L/L_\odot)=4.08$, d=9.7\,kpc, Sp. Type: B0.5\,V$_0$, H-K$_s$ = 1.25\,mag.
           BP2(1a,1b): 0.11, 0.67\,pc; 80$^\circ$, 260$^\circ$; 1.0, 4.8.          }
  \label{figure:h2_map_A255}
\end{figure}

\setcounter{figure}{273}
\begin{figure}
  \resizebox{0.95\linewidth}{!}{\includegraphics{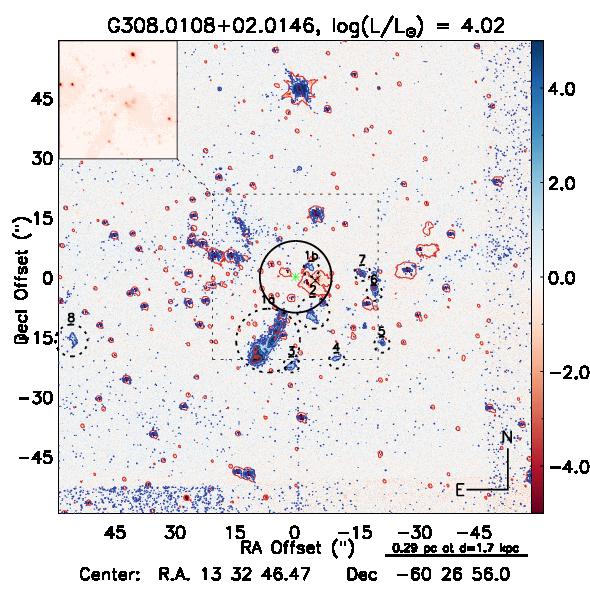}}
  \caption{G308.0108+02.0146, $\log(L/L_\odot)=4.02$, d=1.7\,kpc, Sp. Type: B1\,V$_0$, H-K$_s$ = 3.43\,mag.
           BP2(1a,1b): 0.20, 0.03\,pc; 150$^\circ$, 330$^\circ$; 4.0, 7.5;
           K(2): 0.08\,pc; 205$^\circ$; 
           K(3): 0.18\,pc; 175$^\circ$; 
           K(4): 0.18\,pc; 205$^\circ$; 
           K(5): 0.21\,pc; 235$^\circ$; 
           K(6): 0.14\,pc; 265$^\circ$; 
           K(7): 0.12\,pc; 280$^\circ$.          }
  \label{figure:h2_map_A274}
\end{figure}

\setcounter{figure}{286}
\begin{figure}
  \resizebox{0.95\linewidth}{!}{\includegraphics{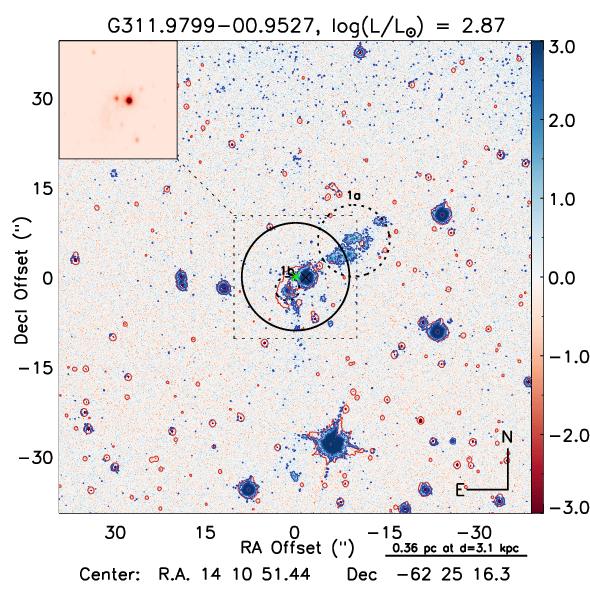}}
  \caption{G311.9799-00.9527, $\log(L/L_\odot)=2.87$, d=3.1\,kpc, Sp. Type: $<$B3\,V$_0$, H-K$_s$ = 0.93\,mag.
           BP2(1a,1b): 0.26, 0.06\,pc; 305$^\circ$, 125$^\circ$; 5.7, 2.0.          }
  \label{figure:h2_map_A287}
\end{figure}

\setcounter{figure}{291}
\begin{figure}
  \resizebox{0.95\linewidth}{!}{\includegraphics{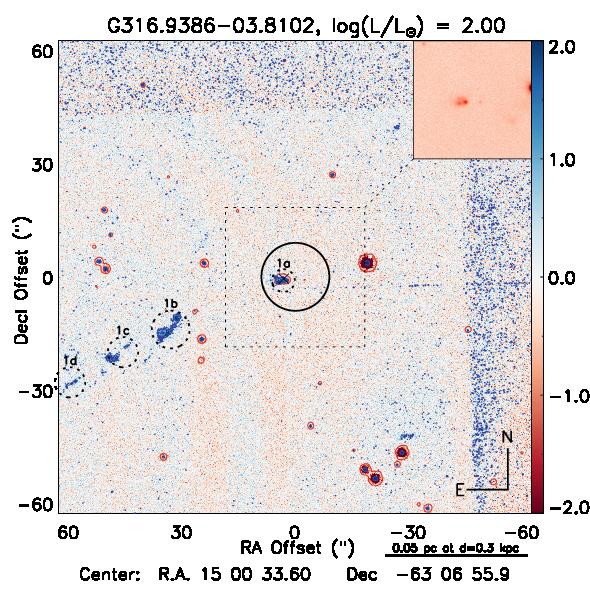}}
  \caption{G316.9386-03.8102, $\log(L/L_\odot)=2.00$, d=0.3\,kpc, Sp. Type: $<$B3\,V$_0$, H-K$_s$ = 1.97\,mag.
           BP1(1a,1b,1c,1d): 0.10\,pc; 115$^\circ$; 15.0.          }
  \label{figure:h2_map_A292}
\end{figure}

\setcounter{figure}{294}
\begin{figure}
  \resizebox{0.95\linewidth}{!}{\includegraphics{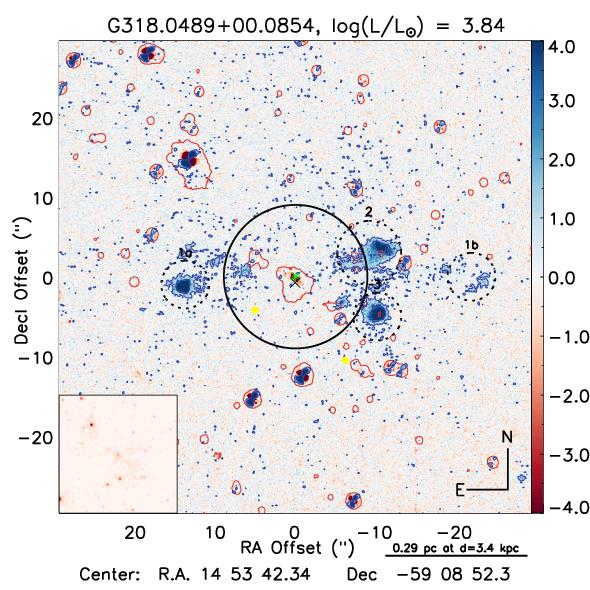}}
  \caption{G318.0489+00.0854, $\log(L/L_\odot)=3.84$, d=3.4\,kpc, Sp. Type: B1\,V$_0$, H-K$_s$ = 3.17\,mag.
           BP4(1a,1b,2,3): 0.23, 0.31\,pc; 90$^\circ$, 270$^\circ$, 290$^\circ$, 245$^\circ$; 7.3, 7.5.          }
  \label{figure:h2_map_A295}
\end{figure}

\setcounter{figure}{295}
\begin{figure}
  \resizebox{0.95\linewidth}{!}{\includegraphics{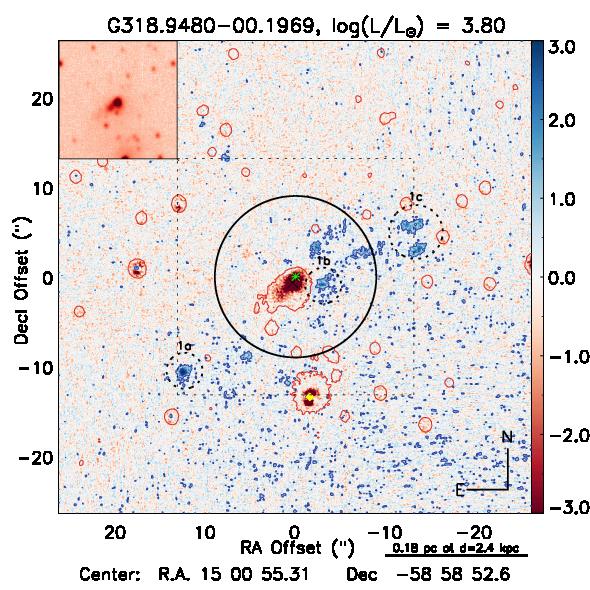}}
  \caption{G318.9480-00.1969, $\log(L/L_\odot)=3.80$, d=2.4\,kpc, Sp. Type: B1\,V$_0$ star, H-K$_s$ = 3.06\,mag.
           BP2(1a,1b,1c): 0.20, 0.18\,pc; 125$^\circ$, 300$^\circ$; 12.7, 4.4.          }
  \label{figure:h2_map_A296}
\end{figure}

\setcounter{figure}{306}
\begin{figure}
  \resizebox{0.95\linewidth}{!}{\includegraphics{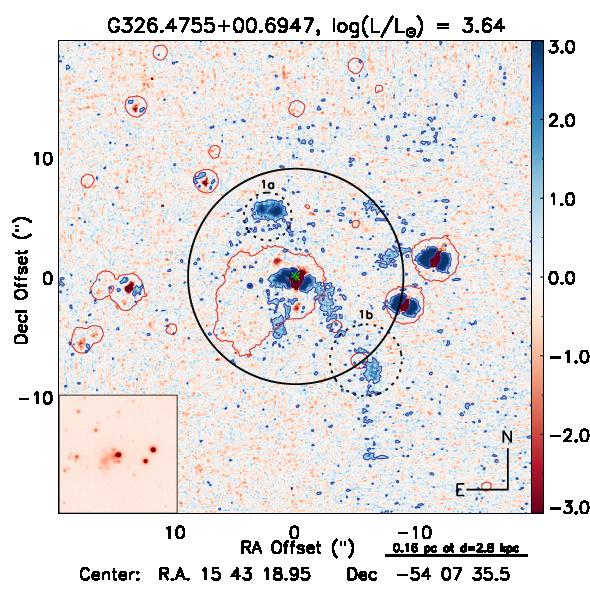}}
  \caption{G326.4755+00.6947, $\log(L/L_\odot)=3.64$, d=2.8\,kpc, Sp. Type: B2\,V$_0$, H-K$_s$ = 3.14\,mag.
           BP2(1a,1b): 0.09, 0.14\,pc; 35$^\circ$, 215$^\circ$; 3.0, 4.8.          }
  \label{figure:h2_map_A307}
\end{figure}

\setcounter{figure}{314}
\begin{figure}
  \resizebox{0.95\linewidth}{!}{\includegraphics{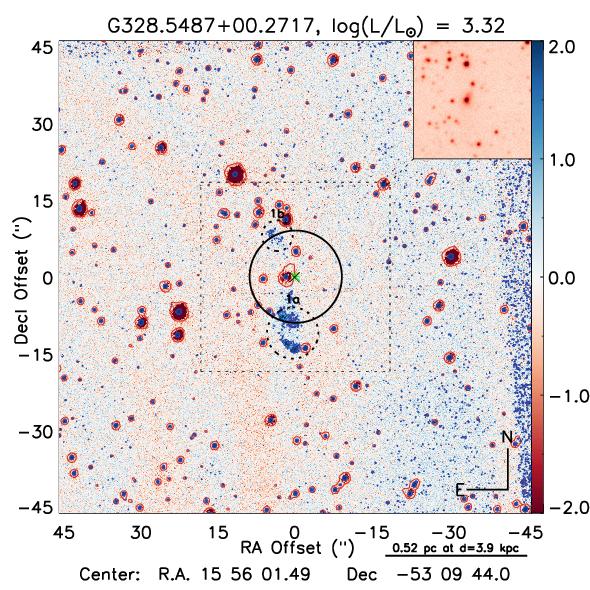}}
  \caption{G328.5487+00.2717, $\log(L/L_\odot)=3.32$, d=3.9\,kpc, Sp. Type: B3\,V$_0$, H-K$_s$ = 2.25\,mag.
           BP2(1a,1b): 0.35, 0.26\,pc; 10$^\circ$, 185$^\circ$; 4.2, 5.3.          }
  \label{figure:h2_map_A315}
\end{figure}

\setcounter{figure}{336}
\begin{figure}
  \resizebox{0.95\linewidth}{!}{\includegraphics{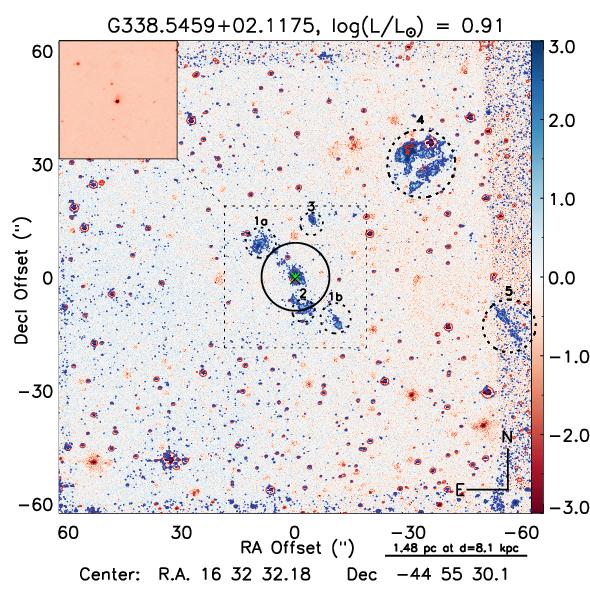}}
  \caption{G338.5459+02.1175, $\log(L/L_\odot)=0.91$, d=8.1\,kpc, Sp. Type: $<$B3\,V$_0$, H-K$_s$ = 1.42\,mag.
           BP2(1a,1b): 0.49, 0.61\,pc; 45$^\circ$, 225$^\circ$; 5.3, 6.7;
           K(2): 0.34\,pc; 205$^\circ$; 
           K(3): 0.62\,pc; 345$^\circ$; 
           D(4): 1.83\,pc; 315$^\circ$; 
           D(5): 2.14\,pc; 255$^\circ$.          }
  \label{figure:h2_map_A337}
\end{figure}

\setcounter{figure}{340}
\begin{figure}
  \resizebox{0.95\linewidth}{!}{\includegraphics{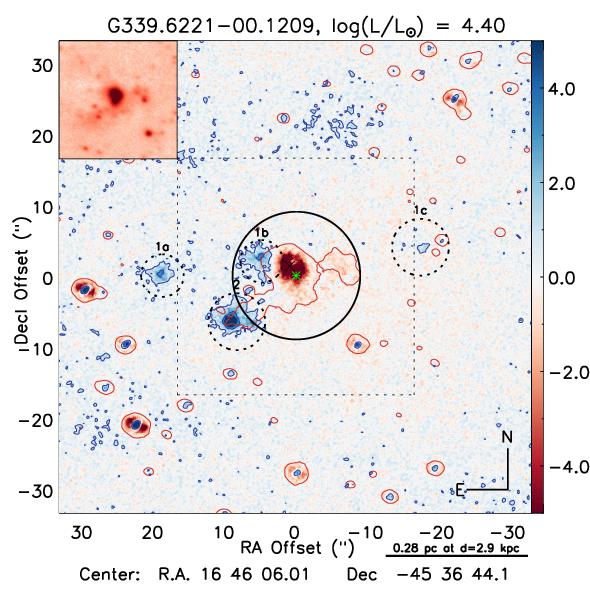}}
  \caption{G339.6221-00.1209, $\log(L/L_\odot) = 4.40$, d=2.9\,kpc, Sp. Type: B0\,V, H-K$_s$ = 2.49\,mag.
           BP2(1a,1b,1c): 0.30, 0.28\,pc;  90$^\circ$, 75$^\circ$, 275$^\circ$; 7.8, 9.3;
           K(2): 0.19\,pc; 135$^\circ$.          }
  \label{figure:h2_map_A341}
\end{figure}

\clearpage

\section{Full tables}
\label{appendix:tables}

\begin{table*}
	\caption{Log of observations.}
\begin{tabular}{clrrrrcc}
\hline\hline
ID	& MSX name	        &	R.A. (J2000)	    &	Decl. (J2000)	    &	d$_{kin}$ (kpc)&$\log(L/L_\odot)$	&	Telescope	&	Semester	\\
\hline\hline
001	&	G016.7981+00.1264	&	18:20:55.28	&	-14:15:30.80	&	14.6	&	5.14	&	SOAR	&	2011A	\\
002	&	G017.0332+00.7476	&	18:19:07.33	&	-13:45:23.60	&	13.8	&	4.64	&	SOAR	&	2011A	\\
003	&	G017.6380+00.1566	&	18:22:26.37	&	-13:30:12.00	&	2.3	&	4.77	&	SOAR	&	2011A	\\
004	&	G018.6608+00.0372	&	18:24:50.24	&	-12:39:22.40	&	10.8	&	4.29	&	CFHT	&	2012A	\\
005	&	G019.7268-00.1132	&	18:27:25.20	&	-11:46:59.10	&	11.7	&	4.53	&	CFHT	&	2012A	\\
006	&	G019.8817-00.5347	&	18:29:14.68	&	-11:50:23.60	&	3.5	&	3.94	&	CFHT	&	2012A	\\
007	&	G019.9224-00.2577	&	18:28:18.96	&	-11:40:36.70	&	11.4	&	4.29	&	CFHT	&	2012A	\\
\hline\hline
	\end{tabular}
	\label{table:observations_full}
\end{table*}

\begin{table*}
	\contcaption{Log of observations.}
	\begin{tabular}{clrrrrcc}
\hline\hline
ID	& MSX name	        &	R.A. (J2000)	    &	Decl. (J2000)	    &	d$_{kin}$ (kpc)&$\log(L/L_\odot)$	&	Telescope	&	Semester	\\
\hline\hline
065	&	G041.0780-00.6365	&	19:08:39.23	&	6:57:07.20	&	7.0	&	3.57	&	CFHT	&	2011B	\\
066	&	G042.0977+00.3521	&	19:07:00.53	&	8:18:44.30	&	11.2	&	4.81	&	SOAR	&	2011A	\\
067	&	G042.1099-00.4466	&	19:09:53.57	&	7:57:14.50	&	8.7	&	3.95	&	CFHT	&	2011B	\\
068	&	G043.0786+00.0033	&	19:10:05.01	&	9:01:15.60	&	0.9	&	1.63	&	CFHT	&	2011B	\\
069	&	G043.1497+00.0272	&	19:10:05.01	&	9:01:15.60	&	12.3	&	4.67	&	CFHT	&	2011B	\\
070	&	G043.1650-00.0285	&	19:10:21.80	&	9:05:02.90	&	11.6	&	6.28	&	CFHT	&	2011B	\\
071	&	G043.1679+00.0095	&	19:10:13.45	&	9:06:12.90	&	11.7	&	5.87	&	CFHT	&	2011B	\\
\hline\hline
	\end{tabular}
\end{table*}

\begin{table*}
	\contcaption{Log of observations.}
	\begin{tabular}{clrrrrcc}
\hline\hline
ID	& MSX name	        &	R.A. (J2000)	    &	Decl. (J2000)	    &	d$_{kin}$ (kpc)&$\log(L/L_\odot)$	&	Telescope	&	Semester	\\
\hline\hline
129	&	G109.8715+02.1156	&	22:56:17.97	&	62:01:54.40	&	1.5	&	4.58	&	CFHT	&	2010B	\\
130	&	G110.0931-00.0641	&	23:05:25.08	&	60:08:16.00	&	5.7	&	4.30	&	CFHT	&	2010B	\\
131	&	G110.1082+00.0473	&	23:05:10.15	&	60:14:42.70	&	5.6	&	4.69	&	CFHT	&	2010B	\\
132	&	G110.4771+01.4803	&	23:03:12.86	&	61:42:24.10	&	1.5	&	2.59	&	CFHT	&	2010B	\\
133	&	G111.5234+00.8004	&	23:13:32.39	&	61:29:06.20	&	6.2	&	4.56	&	CFHT	&	2010B	\\
134	&	G111.5320+00.7593	&	23:13:43.89	&	61:26:58.20	&	6.1	&	4.33	&	CFHT	&	2010B	\\
135	&	G111.5423+00.7776	&	23:13:45.36	&	61:28:10.30	&	6.1	&	5.90	&	CFHT	&	2010B	\\
\hline\hline
	\end{tabular}
\end{table*}

\begin{table*}
	\contcaption{Log of observations.}
	\begin{tabular}{clrrrrcc}
\hline\hline
ID	& MSX name	        &	R.A. (J2000)	    &	Decl. (J2000)	    &	d$_{kin}$ (kpc)&$\log(L/L_\odot)$	&	Telescope	&	Semester	\\
\hline\hline
193	&	G225.3266-00.5318	&	7:12:24.40	&	-11:15:34.90	&	1.6	&	2.87	&	CFHT	&	2011B	\\
194	&	G231.7986-01.9682	&	7:19:35.92	&	-17:39:18.00	&	4.3	&	3.74	&	CFHT	&	2011B	\\
195	&	G231.9005-02.1378	&	7:19:10.17	&	-17:49:27.40	&	4.4	&	2.96	&	CFHT	&	2011B	\\
196	&	G232.0766-02.2767	&	7:18:59.92	&	-18:02:41.20	&	4.1	&	3.67	&	CFHT	&	2011B	\\
197	&	G232.6207+00.9959	&	7:32:09.66	&	-16:58:12.80	&	1.6	&	4.00	&	CFHT	&	2011B	\\
198	&	G233.8306-00.1803	&	7:30:16.65	&	-18:35:48.80	&	4.4	&	4.27	&	CFHT	&	2011B	\\
199	&	G242.9402-00.4501	&	7:48:43.12	&	-26:39:31.30	&	6.7	&	3.71	&	SOAR	&	2011B	\\
\hline\hline
	\end{tabular}
\end{table*}

\begin{table*}
	\contcaption{Log of observations.}
	\begin{tabular}{clrrrrcc}
\hline\hline
ID	& MSX name	        &	R.A. (J2000)	    &	Decl. (J2000)	    &	d$_{kin}$ (kpc)&$\log(L/L_\odot)$	&	Telescope	&	Semester	\\
\hline\hline
257	&	G301.8147+00.7808	&	12:41:53.86	&	-62:04:14.60	&	4.4	&	4.33	&	SOAR	&	2012A	\\
258	&	G302.1515-00.9488	&	12:44:21.93	&	-63:48:36.00	&	12.1	&	3.83	&	SOAR	&	2012A	\\
259	&	G302.4546-00.7401	&	12:47:08.62	&	-63:36:30.00	&	12.6	&	4.36	&	SOAR	&	2012A	\\
260	&	G302.4867-00.0308	&	12:47:31.72	&	-62:53:57.40	&	4.5	&	4.10	&	SOAR	&	2012A	\\
261	&	G302.5005-00.7701	&	12:47:33.11	&	-63:38:19.30	&	12.1	&	3.94	&	SOAR	&	2012B	\\
262	&	G304.7592-00.6299	&	13:07:47.42	&	-63:26:37.00	&	12.2	&	3.64	&	SOAR	&	2012A	\\
263	&	G305.4748-00.0961	&	13:13:45.78	&	-62:51:27.60	&	6.1	&	3.57	&	SOAR	&	2012A	\\
\hline\hline
	\end{tabular}
\end{table*}

\begin{table*}
	\contcaption{Log of observations.}
	\begin{tabular}{clrrrrcc}
\hline\hline
ID	& MSX name	        &	R.A. (J2000)	    &	Decl. (J2000)	    &	d$_{kin}$ (kpc)&$\log(L/L_\odot)$	&	Telescope	&	Semester	\\
\hline\hline
321	&	G331.6191-00.0442	&	16:12:24.61	&	-51:21:42.40	&	10.2	&	3.98	&	SOAR	&	2012A	\\
322	&	G332.0939-00.4206	&	16:16:16.46	&	-51:18:25.20	&	11.2	&	5.88	&	SOAR	&	2011A	\\
323	&	G332.4683-00.5228	&	16:18:26.53	&	-51:07:12.00	&	11.5	&	4.53	&	SOAR	&	2011A	\\
324	&	G332.7013-00.5874	&	16:19:47.46	&	-51:00:06.60	&	11.3	&	4.63	&	SOAR	&	2011A	\\
325	&	G333.1075-00.5020	&	16:21:14.22	&	-50:39:12.60	&	3.7	&	3.50	&	SOAR	&	2012A	\\
326	&	G333.3151+00.1053	&	16:19:29.06	&	-50:04:40.80	&	3.3	&	3.67	&	SOAR	&	2012A	\\
327	&	G333.4747-00.2366	&	16:21:41.68	&	-50:12:30.90	&	11.7	&	4.52	&	SOAR	&	2011A	\\
\hline\hline
	\end{tabular}
\end{table*}

\begin{table*}
	\caption{General classification of the sources.}
    \begin{center}
	\begin{tabular}{cccccl}
\hline\hline
ID	&	Ext. Continuum	&	Stellar 	&	Primary H$_2$	&	Secondary H$_2$	&	Components	\\
	&	 Emission	&	Cluster		&	Classification	&	Classification	&			\\
\hline
001	&	Y	&	N	&	N	&	-	&	-	\\
002	&	N	&	N	&	BP2	&	D	&	BP2(1a,b), D(2a,b)	\\
003	&	Y	&	N	&	D	&	-	&	D(1)	\\
004	&	N	&	N	&	N	&	-	&	-	\\
005	&	N	&	N	&	N	&	-	&	-	\\
006	&	Y	&	N	&	BP2	&	K	&	BP2(1a-f),K(2,3,4)	\\
\hline\hline
	\end{tabular}
	\label{table:structures_full}
    \end{center}
\end{table*}

\begin{table*}
	\contcaption{General classification of the sources.}
    \begin{center}
	\begin{tabular}{cccccl}
\hline\hline
ID	&	Ext. Continuum	&	Stellar 	&	Primary H$_2$	&	Secondary H$_2$	&	Components	\\
	&	 Emission	&	Cluster		&	Classification	&	Classification	&			\\
\hline
064	&	N	&	N	&	N	&	-	&	-	\\
065	&	N	&	N	&	N	&	-	&	-	\\
066	&	N	&	Y	&	N	&	-	&	-	\\
067	&	N	&	N	&	N	&	-	&	-	\\
068	&	N	&	N	&	N	&	-	&	-	\\
069	&	N	&	N	&	N	&	-	&	-	\\
\hline\hline
	\end{tabular}
    \end{center}
\end{table*}

\begin{table*}
	\contcaption{General classification of the sources.}
    \begin{center}
	\begin{tabular}{cccccl}
\hline\hline
ID	&	Ext. Continuum	&	Stellar 	&	Primary H$_2$	&	Secondary H$_2$	&	Components	\\
	&	 Emission	&	Cluster		&	Classification	&	Classification	&			\\
\hline
127	&	Y	&	N	&	BP2	&	K,D	&	BP2(1a,b),K(2,3,4),BP2(128-1a,b),D(128-2)	\\
128	&	Y	&	Y	&	BP2	&	D,K	&	BP2(1a,b),D(2),K(3)	\\
129	&	Y	&	Y	&	BP2	&	-	&	BP2(1a,b)	\\
130	&	Y	&	Y	&	BP2	&	K	&	BP2(1a,b), K(2)	\\
131	&	Y	&	Y	&	D	&	-	&	D(1a,b,c), D(2a,b,c)	\\
132	&	Y	&	N	&	BP2	&	-	&	BP2(1a,b,c)	\\
\hline\hline
	\end{tabular}
    \end{center}
\end{table*}

\begin{table*}
	\contcaption{General classification of the sources.}
    \begin{center}
	\begin{tabular}{cccccl}
\hline\hline
ID	&	Ext. Continuum	&	Stellar 	&	Primary H$_2$	&	Secondary H$_2$	&	Components	\\
	&	 Emission	&	Cluster		&	Classification	&	Classification	&			\\
\hline
189	&	N	&	N	&	BP2	&	-	&	BP2(1a,b)	\\
190	&	Y	&	N	&	K	&	-	&	K(1,2,3)	\\
191	&	N	&	N	&	N	&	-	&	-	\\
192	&	Y	&	N	&	K	&	-	&	K(1,2)	\\
193	&	Y	&	N	&	BP2	&	K	&	BP2(1a,b),K(2)	\\
194	&	N	&	N	&	N	&	-	&	-	\\
\hline\hline
	\end{tabular}
    \end{center}
\end{table*}

\begin{table*}
	\contcaption{General classification of the sources.}
    \begin{center}
	\begin{tabular}{cccccl}
\hline\hline
ID	&	Ext. Continuum	&	Stellar 	&	Primary H$_2$	&	Secondary H$_2$	&	Components	\\
	&	 Emission	&	Cluster		&	Classification	&	Classification	&			\\
\hline
252	&	Y	&	Y	&	N	&	-	&	-	\\
253	&	Y	&	Y	&	N	&	-	&	-	\\
254	&	N	&	Y	&	N	&	-	&	-	\\
255	&	N	&	N	&	BP2	&	-	&	BP2(1a,b)	\\
256	&	Y	&	N	&	N	&	-	&	-	\\
257	&	N	&	Y	&	N	&	-	&	-	\\
\hline\hline
	\end{tabular}
    \end{center}
\end{table*}

\begin{table*}
	\contcaption{General classification of the sources.}
    \begin{center}
	\begin{tabular}{cccccl}
\hline\hline
ID	&	Ext. Continuum	&	Stellar 	&	Primary H$_2$	&	Secondary H$_2$	&	Components	\\
	&	 Emission	&	Cluster		&	Classification	&	Classification	&			\\
\hline
315	&	Y	&	N	&	BP2	&	-	&	BP2(1a,b)	\\
316	&	N	&	N	&	N	&	-	&	-	\\
317	&	N	&	N	&	N	&	-	&	-	\\
318	&	N	&	N	&	N	&	-	&	-	\\
319	&	Y	&	Y	&	D	&	-	&	D(1,2)	\\
320	&	Y	&	N	&	N	&	-	&	-	\\
321	&	N	&	N	&	N	&	-	&	-	\\
322	&	Y	&	N	&	N	&	-	&	-	\\
323	&	Y	&	N	&	D	&	-	&	D(1)	\\
324	&	Y	&	N	&	N	&	-	&	-	\\
325	&	Y	&	N	&	K	&	D	&	K(1),D(2a,b,c,d)	\\
326	&	Y	&	N	&	K	&	-	&	K(1,2)	\\
327	&	N	&	N	&	N	&	-	&	-	\\
328	&	N	&	N	&	N	&	-	&	-	\\
329	&	N	&	Y	&	N	&	-	&	-	\\
330	&	N	&	N	&	N	&	-	&	-	\\
331	&	N	&	N	&	N	&	-	&	-	\\
332	&	Y	&	N	&	N	&	-	&	-	\\
333	&	N	&	N	&	N	&	-	&	-	\\
334	&	N	&	N	&	N	&	-	&	-	\\
335	&	N	&	N	&	N	&	-	&	-	\\
336	&	Y	&	N	&	K	&	-	&	BP2(1a,b,c),K(2)	\\
337	&	N	&	N	&	BP2	&	K,D	&	BP2(1a,b),K(2,3),D(4,5)	\\
338	&	Y	&	N	&	K	&	-	&	K(1)	\\
339	&	Y	&	N	&	K	&	-	&	K(1,2,3)	\\
340	&	N	&	N	&	N	&	-	&	-	\\
341	&	Y	&	N	&	BP2	&	K	&	BP2(1a,b,c),K(2)	\\
342	&	N	&	N	&	N	&	-	&	-	\\
343	&	N	&	N	&	D	&	-	&	D(not labelled)	\\
344	&	Y	&	N	&	N	&	-	&	-	\\
345	&	N	&	N	&	K	&	-	&	K(1)	\\
346	&	N	&	N	&	N	&	-	&	-	\\
347	&	Y	&	N	&	D	&	K	&	D(1)	\\
348	&	Y	&	N	&	N	&	-	&	-	\\
349	&	N	&	N	&	N	&	-	&	-	\\
350	&	Y	&	N	&	N	&	-	&	-	\\
351	&	Y	&	N	&	D	&	-	&	D(1)	\\
352	&	N	&	N	&	N	&	-	&	-	\\
353	&	N	&	N	&	N	&	-	&	-	\\
\hline\hline
	\end{tabular}
    \end{center}
\end{table*}

\clearpage

\begin{table*}
	\label{table:structures_parameters}
	\begin{scriptsize}
	\begin{center}
	\caption{Properties of the structures identified on each field.}
	\begin{tabular}{ccccc|ccccc}
\hline\hline
ID  & Structure & $\ell_{proj}$ (pc) & PA ($^\circ$) & R & ID  & Structure & $\ell_{proj}$ (pc) & PA ($^\circ$) & R \\
\hline
002	&	BP2(1a,b)	&	1.26, 0.91	&	50, 265	&	11.0, 8.0	&	108	&	D(2)	&	1.87	&	30-200	&	-	\\
002	&	D(2a,b)	&	3.96	&	200-320	&	-	&	108	&	D(3)	&	0.67	&	300-320	&	-	\\
003	&	D(1)	&	0.51	&	0-360	&	-	&	109	&	K(1)	&	0.16	&	115	&	-	\\
006	&	BP2(1a-f)	&	0.68, 0.74	&	70, 285	&	5.5, 5.9	&	110	&	K(1)	&	0.08	&	135	&	-	\\
006	&	K(2)	&	0.84	&	50	&	-	&	110	&	D(2)	&	0.10	&	45	&	-	\\
006	&	K(3)	&	0.68	&	290	&	-	&	110	&	K(3)	&	0.31	&	90	&	-	\\
006	&	K(4)	&	0.96	&	330	&	-	&	111	&	BP2(1a,b)	&	0.45, 0.84	&	45, 240	&	3.5, 7.4	\\
\hline\hline
	\end{tabular}
	\end{center}
\end{scriptsize}
\end{table*}

\begin{table*}
	\label{table:structures_parameters}
	\begin{scriptsize}
	\begin{center}
	\contcaption{Properties of the structures identified on each field.}
	\begin{tabular}{ccccc|ccccc}
\hline\hline
ID  & Structure & $\ell_{proj}$ (pc) & PA ($^\circ$) & R & ID  & Structure & $\ell_{proj}$ (pc) & PA ($^\circ$) & R \\
\hline
145	&	K(4)	&	1.35	&	235	&	-	&	171	&	K(2-12)	&	-	&	-	&	-	\\
145	&	K(5)	&	1.13	&	305	&	-	&	171	&	D(13a,b,c)	&	3.15	&	-	&	-	\\
146	&	BP1(1a,b)	&	2.51, 2.01	&	350, 340	&	6.3, 8.0	&	175	&	K(1a)	&	0.38	&	280	&	-	\\
146	&	D(2a,b,c)	&	5.24	&	100-225	&	-	&	175	&	K(1b)	&	0.13	&	95	&	-	\\
149	&	BP1(1)	&	0.38	&	310	&	2	&	175	&	K(2a)	&	0.38	&	125	&	-	\\
149	&	K(2)	&	0.92	&	320	&	-	&	175	&	K(2b)	&	0.27	&	260	&	-	\\
149	&	D(3a,b,c,d)	&	4.88	&	-	&	-	&	175	&	K(3)	&	0.46	&	200	&	-	\\
\hline\hline
	\end{tabular}
	\end{center}
\end{scriptsize}
\end{table*}

\begin{table*}
	\label{table:structures_parameters}
	\begin{scriptsize}
	\begin{center}
	\contcaption{Properties of the structures identified on each field.}
	\begin{tabular}{ccccc|ccccc}
\hline\hline
ID  & Structure & $\ell_{proj}$ (pc) & PA ($^\circ$) & R & ID  & Structure & $\ell_{proj}$ (pc) & PA ($^\circ$) & R \\
\hline
214	&	K(1c-f)	&	0.53	&	90	&	-	&	269	&	K(1)	&	0.51	&	155	&	-	\\
214	&	K(2)	&	0.48	&	70	&	-	&	274	&	BP2(1a,b)	&	0.20, 0.03	&	150, 330	&	4.0, 7.5	\\
214	&	K(3)	&	0.18	&	10	&	-	&	274	&	K(2)	&	0.08	&	205	&	-	\\
214	&	K(4)	&	0.30	&	315	&	-	&	274	&	K(3)	&	0.18	&	175	&	-	\\
215	&	K(1)	&	0.20	&	80	&	-	&	274	&	K(4)	&	0.18	&	205	&	-	\\
215	&	D(2a,b)	&	0.68	&	260, 290	&	-	&	274	&	K(5)	&	0.21	&	235	&	-	\\
216	&	K(1)	&	0.14	&	90	&	-	&	274	&	K(6)	&	0.14	&	265	&	-	\\
\hline\hline
	\end{tabular}
	\end{center}
\end{scriptsize}
\end{table*}

\clearpage

\begin{table}
	\caption{Properties of the stellar clusters identified on each field.}
	\begin{tabular}{cccccc}
\hline\hline
ID  & RA            &	Dec     & d     & $\ell_{proj}$ & $\Delta\theta$ \\
    & (J2000)       &	(J2000)	& (kpc) & (pc)          & ('') \\
\hline
22	&	18 36 47	&	-07 35 39	&	9.6	&	0.85	&	20	\\
34	&	18 43 46	&	-03 35 29	&	8.7	&	1.35	&	45	\\
39	&	18 47 36	&	-02 01 50	&	9.1	&	0.95	&	25	\\
66	&	19 07 01	&	08 18 44	&	11.2	&	0.47	&	20	\\
81	&	19 44 24	&	25 48 43	&	15.9	&	0.96	&	15	\\
85	&	20 21 55	&	39 59 45	&	4.5	&	0.70	&	35	\\
\hline\hline
	\end{tabular}
	\label{table:clusters}
\end{table}

\begin{table}
	\contcaption{Properties of the stellar clusters identified on each field.}
	\begin{tabular}{cccccc}
\hline\hline
ID  & RA            &	Dec     & d     & $\ell_{proj}$ & $\Delta\theta$ \\
    & (J2000)       &	(J2000)	& (kpc) & (pc)          & ('') \\
\hline
199	&	07 48 43	&	-26 39 31	&	6.7	&	1.42	&	45	\\
200	&	08 02 43	&	-34 31 49	&	5.9	&	0.95	&	35	\\
201	&	08 15 57	&	-36 08 08	&	4.1	&	0.65	&	35	\\
202	&	08 26 17	&	-40 48 40	&	7.0	&	1.17	&	35	\\
205	&	08 29 14	&	-41 10 48	&	7.0	&	0.85	&	25	\\
211	&	08 48 49	&	-43 32 29	&	2.6	&	1.10	&	90	\\
215	&	08 59 28	&	-43 45 18	&	2.0	&	1.38	&	150	\\
217	&	08 58 04	&	-47 23 10	&	2.1	&	1.1	&	115	\\
219	&	09 00 41	&	-47 26 02	&	2.1	&	0.74	&	75	\\
224	&	09 04 29	&	-48 49 25	&	2.3	&	0.3	&	30	\\
226	&	09 11 08	&	-48 15 56	&	10.2	&	0.97	&	20	\\
229	&	09 59 16	&	-56 54 45	&	1.6	&	0.48	&	60	\\
230	&	10 01 34	&	-56 47 18	&	0.2	&	0.03	&	25	\\
235	&	10 26 36	&	-58 00 30	&	6.2	&	0.41	&	15	\\
237	&	10 42 49	&	-59 25 29	&	2.5	&	0.82	&	70	\\
245	&	11 30 07	&	-62 03 13	&	3.3	&	0.47	&	50	\\
246	&	11 34 33	&	-63 58 03	&	0.8	&	0.17	&	45	\\
250	&	11 53 08	&	-63 08 56	&	10.4	&	0.4	&	10	\\
252	&	11 56 50	&	-63 32 05	&	9.9	&	1.50	&	25	\\
253	&	12 15 25	&	-63 01 17	&	11.1	&	1.68	&	30	\\
254	&	12 28 36	&	-62 58 33	&	4.2	&	0.84	&	40	\\
257	&	12 41 54	&	-62 04 15	&	4.4	&	0.63	&	30	\\
259	&	12 47 09	&	-63 36 29	&	12.6	&	0.92	&	15	\\
266	&	13 14 26	&	-62 44 30	&	5.9	&	0.50	&	20	\\
274	&	13 32 47	&	-60 26 56	&	1.7	&	0.73	&	90	\\
305	&	15 31 44	&	-56 30 27	&	10.4	&	3.45	&	65	\\
307	&	15 43 19	&	-54 07 35	&	2.8	&	0.42	&	30	\\
319	&	16 06 26	&	-50 43 22	&	4.8	&	0.93	&	40	\\
329	&	16 25 40	&	-48 55 16	&	10.4	&	0.45	&	10	\\
\hline\hline
	\end{tabular}
\end{table}

\clearpage
\bsp

\label{lastpage}

\end{document}